\documentclass[review,3p]{elsarticle}

\usepackage[T1]{fontenc}
\usepackage{makecell}
\usepackage{amsmath,amssymb}
\usepackage{amsfonts}        
\usepackage{mathtools}
\usepackage{color}
\usepackage{graphicx}
\usepackage{dcolumn}
\usepackage{bm}
\usepackage{multirow}
\usepackage{nicefrac}
\usepackage{soul}
\usepackage{booktabs}
\usepackage{blindtext}
\usepackage{todonotes}
\usepackage{lineno}
\usepackage{comment}
\usepackage{physics}
\usepackage{tensor}
\usepackage[colorlinks=true,citecolor=blue,linkcolor=blue,urlcolor=blue]{hyperref}
\usepackage{tabularx}
\usepackage{gensymb}
\biboptions{sort&compress}
\usepackage{tikz}
\usetikzlibrary{mindmap, backgrounds}
\usepackage{xcolor}
\newcommand{\whitelink}[2]{\hypersetup{linkcolor=white}\hyperref[#1]{\textcolor{white}{#2}}\hypersetup{linkcolor=blue}} 
\newcommand{\blacklink}[2]{\hypersetup{linkcolor=black}\hyperref[#1]{\textcolor{black}{#2}}\hypersetup{linkcolor=blue}} 

\usepackage{babel}
\makeatother

\begin{document}


\title{Roto-translational levitated optomechanics}

\author[UCL]{M. Rademacher\corref{mycorrespondingauthor}}
\cortext[mycorrespondingauthor]{m.rademacher.18@ucl.ac.uk}
\author[CNR-INO]{A. Pontin}
\author[UCL]{J. M. H. Gosling}
\author[UCL]{P. F. Barker}
\author[UNI-LJ]{M. Toro\v{s}\corref{markocorrespondingauthor}}
\cortext[markocorrespondingauthor]{marko.toros@fmf.uni-lj.si}

\address[UCL]{Department  of  Physics  \&  Astronomy,University  College  London,  WC1E  6BT, United Kingdom}
\address[CNR-INO]{CNR-INO, largo Enrico Fermi 6, I-50125 Firenze, Italy}
\address[UNI-LJ]{Faculty of Mathematics and Physics, University of Ljubljana, Jadranska 19, SI-1000 Ljubljana, Slovenia}

\begin{abstract}

Levitated optomechanics, the interaction between light and small levitated objects, is a new macroscopic quantum system that is being used as a testing ground for fundamental physics and for the development of sensors with exquisite sensitivity. The utility of this system, when compared to other quantum optomechanical systems, is its extreme isolation from the environment and, by the relatively few degrees of freedom that a levitated object has. While work in the field has strongly focused on the three translational degrees of freedom of this system, it has become increasingly important to understand the induced rotational motion of levitated objects, particularly in optical trapping fields, but also in magnetic and electric traps. These additional three degrees of freedom, which are intrinsic to levitated systems, offer a new set of optomechanical nonlinear interactions that lead to a rich and yet largely unexplored roto-translational motion. The control and utilization of these interactions promise to extend the utility of levitated optomechanics in both fundamental studies and applications.  

In this review, we provide a brief overview of levitated optomechanics, before focusing on the roto-translational motion of optically levitated anisotropic objects. We first present a classical treatment of this induced motion, bridging the gap between classical and quantum formalisms. We describe the different types of roto-translational motion for different particle shapes via their interaction with polarized optical trapping fields. Subsequently, we provide an overview of the theoretical and experimental approaches as well as applications that have established this new field. The review concludes with an outlook of promising experiments and applications as well as the outstanding theoretical and experimental challenges that must be overcome to meet them. This includes the creation of non-classical states of roto-translational motion, quantum-limited torque sensing and particle characterization methods.  

\end{abstract}
\maketitle
\newpage

\newpage

\tableofcontents

\newpage
\begin{table}[htbp]
  \centering
  \label{tab:symbols}
  \setlength{\tabcolsep}{8pt}
  \renewcommand{\arraystretch}{1.3}
  \begin{tabular}{@{}ll@{}}
    \toprule
    \textbf{Symbol} & \textbf{Description} \\
    \midrule
    $\bm{E}_d$ & Driving electric field (tweezer field) \\
    $\bm{E}_f$ & Scattered electric field from the nanoparticle \\
    $u(\bm{r})$ & Gaussian beam mode function \\
    $b_x,\,b_y$ & Polarization coefficients ($b_x^2+b_y^2=1$) \\
    $\bm{\epsilon}_d$ & Polarization vector, defined as $(b_x,\, i\,b_y,\,0)^\top$ \\
    $P$ & Laser power \\
    $\sigma_L$ & Effective beam cross-section, $\pi w_0^2$ \\
    $w_0$ & Beam waist \\
    $z_R$ & Rayleigh range \\
    $k_L$ & Laser wave number, $2\pi/\lambda$ \\
    $\lambda$ & Laser wavelength \\
    $V$ & Nanoparticle volume \\
    $\epsilon_0$ & Vacuum permittivity \\
    $c$ & Speed of light \\
    $\bm{r}=(x,y,z)$ & Center-of-mass position (lab frame) \\
    $\bm{p}=(p_x,p_y,p_z)$ & Conjugate momentum for $\bm{r}$ \\
    $\bm{\phi}=(\alpha,\beta,\gamma)$ & Euler angles (orientation in the z–y'–z" convention) \\
    $\bm{\pi}=(\pi_\alpha,\pi_\beta,\pi_\gamma)$ & Conjugate angular momenta for $\bm{\phi}$ \\
    $\mathcal{R}$ & Rotation matrix (body \(\to\) lab frame) \\
    $I$ & Moment of inertia tensor (body frame), \(I=\mathrm{diag}(I_1,I_2,I_3)\) \\
    $\chi$ & Susceptibility tensor (body frame), \(\chi=\mathrm{diag}(\chi_1,\chi_2,\chi_3)\) \\
    $\chi_{\rm lab}$ & Susceptibility tensor in the lab frame, given by $\chi_{\rm lab}=\mathcal{R}\chi\,\mathcal{R}^\top$ \\
    $H_{\rm gradient}$ & Optical gradient potential, 
    $H_{\rm gradient}=-\tfrac{1}{2}V\epsilon_0\,\bm{E}_d^\top \chi_{\rm lab}\,\bm{E}_d$ \\
    $\gamma_s$ & Photon scattering rate \\
    $\gamma_c$ & Gas-collision damping rate \\
    $F^{(t)}$ & Translational friction tensor \\
    $F^{(r)}$ & Rotational friction tensor \\
    \midrule
    \multicolumn{2}{l}{\textbf{Acronyms:}} \\
    CoM   & Center of mass \\
    DoF   & Degrees of Freedom \\
    NV    & Nitrogen–vacancy center \\
    SQUID & Superconducting Quantum Interference Device \\
    PSD   & Power Spectral Density \\
    PLL   & Phase-Locked Loop \\
    SME   & Stochastic master equation \\
    \bottomrule
  \end{tabular}
  \caption{List of symbols and acronyms used in this review.}
\end{table}
\clearpage



%
\section{Introduction}
\label{sec:introduction}
Optomechanics explores the interaction between light and the mechanical motion of objects from the centimeter scale down to atomic scales~\cite{barzanjeh2022optomechanics,gonzalez2021levitodynamics}. Optomechanical systems are typically high Q mechanical oscillators whose interaction with light is enhanced by coupling their motion to an optical cavity in which the mechanical motion modulates the properties of the cavity while back action of the light can be used to control and even cool the mechanical motion of the oscillator~\cite{aspelmeyer2014cavity}. Optomechanical systems have now been developed into sensitive and versatile platforms for investigating fundamental physics~\cite{bowen2015quantum}, and particularly for studying quantum mechanics in the macroscopic limit~\cite{bild2023scat}. Increasingly, the ability to control and measure small perturbations to their motion has led to the development of sensors. For example, the ability to engineer the properties of these macroscopic systems allows them to be designed for optimal sensitivity to electrical~\cite{Bagci2014optical}, magnetic~\cite{Li2018quantum}  or inertial forces~\cite{Krause2012ahigh,Gavartin2012ahybrid}. They also show promise for transduction between microwave and optical fields~\cite{jiang2020efficient}, even to the single photon level~\cite{nunnenkamp2011singlephoton}. 

Almost all of these oscillators are clamped systems in which the oscillator of interest is physically attached to a larger system. This not only often reduces the mechanical and thermal isolation of the oscillator from the environment, but also can lead to a system with a large number of coupled mechanical modes of which only a few can interact and be controlled by an optical field.  Levitated optomechanics ameliorates some of these effects by forming and suspending oscillators in electromagnetic fields within a vacuum.  This approach further isolates the oscillator from the thermal and mechanical environment and, like in the case of trapped atoms, ions or molecules, the number of mechanical degrees of freedom of these oscillators is drastically reduced. For nanoscale and microscale systems that can be levitated, there are now only six important degrees of freedom that must be considered, since vibrational motion is at significantly higher frequencies~\cite{gonzalez2021levitodynamics,chang2010cavity,barker2010cavity}.  In such systems, the translational and rotational degrees of freedom can be optically cooled and controlled, enabling precise manipulation of the position, orientation and spin of the particles~\cite{pontin2023simultaneous,piotrowski2023simultaneous,magrini2021realtime,delic2020cooling}. Although to date the field of levitated optomechanics has mainly concentrated on the translational motion of these systems, this review will focus on the developing understanding and outlining the applications of the simultaneous translational and rotational motion of these levitated systems. This more detailed approach is often required to understand the motion of these systems, but we show that it also opens up new avenues and opportunities in both quantum and fundamental physics, as well as in sensing and particle characterization~\cite{stickler2021quantum,stickler2018probing,perdenales2020decoherence}.

\subsection{Purpose of this review}
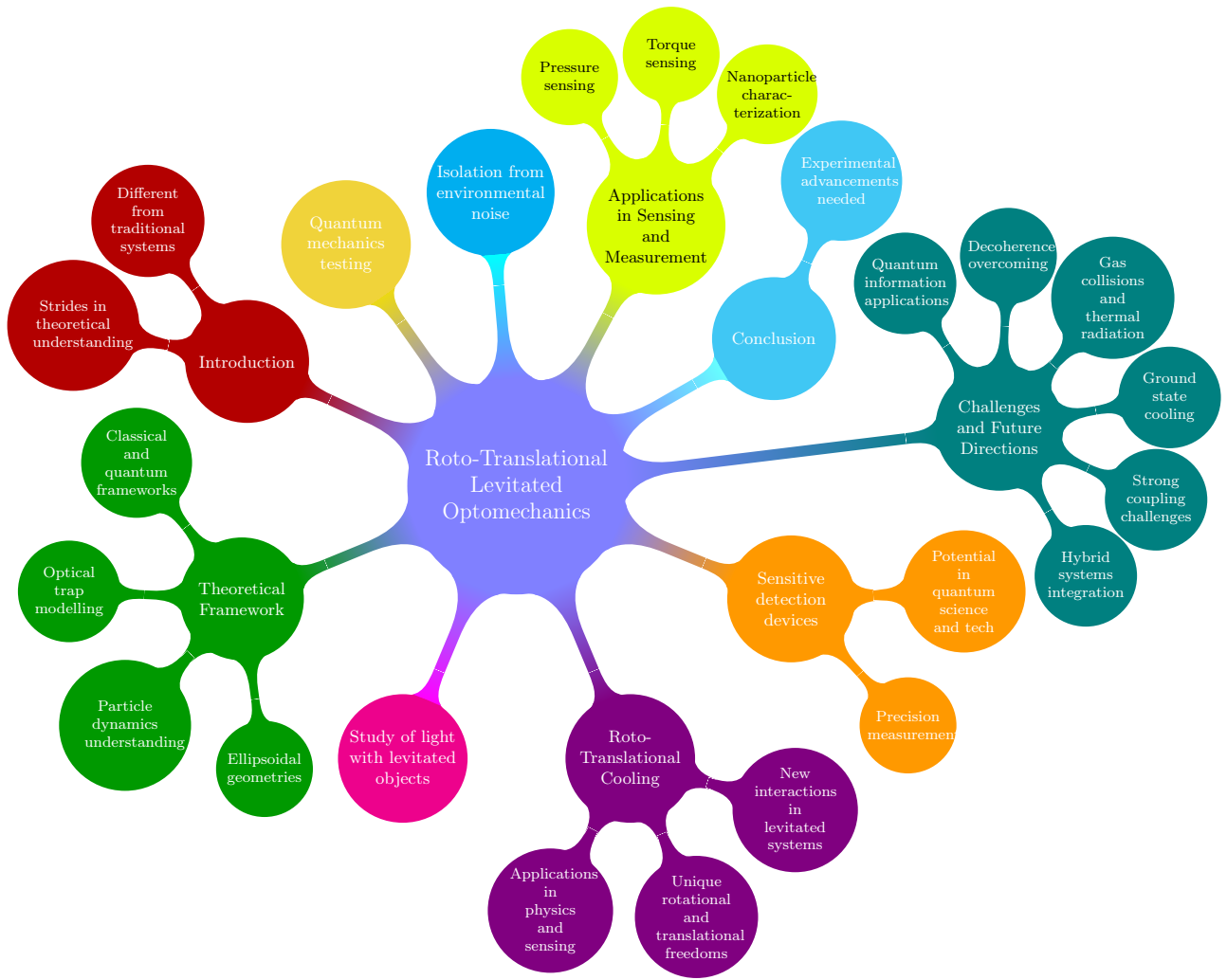
\begin{figure}
    \centering 
    \scalebox{0.75}{
    \begin{tikzpicture}[mindmap, grow cyclic, every node/.style=concept,
  concept color=blue!50, text=white, level 1/.append style={level distance=5.5cm, sibling angle=45},
  level 2/.append style={level distance=3.2cm, sibling angle=50}]

\node[concept] {Roto-Translational \\ Levitated \\ Optomechanics}
  child[concept color=red!70!black, grow=155] { node {\whitelink{sec:introduction}{Introduction}}
    child[grow=125] { node {\whitelink{sec:techniques}{Different from \\ traditional systems}} }
    child[minimum size=2.4cm] { node {\whitelink{sec:Levitation}{Strides in \\ theoretical understanding}} }
  }
  child[concept color=green!60!black] { node {\whitelink{sec:theory}{Theoretical \\ Framework}}
    child { node {\whitelink{sec:quantum_model}{Classical and \\ quantum frameworks}} }
    child { node {\whitelink{sec:model}{Optical trap modelling}} }
    child[minimum size=2.4cm] { node {\whitelink{sec:Numerical-analysis}{Particle dynamics \\ understanding}} }
    child { node {\whitelink{sec:top}{Ellipsoidal \\ geometries}} }
  }
  child[concept color=magenta] { node {\whitelink{sec:det_contr_man}{Study of light \\ with levitated objects}} }
  child[concept color=violet] { node {\whitelink{controlANDmanipualtion}{Roto-Translational \\ Cooling}}
    child { node {\whitelink{sec:sensing}{Applications in \\ physics and sensing}} }
    child { node {\whitelink{othersystems}{Unique rotational and \\ translational freedoms}} }
    child { node {\whitelink{sec:rot_intro}{New interactions in \\ levitated systems}} }
  }
  child[concept color=orange!80!yellow] { node {\whitelink{sec:sensing}{Sensitive \\ detection devices}}
    child { node {\whitelink{sec:inertial_sensing}{Precision \\ measurements}} }
    child { node {\whitelink{sec:roto_quantum}{Potential in quantum \\ science and tech}} }
  }
  child[concept color=cyan!60,grow=30] { node {\whitelink{sec:conclusion}{Conclusion}}
    child[grow=67, minimum size=2.2cm] { node {\whitelink{sec:discussion}{Experimental \\ advancements needed}} }
  }
  child[concept color=lime!60!yellow, text=black, grow=62] { node {\blacklink{sec:applications}{Applications in Sensing \\ and Measurement}}
    child[grow=120] { node {\blacklink{sec:pressuresensing}{Pressure sensing}} }
    child[grow=85] { node {\blacklink{sec:Force}{Torque sensing}} }
    child[grow=50] { node {\blacklink{sec:part_characterization}{Nanoparticle \\ charac-\\terization}} }
  }
  child[concept color=cyan, grow=95] { node {\whitelink{sec:comparison}{Isolation from \\ environmental noise}} }
  child[concept color=yellow!70!brown, grow=125] { node {\whitelink{sec:Force}{Quantum mechanics \\ testing}} }
  child[concept color=teal, grow=7, level distance=9cm] { node {\whitelink{sec:challenges}{Challenges and Future \\ Directions}}
    child[grow=300] { node {\whitelink{sec:NV}{Hybrid systems \\ integration}} }
    child[grow=335] { node {\whitelink{sec:strong}{Strong coupling \\ challenges}} }
    child[grow=10] { node {\whitelink{sec:ground}{Ground state \\ cooling}} }
    child[grow=49] { node {\whitelink{sec:discussion}{Gas collisions and \\ thermal radiation}} }
    child[grow=87] { node {\whitelink{sec:discussion}{Decoherence \\ overcoming}} }
    child[grow=123] { node {\whitelink{sec:Non-classical}{Quantum information \\ applications}} }
  };

\end{tikzpicture}
}
    \caption{Interactive mind map overview of the main topics covered in this review on roto-translational levitated optomechanics. Each bubble corresponds to a distinct section or subtopic of the article, structured around key themes: theoretical frameworks (green), experimental techniques and cooling (purple and magenta), sensing and measurement applications (lime), quantum and classical testing (yellow and cyan), technological challenges and future directions (teal), and contextual framing including the introduction and conclusion (red and blue). The colors are used to group related topics and visually organize the scope of the review. Text elements within the bubbles are hyperlinked in the digital version of the article and will take the reader directly to the corresponding section, facilitating quick navigation.}
    \label{fig:mindmap}
\end{figure}
In this paper, we describe the experimental and theoretical methods that have been developed to study the roto-translational motion of nanoparticles in optomechanics. We provide an outline of experiments that demonstrate the utility of these six degrees of freedom and describe how they can be cooled and controlled. We also highlight proposals that aim to observe rotational quantum mechanics in these systems and discuss applications such as torque sensing.  The experimental requirements needed to perform these experiments are outlined as well as the many promising proposals that aim to exploit roto-translational motion and its control. An overview of the topics and structure of this review is presented in the interactive diagram (mind map) in Fig.~\ref{fig:mindmap}. This visually outlines the thematic organization of this review and allows the reader to access these parts of the review through clickable navigation paths to the relevant section in the manuscript.\\

\noindent\textbf{Key questions}\\
This review seeks to address the following questions. 
\begin{enumerate}
\item Why is roto-translational motion central to all levitated optomechanics and how can we use forces and torques to obtain motional control of all degrees of freedom? 
\item What does an understanding and control of this motion offer for studies in fundamental physics, and how can motional dynamics and light scattering be used for sensing applications and nanoparticle characterization?
\item How can this much richer roto-translational motion be used to exploit these mesoscopic scale systems to create and verify quantum phenomena? 
\end{enumerate}

In exploring each of these questions, we aim to provide an overview of the current scientific landscape, encompassing the tools, viewpoints, and interpretations proposed thus far.

\subsection{Levitated optomechanical systems}
Optomechanical systems formed by the levitation of small particles in vacuum were first explored by Ashkin \textit{et al.} \cite{ashkin1971optical,ashkin1975optical} using tailored optical fields in the late sixties and early seventies. Even before this, the levitation of microparticles and nanoparticles was accomplished in electric traps~\cite{wuerker1959electrodynmaic} and magnetic traps~\cite{Prothero1968Superconducting}.  However, it has only been over the last fifteen years that methods to cool and control the motion of these particles in the undamped environment of vacuum have been well developed enough to allow them to become a new macroscopic quantum system. Interest in the field has increased with the desire to now create non-classical states of motion~\cite{bateman2014near,arndt2014testing,moore2021searching} and even to investigate the quantum nature of gravity~\cite{bose2017spin}. The ability to measure the small displacements and rotations of these trapped systems has allowed the measurement of exquisitely small forces and torques, opening up a range of applications~\cite{jin2024towards}. Like cold trapped atomic systems, these relatively massive particles are well isolated from the environment with their motion only limited by the control over classical and quantum noises in or induced by the levitating fields and background gas. These macroscopic levitated quantum systems, which can be trapped for long periods of time, represent an exciting new frontier allowing the exploration of new regimes and applications in quantum physics, optomechanics and fundamental physics. The rotational and translation control of nano and microparticles in liquids and gas in the overdamped regime is well established and has been used for imaging, rheology, and biological applications~\cite{Marago2013Optical, Ying2024Single}. However, in the last 15 years techniques for cooling and control of both translational and rotational motion in optical, magnetic and electrical traps within the underdamped regime have been established. An important milestone here has been the demonstration of cooling to the ground state of translational motion, and only very recently for librational/torsional motion, within confining optical potentials.

Most early work in optomechanics has focused primarily on treating the translational motion of near-spherical levitated systems.  Approximating the shape as a sphere has generally worked well, enabling the general features of dynamical motion to be modeled. However, as the understanding and technical refinement of experiments has progressed beyond initial demonstrations and milestones, there has been an increasing need for a more refined description to find quantitative agreement with experiments that take into account the unavoidable rotational degrees of freedom. Rotational dynamics also offers unique opportunities and challenges both from the applied and foundational application of levitated optomechanics.   These effects are not only relevant for future experiments by they are likely to take a disruptive role in ever more refined experiments which focus on the center-of-mass dynamics alone. Ultimately, understanding and subsequent control of the rotational motion within all levitated systems are needed to successfully realize some of the most pivotal experiments proposed to date~\cite{bateman2014near,bose2017spin}.

\subsection{Structure of this review}

This review has begun by providing a brief overview of levitated optomechanics (Sec.~\ref{sec:introduction}), introducing key concepts and highlighting the importance of roto-translational optomechanics as an emerging topic. Our aim is not to provide a comprehensive review of general levitated optomechanics, as several excellent reviews already cover the subject~\cite{aspelmeyer2014cavity, millen2020optomechanics, cronin2009optics, winstone2024optomechanics, jin2024towards, bose2025massive, rademacher2020quantum, BELENCHIA20221Quantum, Kilian2024Dark, gieseler2018levitated,gonzalez2021levitodynamics,carney2021mechanical,moore2021searching}. Instead, we focus on roto-translational optomechanics and its theoretical and experimental developments.

We begin by reviewing the techniques used in roto-translational levitated optomechanics (Sec.~\ref{sec:techniques}), including various levitation methods (optical, magnetic, and electric) and detection, control, and feedback techniques for rotational degrees of freedom. This section provides a foundation for understanding the experimental capabilities available for controlling and manipulating levitated particles.

Following this, we develop a detailed theoretical framework for modeling roto-translational motions (Sec.~\ref{sec:theory}). We present both classical (Sec.~\ref{sec:model}) and quantum (Sec.~\ref{sec:quantum_model}) descriptions, covering fundamental dynamical equations, optical and collisional forces, as well as numerical simulations exploring different particle morphologies such as spherical, prolate, and oblate ellipsoids, and asymmetric tops. A treatment of stochastic forces, including background gas collisions and photon recoil, is also included.

With this theoretical and numerical framework established, we review the current experimental progress in roto-translational optomechanics (Sec.~\ref{sec:applications}). This includes sensing (Sec.~\ref{sec:sensing}) and characterization techniques (Sec.~\ref{sec:part_characterization}) based on full roto-translational motion, force and torque sensing, inertial and gravitational sensing. Furthermore, we discuss recent advances in achieving quantum control of roto-translational motion, including ground-state cooling (Sec.~\ref{sec:ground}), strong coupling regimes (Sec.~\ref{sec:strong}), non-classical states of motion and entanglement in multimode systems (Sec.~\ref{sec:Non-classical}).

The discussion section (Sec.~\ref{sec:discussion}) outlines future directions in the field, addressing key challenges and identifying situations where rotational degrees of freedom can be beneficial or problematic for specific applications. We also highlight unresolved challenges that need to be addressed for further advancements.

Finally, we conclude with a summary and outlook (Sec.~\ref{sec:conclusion}). To complement the main text, several appendices provide additional details on specific aspects of roto-translational optomechanics. \ref{app:Derivation} discusses the connection between classical roto-translational motion and quantum mechanical description, \ref{app:Correlated-noises} covers correlated noise effects and \ref{app:gauss} explores higher-order approximations for calculation for the trap frequencies in an optical tweezer field.

%

%
\section{Techniques in roto-translational levitated optomechanics}
\label{sec:techniques}

\begin{figure}[h!]
\centering
\includegraphics[width=\textwidth]{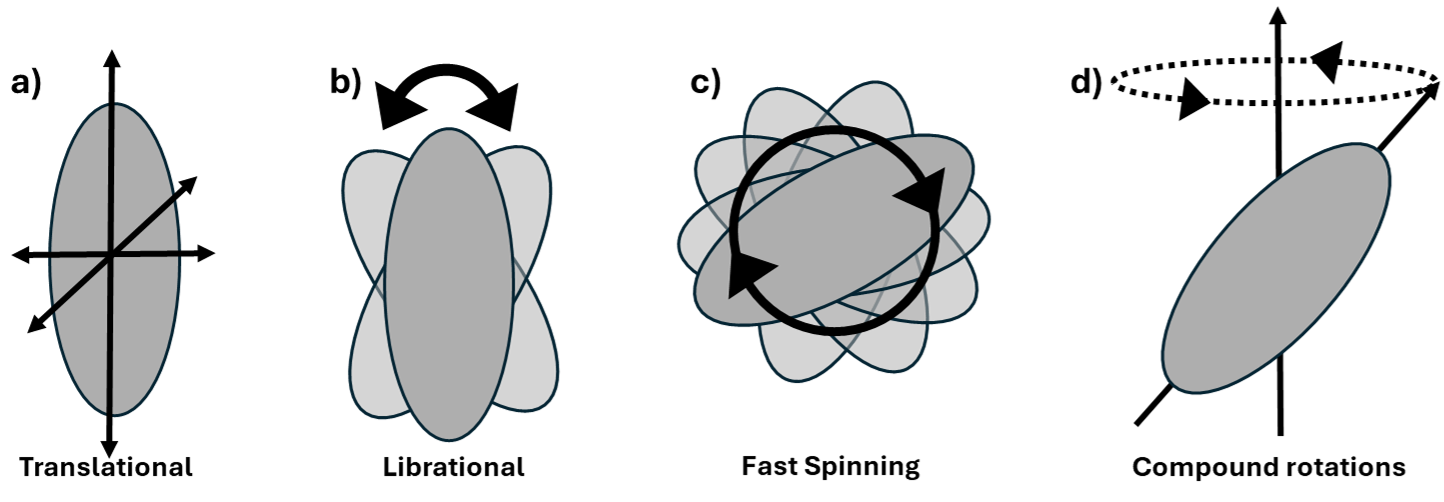}
\caption{Illustration of different types of motion in levitated optomechanics. a) \textbf{Translational motion}: The center of mass of the nanoparticle moves without rotation. b) \textbf{Librational motion}: The particle undergoes small oscillations around a fixed axis due to torque-induced restoring forces. c) \textbf{Fast spinning}: The particle rotates rapidly around a well-defined axis, leading to gyroscopic stability. d) \textbf{Compound rotations}: A combination of rotation, precession, and nutation, characteristic of asymmetric top rotors.}
\label{fig:rotations} 
\end{figure}

A micro/nanoparticle confined in an extrinsic potential allows the exploration of a range of dynamical regimes when the rotational degrees of freedom (DoFs) are considered. In simple terms, these can be broken down into four types of motion, which are illustrated in Fig.~\ref{fig:rotations}, and that are typical of a rigid body kinematics. Like the CoM, the angular DoFs can be confined in a harmonic potential, the resulting small oscillations of the orientation around a given axis are usually referred to as librations. Particles can be set to spin around an axis and can exhibit compound rotations, like precession and nutation, which are direct consequence of the nonlinear coupling of the angular DoFs. Depending of the particle and trapping field parameters, all these regimes can be explored and we will begin, in Sec.~\ref{sec:rot_intro}, with a brief overview of their first observation with micro/nano-objects.

We will then discuss the main experimental techniques for trapping, their characteristics, and advantages / disadvantages in Sect.~\ref{sec:Levitation}. However, the main focus will be kept on optical tweezers, in particular, its theoretical description in Chap.~\ref{sec:theory}. 

Like for the center of mass motion, a central aspect is the ability to measure and control the rotational DoFs. This is discussed in Sec.~\ref{sec:det_contr_man}, where we first introduce the main detection techniques and the associated state-of-the-art sensitivities to then provide a description of the standard techniques to cool the particle motion. These have been developed and refined to address the CoM, or in general a single harmonic oscillator, but the general principles and mathematical description can be adapted to address rotational DoFs. These include cold damping, parametric feedback, and cavity cooling, in particular, in its coherent scattering variant. In this chapter, we will also touch on a line of inquiry which aims at coupling internal degrees of freedom to external ones. We will conclude with a discussion of the challenges and limitations towards applications and the exploration of quantum phenomena with these systems.

\subsection{Rotational motion in trapping fields}\label{sec:rot_intro}
Rotational motion is induced in trapped particles via anisotropic interactions between the morphology of the levitated system and the trapping fields used for levitation. Torques result from differential forces on levitated bodies either through nonspherical mass distribution or via nonuniform material properties such as density or refractive index. As a result, different types of rotational dynamics can emerge, and a schematic view is shown in Fig.~\ref{fig:rotations}. Notable examples are the realization of Beth's experiment with optical tweezers in liquids and the alignment of shaped particles along the polarization axis of linear polarized optical fields~\cite{Friese1998Optical}. Alignment and rotation of particles has also been demonstrated with magnetic tweezers~\cite{Lipfert2010magnetic} and electric tweezers~\cite{Fan2011electric}. 

The first experiments to study rotational and translational motion in the undamped regime in vacuum (optomechanics) were performed using near spherical birefringent vaterite particles~\cite{arita2013laser}. Here trapping in circularly polarized fields resulted in the rotation of the particles via transfer of spin angular momentum. The gyroscopic coupling between the rotational motion and translation was shown to stabilize the particle with rotational rates exceeding MHz frequencies, only limited by the damping from residual gases. Since this first experiment, rotational frequencies exceeding GHz have been demonstrated for silica nanoparticles in higher vacuum~\cite{jin20216,reimann2018ghz,ahn2020ultrasensitive,Monteiro2018optical,zielinska2024longaxis}.  Confined rotational motion in the form of torsional/libration trapped motion was also demonstrated for the first time demonstrating the significantly higher frequency of this harmonic motion when compared to translation~\cite{hoang2016torsional}.  Other early experiments not only observed the rotational motion, but demonstrated moderate cooling of this degree of freedom for the first time. Here cavity cooling of silicon nanorods was observed as they briefly traversed through a high-finesse cavity~\cite{kuhn2015cavity}. A systematic study of the full roto-translational motion, including precession, was observed over frequencies from a few kHz up to 100's of kHz~\cite{rashid2018precession}.

\subsection{Levitation techniques}\label{sec:Levitation}
In this section, we will provide a brief overview of the methods of levitating objects with a discussion of the advantages of using each method.

\begin{figure*}
\centering
\includegraphics[width=0.75\textwidth]{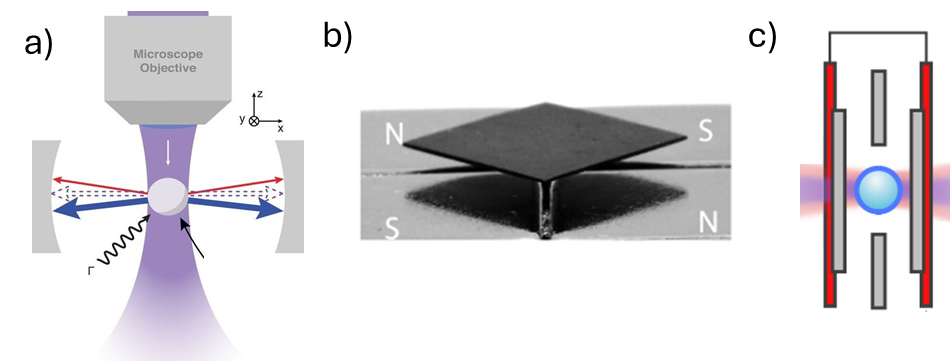}
\caption{
\textbf{Overview of typical levitation platforms used for rotational optomechanics: optical, magnetic, and electric trapping.} 
\textbf{a}) Optical levitation and cavity cooling scheme adapted from Delić \textit{et al.},~\cite{delic2020cooling}. A silica nanoparticle is trapped by a tightly focused optical tweezer (purple) and placed inside a high-finesse optical cavity (gray). The scattering of light from the particle into the cavity mode enables sideband-resolved cavity cooling by coherent scattering. The interplay between laser detuning and cavity positioning facilitates cooling along the cavity axis. This scheme enabled motional ground-state cooling of a nanoparticle at room temperature.
\textbf{b}) Diamagnetic levitation setup adapted from Chen \textit{et al.},~\cite{chen2024nonlinear}. A pyrolytic graphite plate levitates passively above a set of four permanent NdFeB magnets arranged in a checkerboard configuration. The setup operates in vacuum below $10^{-4}$ and passive, zero-power levitation.
\textbf{c}) Paul trap-based electric levitation and feedback cooling adapted from Dania \textit{et al.},~\cite{dania2021optical}. A silica nanoparticle is levitated in a four-rod linear Paul trap, and its motion is detected via homodyne interferometry of forward-scattered light. Feedback cooling is achieved either optically—via an acousto-optic modulator controlling a 1064-nm laser—or electrically—via voltages applied to feedback electrodes. Cooling along all three motional axes is implemented simultaneously, achieving millikelvin-scale cooling of all motional modes.
}
\label{fig:figureExpSchematic}
\end{figure*}
\subsubsection{Optical Levitation\label{sec:optical}}
The preliminary focus of the field of levitated optomechanics has been the levitation of dielectric particles using optical forces. The first demonstration of optical levitation was in 1971 with the seminal work of Arthur Ashkin levitating a dielectric particle in vacuum~\cite{ashkin1971optical}.  Levitation via optical forces does not require the particle to be charged, enabling the trapping of electrically neutral particles. Here levitation primarily occurs due to the creation of a macroscopically induced dipole moment which interacts with the same field to create a force that acts towards the point of maximum intensity. The scattering force can also be used which acts in the direction of propagation. 

A variety of experimental configurations have been utilized to levitate via optical forces. One of the most popular methods for levitating particles is an optical tweezer, shown in Fig.~\ref{fig:figureExpSchematic}~a), formed by tightly focusing Gaussian beams via high numerical aperture (NA) lenses~\cite{gieseler2012subkelvin,delic2020cooling,rahman2017laser}. The appeal of levitating via high NA lenses is that it enables the trapping of submicron particles with high oscillation frequencies in the order of 100 - 500 kHz. To trap larger micron-sized particles, other techniques developed by Ashkin \textit{et al.} are employed. Here, the larger surface area of the micron-sized particles increases the scattering force which pushes the particle out of the tweezer focus. To combat this, the trap is aligned vertically so that the gravitational force opposes the scattering force enabling stable trapping. This is known as a Gravito-optical trap ~\cite{moore2014search,monteiro2020search,arita2022all}. 

An additional method to reduce the scattering force is to use a so-called standing wave trap~\cite{kuhn2017optically,winstone2022optical,hu2023structured,ranjit2016zeptonewton}. Here two counter-propagating beams are focused and aligned to form a standing wave. The particle is able to be trapped in the antinode of the standing wave. As the scattering force acts along the direction of propagation, by having two counter-propagating beams the scattering force from each beam cancels out. This allows micron sized particles to be trapped. This standing wave can also be formed from retro-reflecting the laser beam off a mirror~\cite{kamba2021recoil}. Due to the cancellation of the scattering force in this mode of operation, standing wave traps can have significantly higher trap frequencies, particularly along the direction of propagation.

The previously outlined techniques utilize an optical lens in some shape or form to generate an optical trap. However, a parabolic mirror can also be used to make a high numerical aperture optical trap~\cite{vovrosh2017parametric,dawson2019spectral,Laplane2024Inert}. Here the laser light is focused by a concave mirror to a diffraction limited spot in which the nanoparticle is trapped. This focusing does not suffer from the chromatic aberrations. 

Whilst significant progress has been made with optical levitation in vacuum, one of the major issues with this technique is the use of highly focused laser beams.  Absorption of laser light by the nanoparticles increases the internal temperature to a point at which the nanoparticles can melt or even burn~\cite{rahman2016burning}. Even materials with low absorption at typical laser wavelength, temperatures on the order of 1000 K can be reached at vacuum pressures below $10^{-3}$ mbar~\cite{hebestreit2018measuring}. This increase in the internal temperature limits the types of materials that can be stably confined in an optical tweezer at high vacuum.

In addition to the absorption of the laser beam, the scattering of photons from the trapping beam is also a source of concern. At ultrahigh vacuum $\approx10^{-8}$ mbar, the dominant noise that a levitated nanoparticle experiences is the recoil from photons scattering off the nanoparticle~\cite{jain2016direct}. This represents a fundamental floor in force and torque sensing with levitated nanoparticles and a significant source of decoherence for future quantum mechanical experiments.

\subsubsection{Magnetic Levitation}\label{sec:magnetictraps}

More recently, interest has grown in levitation without the use of laser light for trapping to avoid the associated recoil heating. This can be accomplished via magnetic and electric levitation that allows the levitation of much larger objects.

Levitation via magnetic fields works on the principles of diamagnetism, in which applied magnetic fields create an induced magnetic field that acts in the opposing direction. Levitation via diamagnetism has taken three forms, levitation of a diamagnetic material above a permanent magnet, levitation of a magnet above a superconducting magnetic trap and levitation of a superconductors.

\begin{figure*}
\centering
\includegraphics[width=0.65\textwidth]{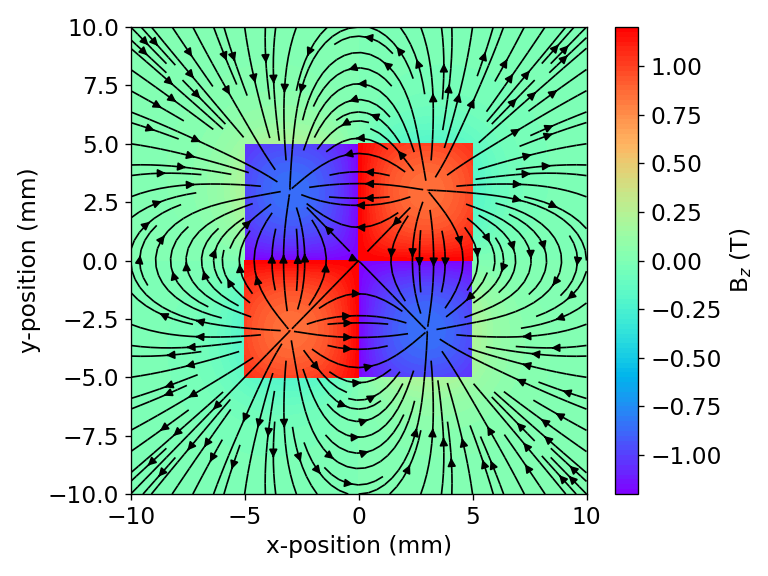}
\caption{Layout of a 2x2 magnetic trap array showing the magnetic fields generated. The red and blue coloring represent the magnetic field within the magnets in the z-direction and the arrows are the magnetic field vectors in the x-y plane.}
\label{fig:figureMag}
\end{figure*}

The simplest form of magnetic levitation is a diamagnetic material above a permanent magnet~\cite{Romagnoli2023Controlling}, an example of which is shown in Fig.~\ref{fig:figureExpSchematic}~b). Here the induced magnetic field opposes the field from permanent magnet below. To form a stable trap in the horizontal plane the magnets are arranged in such a way that the neighboring magnet is of the opposing polarity, for example a 2x2 array as shown in Fig. \ref{fig:figureMag}. This creates large gradients in the magnetic field in the vertical direction to create a force that opposes gravity and also confinement in the horizontal plane forming a stable trap 3-D trap. These forces also lead to restoring torques that add an additional three degrees of freedom to the motion~\cite{Chen2020Rigid}. The size of the object that can be levitated is significantly larger than for optical tweezers with the levitation and cooling of centimeter sized objects~\cite{Tian2024Feedback}. The appeal of a permanent magnetic trap is that the trapping is passive, requiring no active power to trap stably for long periods of time. The trap design can be modified to provide stronger trapping potentials~\cite{lewandowski2021high,OBrien2019Magneto} and trapping of multiple objects~\cite{elahi2024diamagnetic}. However, the potential is harder to modify compared to optical and electrical traps. Whilst all materials demonstrate diamagnetism, the strongest diamagnetic susceptibility is observed in superconductors. This utilises the same principles as before with the cryogenic cooling of a material down to the point where it becomes a superconductor and levitates \cite{hofer2022high}. 

This principle can be reversed, instead of levitating a diamagnetic material, a magnet can be levitated above a superconducting trap \cite{vinante2020ultralow,gieseler2020single}. This can be even extended to levitating superfluid helium in a superconducting trap \cite{brown2021characterization}. The appeal of working with magnetic traps, in particular superconducting traps, is that it opens up the possibilities of light-free experiment setups where the motion of the particle is detected via superconducting quantum interference devices (SQUIDS)~\cite{latorre2023superconducting}. Alternatively, a rotating magnetic array can form a planar magnetic Paul trap which can levitate ferromagnetic particles~\cite{Perdriat2023Planar,schreckenberg2025study} or by an alternating current~\cite{Janse2024Characterization}. Theoretical proposals have also discussed how a single magnetic domain nanoparticle can be levitated in an external static magnetic field despite Earnshaw's theorem, with stable trapping relying on the gyromagnetic effect~\cite{rusconi2017quantum}.

\subsubsection{Electric Levitation}\label{sec:electrodynamictraps}

Whilst magnetic levitation offers the possibility of levitating large objects, the materials that can be levitated are limited and trap frequencies are typically in the Hz to 100 Hz range. Levitation in electric fields via electrodynamic traps such as the Paul trap can also be used. These are typically used for the trapping of atomic ions, but can be applied to larger charged objects with trap frequencies up to the 10 kHz range demonstrated~\cite{bullier2021quadratic,alda2016trapping,dania2022position,delord2020spin}. Additionally, the trapping potential can be easily tuned by the variation in the applied voltages to the trap. As dictated by Earnshaw's theorem in classical electromagnetism, a three dimensional trapping potential for charged particles cannot be formed by static fields only.  To circumvent this,  oscillating electric fields are used which rapidly switch sign at a well-defined (drive) frequency that leads to an effective stable trapping potential, often called a psuedo-potential. Well-defined regions of stable trapping occur for particular combinations of charge-to-mass ratios, applied voltage and drive frequency. In addition to the harmonic motion in these types of traps, the drive frequency leads to additional motion at that frequency which can lead to heating the motion. Torques can also be generated on trapped particles via the anisotropy in charge or charge distribution~\cite{jin2024quantum}. These asymmetric charge distributions allowed rotations of a variety of different shapes in Paul traps~\cite{perdriat2024Rotational}. A variety of different materials can also be levitated as long as the object is charged.

A variety of electrode configurations can provide stable trapping, the most commonly used is a linear trap, an example of which is shown in Fig.~\ref{fig:figureExpSchematic}~c). Regardless of the geometrical configuration, the trap depth of a Paul trap is significantly deeper and wider than that of an optical tweezer. This enables levitation of multiple particles within the same trap~\cite{bykov20233d,penny2023sympathetic}. Here, particles interact with each other in the trap via strong Coulomb repulsion with trap frequencies defined by their normal modes of oscillation. This type of multiple particle trapping is appealing towards the goal of trapping an ion and a particle within the same trap~\cite{Hopper2024a}. This was recently demonstrated~\cite{bykov2024nanoparticle} with the view towards mediating interactions between the ion and the nanoparticle by a cavity to generate non-classical states of motion of nano-sized objects.

There has been growing interest in combining both optical and electrical traps into a hybrid trap~\cite{Conangla2018Motion,Bonvin2023Hybrid}. This utilizes the advantages of both systems with higher trap frequencies than just electrical trapping, whilst using less laser light and a deeper trap potential compared to just optical trapping. The different trapping potentials have also been proposed as a protocol of state expansion for generating macroscopic quantum superpositions~\cite{bonvin2023state,tomassi2025accelerated}. The trap potentials have different trap depths and spatial spread which results in vastly different trap frequencies. This protocol utilizes this for state expansion. In addition to the hybrid trapping potentials, Paul traps when combined with cavities offer enhanced cooling of the nanoparticle motion with a view to ground state cooling~\cite{millen2015cavity,pontin2020ultranarrow}.

\subsection{Detection, Control and Manipulation}\label{sec:det_contr_man}

Levitated optomechanics has seen rapid progress during the last decade and proven to be a very flexible platform. Initial efforts were particularly focused on bringing the centre of mass motion (CoM) of these systems into the quantum regime. As such, techniques to cool the particles degrees of freedom have played a central role in the development of the field since one always needs to battle against thermal noise. In the following, we will give an overview of the most important techniques implemented. We will give a general description that applies to both the CoM and rotational DoFs when confined by a potential. The main differences appear only in the actual implementations which will be discussed in \ref{controlANDmanipualtion}.

The basic concepts of the various techniques can be discussed by considering as starting point an underdamped harmonic oscillator with inertial property $\mu=(m,I)$, i.e., mass or moment of inertia depending on the DoF, and resonance frequency $\omega_0$ under the effect of a thermal force $f_{\text{th}}(t)$ with a Markovian autocorrelation function $\langle f_{\text{th}}(0)f_{\text{th}}(\tau)\rangle=2 k_B T \mu \gamma \delta(\tau)$. Here, $k_B$ is the Boltzmann constant, $T$ the temperature of the system and $\gamma$ the damping acting on the oscillator. The equation of motion can then be written as 
\begin{equation}
    \ddot{q}+\gamma\dot{q}+\omega_0^2 q = f_{\text{th}}(t)/\mu
\end{equation}
\noindent where $q$ identifies the general coordinate of the oscillator. 
Typically, one has two handles to reduce the mean energy of the motion, reduce the temperature or the dissipation. Among the appealing factors of levitated optomechanics is that the dissipation $\gamma$ is due to collisions with the background gas~\cite{epstein1924On,Cavalleri2010Gas} and, as such, it is proportional to pressure. All cooling techniques can be summarized as the introduction of an additional dissipation channel $\gamma_{\text{c}}$ by external means so that when the pressure is reduced the oscillator quickly reaches an out of equilibrium steady state with a lower mean energy.  

\subsubsection{Detection of translational and rotational motion}\label{sec:detection}
The sensitive detection of the rotational and translational motion of levitated systems is critical for achieving the  control and cooling required to bring these systems into the quantum regime. It is also critically important for utilizing them as sensors. For optically levitated systems these measurements have been well developed within optical trapping community over the last 40 years \cite{gittes1998intreferometry}. Of particular importance is efficient low-noise collection of the scattered light. Almost all methods use interferometric techniques which interfere the scattered light with a local oscillator usually derived from the trapping laser.  Balanced homodyne and heterodyne detection are most commonly employed. For levitation in optical tweezers, the light scattered from the particle is interfered with a local oscillator at the same frequency (homodyne) or with a different frequency (heterodyne) with the resulting intensity modification to the interference pattern proportional to displacement. As the scattered light, even for very small particles, is not isotropically distributed, the sensitivity to displacement in any direction can be enhanced by using strategies to efficiently collect and detect light over large solid angles. This maximizes the information content of the scattering field in the particular direction/axis of interest~\cite{tebbenjohanns2019optimal}. Using these technique, sensitivities down to $10^{-27}$ m$^2$Hz$^{-1}$, limited by measurement imprecision and backaction have been demonstrated. These were central to achieving ground-state cooling via feedback \cite{tebbenjohanns2019cold}.  The motion of the particle within an optical cavity, as utilized in cooling via coherent scattering, has an approximate detection sensitivity of $10^{-26}$m$^2$Hz$^{-1}$ \cite{deplano2025high}. 
Rotational motion in optically trapped particles is typically detected by the time-dependent change in polarization of the scattered light as the particle rotates or librates in the fixed field. The change in polarization is induced by either the orientation of birefringence of the particle, or by the shape birefringence from the non-spherical nature of the trapped particle.  The scattered light sent through a polarization sensitive optical element, such as a beam splitter and coupled with balanced detection~\cite{arita2013laser,seberson2019parametric} appears to be capable of measurements in angular displacement with sensitivities in the range of $10^{-11}$ rad$^2$Hz$^{-1}$~\cite{bang2020five}. This however is to be confirmed experimentally.

\subsubsection{Feedback cooling\label{sec:colddamping}}
Also known as cold damping or velocity feedback, it is a very old but also a very successful approach~\cite{ashkin1977feedback,Poggio2007Feedback}. Here a feedback force  which opposes the instantaneous velocity of the oscillator is used to damp the motion. Typically this is implemented through an optical~\cite{li2011millikelvin,slezak2018cooling,ranjit2015attonewton,kamba2022optical,bang2020five} or a Coulomb interaction~\cite{tebbenjohanns2019cold,tebbenjohanns2020motional,tebbenjohanns2021quantum,kamba2021recoil,dania2021optical,kamba2023nanoscale,gosling2025feedback} but the nature of the force/torque is not particularly important for the technique and the choice often reflects more the experimental objectives than technical limitations. In order to apply the feedback the position/orientation of the oscillator must be measured so that an estimate of the velocity can be obtained. Several approaches are available, from direct derivative to optimal control
~\cite{ferialdi2019optimal, setter2018real,conangla2019optimal,magrini2021realtime}. However, apart from the specific implementation, the main limitation is that the measured position $q_{m}$ is not exactly the true position $q$ since the measurement is affected by an imprecision noise $q_n$, i.e., $q_m=q+q_n$. Thus, the displacement sensitivity ultimately determines the lowest motional effective temperature that can be reached. The equation of motion in the presence of cold damping can be written as 
\begin{equation}
    \ddot{q}+(\gamma+\gamma_{fb})\dot{q}+\omega_0^2 q = f_{\text{th}}(t)/\mu-\gamma_{fb}\delta\dot{q}
\end{equation}
\noindent where $\gamma_{fb}$ is the damping due to the feedback, and represents the closed loop gain, and $\delta\dot{q}$ is a stochastic, additive noise driving the motion which arises from the detection noise $q_n$. Thus, the imprecision noise becomes a heating term limiting the lowest effective temperature attainable. This balance between imprecision noise and backaction of the measurement can be directly cast in terms of an Heisenberg uncertainty relation~\cite{Braginsky1992Quantum,Gieseler2021Optical}. In the case of levitation, where thermal noise depends on residual gas pressure $P$, one finds a linear relationship between effective temperature and pressure, i.e., $T_{eff}\propto P$.

\subsubsection{Parametric feedback\label{sec:parametricfeedback}}
An alternative and widely used approach to cool the nanoparticle motion is parametric feedback cooling~\cite{gieseler2012subkelvin,vovrosh2017parametric,vanderLaan2021subkelvin}. This technique relies on a modulation at the oscillator trap frequency synchronized with the particle motion such that the stiffness is increased when the oscillator is moving away from the trap center and is decreased when moving towards the center. This results in a modulation at twice the trap frequency with a modulation depth which depends on the motional amplitude. The equation of motion of an oscillator under parametric feedback can be written as 
\begin{equation}\label{parfeedback}
    \ddot{q}+\gamma\dot{q}+\omega_0^2 (1+\eta \frac{q\dot{q}}{\omega_0}) q = f_{\text{th}}(t)/\mu
\end{equation}
\noindent where $\eta$ represents the loop gain.  Eq.~\ref{parfeedback} can be identified as general Van der Pol non-linear oscillator, as such, the resulting dynamics can be drastically different from the cold damping case. 
In particular, the equilibrium energy distribution  predicted by Eq.~\ref{parfeedback} is no longer an exponential Boltzmann-Gibbs distribution since the feedback is much more efficient in suppressing large amplitude fluctuations. A direct consequence is that, as the oscillator motion gets colder, the cooling process becomes less efficient with a resulting effective temperature $T_{eff}\propto \sqrt{P}$~\cite{gieseler2012subkelvin,gieseler2014dynamic}.  To overcome this limitation, an alternative implementation has been developed which relies on a phase-locked loop (PLL) to track the instantaneous phase of the oscillator once it reaches the underdamped regime. In this case, the modulation depth of the trap stiffness is a fixed parameter of the feedback loop and is no longer proportional to the oscillator amplitude. The equation of motion can be expressed as
\begin{equation}\label{parfeedbackPLL}
    \ddot{q}+\gamma\dot{q}+\omega_0^2 \{1-G\sin[2(\omega_0 t+\theta_q)]\} q = f_{\text{th}}(t)/\mu
\end{equation}
\noindent where $G$ represents the loop gain, i.e., the modulation depth, and $\theta_q$ is the instantaneous oscillator phase. This modified parametric feedback allows to recover an effective temperature $T_{eff}\propto P$, however, as the motion gets colder the importance of the imprecision noise becomes more relevant, making the PLL less effective in tracking the oscillator phase~\cite{penny2021performance}. 

Parametric feedback has played a central role in the early stages of levitated optomechanics \cite{gieseler2012subkelvin}, in particular, in the context of optical levitation. The main advantage of the technique is the possibility to address multiple degrees of freedom by controlling a single parameter, i.e., the trapping power, without requiring additional hardware.

\subsubsection{Coherent scattering\label{sec:coherent}}
The use of optical cavities has been an invaluable tool to explore the quantum dynamics of mechanical oscillators. The typical approach in the general optomechanical framework relies on a dispersive coupling between the cavity field and the oscillator motion. The coupling arises from radiation pressure, i.e., power,  and allows sideband cooling~\cite{Marquardt2007Quantum,Schliesser2008Resolvedsideband} to the mechanical ground state~\cite{chan2011laser}. Early demonstrations in levitation followed the same approach~\cite{kiesel2013cavity,millen2015cavity,fonseca2016nonlinear,meyer2019resolved}. However, it is the introduction of coherent scattering that allowed a significant leap forward in the field. This technique adapted from atomic physics~\cite{vuletic2000laser}, considers an optical tweezer confining a Rayleigh nanoparticle and centered in an empty high-finesse optical cavity with the tweezer propagation direction orthogonal to the cavity axis. Contrary to the standard optomechanical approach, the cavity is not externally driven.  If the tweezer field is resonant with the cavity mode the intra-cavity field is then populated by light scattered by the particle in a dipolar pattern, leading to interference between the cavity field and the scattered light and greatly enhancing the coupling between the particle motion and the cavity mode. This implies that the coupling does not occur through power but through the cavity field. The mechanism can also be understood in terms of the Purcell effect~\cite{Kuhn2010Cavitybased} with the cavity modifying the environment seen by the particle or, in other words, the cavity changes the local density of photonic states into which light can be scattered. The coherent scattering approach can provide the means to cool all the degrees of freedom of a levitated (non-spherical) nanoparticle, however, the cooling efficiency cannot be optimal for all simultaneously.  Indeed, the optomechanical couplings are controlled by two angles: $\phi=k q_0$, which encodes the distance, $q_0$, from the cavity anti-node ($k=\frac{2\pi}{\lambda}$ with $\lambda$ the optical wavelength), and $\theta$, which quantifies the angular tilt of the tweezer polarization with respect to the cavity symmetry axis (we will consider $\theta=0$ corresponding to the tweezer polarization being aligned with the cavity symmetry axis). Which and how strongly each DoF is coupled to the cavity depends on the particle mean position along the cavity standing wave with some optimal at a cavity node ($\phi=\pi/2$), while vanishing at an anti-node ($\phi=0$) and vice-versa~\cite{delic2020levitated}. It can be shown that the optimal position to couple the motion in the tweezer polarization plane is at a cavity node while all other DoFs are optimized at a cavity antinode. A detailed theoretical analysis can be found in Refs.~\cite{toros2020quantum,toros2021coherent-scattering,gonzalez2019theory} regarding the CoM  and it has been extended to include rotational DoFs in Refs.~\cite{schafer2021cooling,rudolph2021theory}.  The first experimental demonstrations can be found in Refs.~\cite{delic2019cavity,windey2019cavitybased} for the CoM motion and in Ref.~\cite{pontin2023simultaneous} for all 6 DoFs.

\subsubsection{Coupling internal and external degrees of freedom}
As levitated particles can have embedded atomic-like impurities, such as the spin of an NV centers contained within nano- and micro-diamond, manipulation of these quantum systems and the subsequent coupling to the external rotational and translational motion can be used for detection, control and manipulation. They hold significant promise toward macroscopic quantum superpositions and probing entanglement between these levitated masses~\cite{bose2017spin,Japha2023Quantum,wan2016free,wachter2025gyroscopically,wood2022spin}. NV centers are particularly promising for the field of levitated optomechanics which can leverage the single-spin state embedded in the larger system~\cite{Perdriat2021Spin-Mechanics}. The single-spin state can be addressed magnetically and optically. This enables the smaller quantum spin state to control the larger classical motion of the particle. Observations of the electron spin from NV centers embedded in nanodiamond, their resonances and control has been achieved~\cite{Delord2017Electron,Delord2018Ramsey,Hoang2016Electron} and this has been further extended to the nuclear magnetic resonances using $^{14}$N nuclear spins~\cite{Voisin2024Nuclear}. 
Controlled excitation of Yb ions embedded in nanoparticles have used the preferential anti-Stokes fluorescence for cooling the bulk temperature. This is often termed laser refrigeration and has been used to cool levitated nanoparticles to bulk temperatures in the 100 K range~\cite{rahman2017laser,Laplane2024Inert}. Such a laser refrigeration process is only feasible in materials where there is weak coupling between electronic degrees of freedom and phonons such as crystal of yttrium lithium fluoride. The aim in these experiments is to reduce the motional heating both from collisions with background gases and from black body radiation which impacts both rotational and translational temperatures. The crystalline structure of the host material also enables the growth of a variety of nanoparticle shapes which in turn have their own unique rotational behavior and lower photon-recoil-heating~\cite{winstone2022optical}.

Lastly, cold atoms collisionally interacting with \cite{Zhang2011three}, or even trapped around nanoparticles~\cite{bykov2024a,Hopper2024a,toros2021creating}, could also be used to control their rotational and translational motion. This includes sympathetic cooling and the potential for controlled creation of non-classical states of motion.

\subsubsection{Challenges and limitations}
\label{sec:challenges}
    
A central challenge is to maintain coherence of the nanoparticle quantum state against environmental decoherence. Unlike clamped microresonators, a levitated particle is isolated from solid supports, but decoherence still arises from interactions with the trapping field and residual gas. One well-known issue is that the trapping laser continuously measures the particle’s position (via scattered photons), collapsing the particle’s wavefunction and adding heating (recoil kicks) – effectively a form of measurement back-action. Researchers have proposed clever ways to mitigate such decoherence in the past five years. For instance, \cite{gonzalezballestero2023surpressing} theoretically showed that illuminating the trap with squeezed light can suppress recoil heating in free-space levitation. Squeezed light has reduced quantum fluctuations in one quadrature, and by choosing the squeezing phase and angle appropriately, one can reduce the information carried away by scattered photons about the particle’s motion. This, in turn, reduces measurement-induced decoherence, allowing the particle to remain in a pure quantum state longer. Their calculations predict that state-of-the-art squeezed light sources could reduce laser recoil heating by over an order of magnitude (10$\times$ with one squeezed mode, or even 100$\times$ with an optimally mode-matched squeezed field). Achieving quantum-limited detection of a levitated particle’s motion (beyond the standard quantum limit) thus appears feasible with engineered illumination. This would significantly extend coherence times and enable more precise experimental quantum control.

While control and cooling single degrees of freedom has been demonstrated on many levitation platforms, the development of multi-dimensional and multi-particle optomechanical control is not well developed. For example, when cooling or manipulating more than one degree of freedom, so-called dark modes (uncoupled eigenmodes that do not interact with light) can prevent certain motions from being cooled. Recent studies have provided recipes to avoid this problem, for example Toro\v{s} \textit{et al}. \cite{toros2021coherent-scattering} analyzed two-dimensional cooling in a coherent-scattering setup and identified optimal trap ellipticities and cavity detunings to cool both transverse modes efficiently without exciting dark modes. Similarly, Xu \textit{et al.} \cite{xu2024simultaneous} developed a full multimode cavity theory for two levitated nanoparticles and found that by slightly mismatching the tweezer powers or trap frequencies, one can break the symmetry that causes dark modes, enabling simultaneous ground-state cooling of two particles’ motions. These theoretical insights are crucial for scaling levitated optomechanics to multiple objects or full 3D control. They also inform the design of future experiments aimed at entangling two levitated masses—by ensuring both particles can be cooled and controlled concurrently, a prerequisite for generating entangled states via their mutual light field.

Levitated roto-translational optomechanics must contend with several fundamental challenges that arise from the same light–matter coupling that enables control. One such challenge is optical back-action: the strong cavity field necessary for cooling can significantly perturb the particle’s motion. In the CS setup, while high optical powers yield large optomechanical coupling (facilitating cooling), they unavoidably also induce frequency shifts (optical spring effects) and mixing of the particle’s mechanical modes~\cite{pontin2023controlling}. For a particle trapped in free space (without a fixed oscillator frame), the cavity-mediated coupling can cause different translational modes to hybridize with each other and with the cavity field. In practical terms, the normal eigenmodes of the particle’s motion (e.g. along $x$ and $y$ axes) may no longer remain orthogonal or aligned to the trap coordinates once strong light–matter interaction is introduced. Instead, the modes can rotate in orientation and become combinations of the original degrees of freedom~\cite{pontin2023controlling}. This mode rotation and coupling is not merely a theoretical curiosity; it has direct implications. For example, in multi-axis cooling attempts, the interference of scattering can create “dark modes,” which are motion patterns that do not couple to the cavity light and thus evade cooling~\cite{pontin2023controlling}. Such dark modes (and their counterpart, bright modes) arise from destructive interference in the cavity scattering and can impede the simultaneous cooling of all translational directions. Dealing with these effects requires careful tuning of system parameters. In practice, one may adjust the particle’s position in the standing wave or use slight asymmetries (e.g. elliptic polarization or nonequivalent trap frequencies) to break the symmetry and ensure all modes are optomechanically coupled. Even so, the presence of strong back-action means that achieving stable simultaneous ground-state cooling of multiple degrees of freedom is non-trivial. Researchers have identified special operating points (for instance, a “cancellation point” in the cavity detuning or particle position) where the deleterious optical spring shifts and mode rotations cancel out, but maintaining such conditions is experimentally challenging. Thus, while CS enables powerful multi-dimensional cooling, it also introduces a complex interplay between the light and mechanics which must be tamed to fully exploit the technique.

Another set of limitations is tied to the practical and environmental factors in levitated optomechanics. Foremost, maintaining ultrahigh vacuum is essential for isolating the nanoparticle from gas damping and decoherence. Any residual gas molecules can constantly kick the particle, heating its motion and overwhelming the cooling effects. For rotational degrees of freedom, gas collisions and background torques (from, e.g., random photon scattering or internal molecular outgassing) can quickly randomize the particle’s orientation. However, pushing to lower pressures for better isolation exposes a critical thermal problem: optical levitation inherently carries the risk of laser heating the particle. Even a nominally transparent nanoparticle absorbs a tiny fraction of the intense trapping light, leading to internal heating. In high vacuum, the particle cannot efficiently dissipate this heat to the sparse surrounding gas~\cite{hu2023understanding}. The result is the build-up of thermal gradients within the particle, which can produce a subtle but significant photophoretic force – a light-induced drift that can expel the particle from the trap or shift its equilibrium position unpredictably~\cite{hu2023understanding}. This effect has been identified as a major cause of particle loss when attempting to reach extreme vacuum levels in optical traps~\cite{hu2023understanding}. It imposes a practical limit on how low the pressure can be reduced (or how high the laser power can be increased) before the particle becomes unstable. Mitigating these thermal forces may involve using lower-absorption materials, implementing active cooling of the particle (via, for example, laser refrigeration~\cite{rahman2017laser} or feedback cooling of its motion to reduce laser dwell time on any one spot). Of course switching to alternative trapping methods (such as electric or magnetic traps can avoid continuous or high-intensity laser illumination~\cite{yin2024towards}).

Additional challenges in levitated optomechanics arise from technical noise sources and fundamental limitations in detection and cooling strategies. Active feedback cooling is inherently constrained by detection efficiency, where imprecision noise limits performance regardless of the specific platform~\cite{tebbenjohanns2019optimal,penny2021performance}. Passive cavity cooling, on the other hand, is susceptible to classical and technical noise from the optical field, which remains an issue even in coherent scattering setups~\cite{meyer2019resolved}. More broadly, photon-mediated interactions impose additional constraints. Recoil heating, a well-known consequence of scattering, disrupts cooling and coherence, while blackbody radiation from the particle itself can introduce unwanted thermal noise and decoherence~\cite{jain2016direct,chang2010cavity}. Experiments have not yet to reach this regime where blackbody effects play an important role. These effects underscore the need to improve detection techniques, minimize excess noise in optical fields, and carefully manage the particle interaction with photons to preserve quantum control.

A particularly illustrative case of these challenges is in proposed experiments involving levitated nitrogen-vacancy (NV) centers. The orientation of the NV center plays a crucial role in both its dynamics and measurement processes, making precise control over rotational degrees of freedom essential. Developing techniques to manipulate and cool these degrees of freedom is a critical step toward realizing key experimental goals in levitated optomechanics. These include fundamental tests of quantum mechanics and spin-motional coupling schemes that could unlock new avenues in quantum sensing and hybrid quantum systems. Gaining expertise in these control techniques is critical for implementing some of the most ambitious proposals in the field, such as macroscopic quantum superpositions and gravitationally induced entanglement, further highlighting the importance of systematically addressing these challenges~\cite{bateman2014near,bose2017spin}.

%
\section{Theoretical framework for modelling roto-translational motions}
\label{sec:theory}
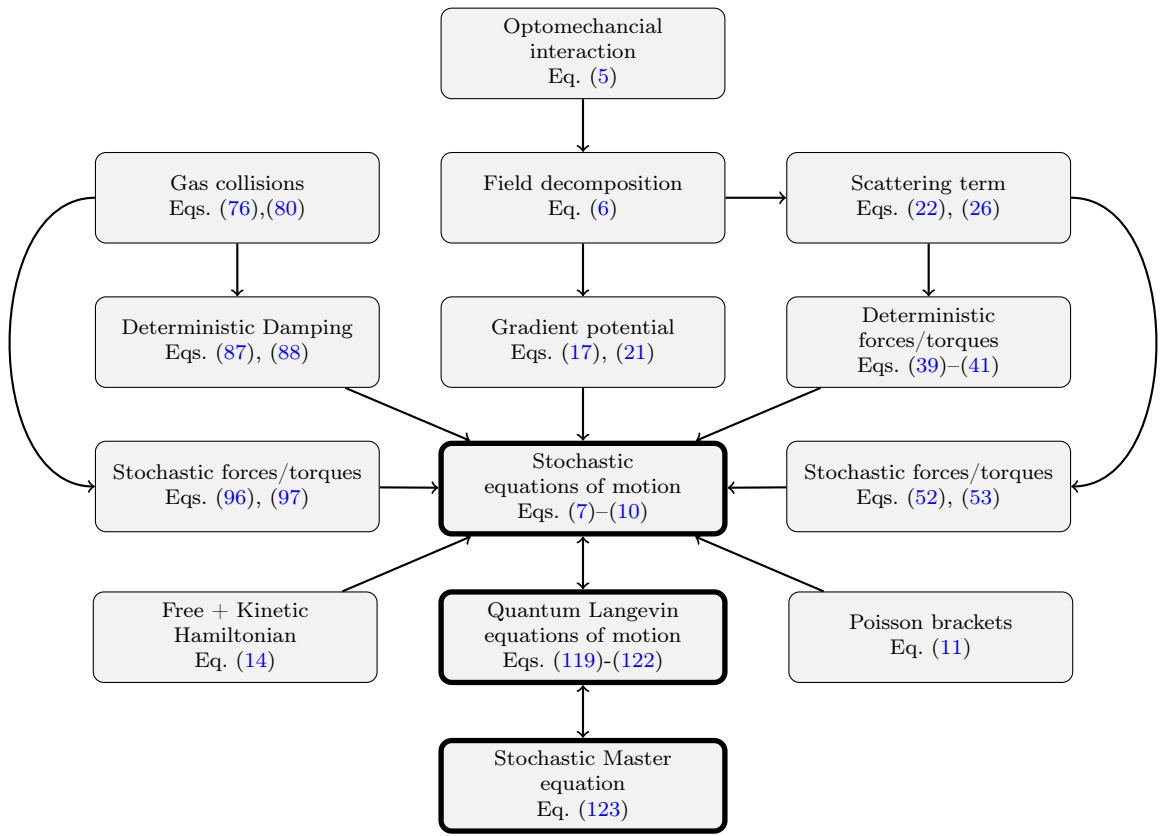
\begin{figure}[h!]
\centering
\begin{tikzpicture}[
    node distance=.7cm and .8cm,
    every node/.style={font=\footnotesize, align=center},
    block/.style={draw, rectangle, rounded corners, text width=3.5cm, minimum height=1.2cm, text centered, fill=gray!10},
    arrow/.style={->, thick},
]

\node[block] (opticalHam) {Optomechancial interaction\\ Eq.~\eqref{eq:opt}};
\node[block, below=of opticalHam] (Efields) {Field decomposition\\ Eq.~\eqref{eq:fields}};
\node[block, left=of Efields] (LELequs) {Gas collisions \\Eqs.~\eqref{eq:plab0},\eqref{eq:E-L}};
\node[block, below=of LELequs] (detdamp) {Deterministic Damping\\Eqs.~\eqref{eq:Pd}, \eqref{eq:Pid}};
\node[block, below=of detdamp] (stochcoll) {Stochastic forces/torques\\Eqs.~\eqref{eq:Pnc3}, \eqref{eq:Pinc3}};
\node[block, below=of Efields] (Hgradient) {Gradient potential\\ Eqs.~\eqref{eq:gradient}, \eqref{eq:Hgradient}};
\node[block, right=of Efields] (Hscattering) {Scattering term\\ Eqs.~\eqref{eq:scattering}, \eqref{eq:magic}};
\node[block, line width=2pt, below=of Hgradient] (dynamics) {Stochastic equations of motion\\ Eqs.~\eqref{eq:dr}–\eqref{eq:dpi}};
\node[block, below=of Hscattering] (deterministic) {Deterministic forces/torques\\ Eqs.~\eqref{eq:ds1}–\eqref{eq:ds3}};
\node[block, below=of deterministic] (stochastic) {Stochastic forces/torques\\ Eqs.~\eqref{eq:Pns}, \eqref{eq:Pins}};
\node[block, line width=2pt, below=of dynamics] (quantum) {Quantum Langevin\\equations of motion\\ Eqs.~\eqref{eq:drQ}-\eqref{eq:dpiQ}};
\node[block, left=of quantum] (Hfree) {Free + Kinetic Hamiltonian\\ Eq.~\eqref{eq:Hfree}};
\node[block, right=of quantum] (PB) {Poisson brackets\\ Eq.~\eqref{eq:Pb}};
\node[block, line width=2pt, below=of quantum] (master) {Stochastic Master\\equation \\ Eq.~\eqref{eq:conditional}};

\draw[arrow] (opticalHam) -- (Efields);
\draw[arrow] (Efields) -- (Hgradient);
\draw[arrow] (Efields) -- (Hscattering);
\draw[arrow] (Hgradient) -- (dynamics);
\draw[arrow] (Hscattering)  to [out=0, in=0]  (stochastic);
\draw[arrow] (Hscattering) -- (deterministic);
\draw[arrow] (stochastic) to  (dynamics);
\draw[arrow] (deterministic) to (dynamics);
\draw[arrow] (PB) -- (dynamics);
\draw[arrow] (Hfree) -- (dynamics);
\draw[arrow] (LELequs) -- (detdamp);
\draw[arrow] (LELequs) to [out=180, in=180] (stochcoll);
\draw[arrow] (stochcoll) to (dynamics);
\draw[arrow] (detdamp) to (dynamics);
\draw[arrow,<->] (dynamics) to (quantum);
\draw[arrow,<->] (quantum) to (master);

\end{tikzpicture}
\caption{Overview diagram of the theoretical framework. Arrows indicate how key equations and components flow into the construction of the full dynamics of the nanoparticle roto-translational motion. We begin by briefly outlining the general modelling strategy for describing roto-translational dynamics of a levitated nanoparticle. At the core of the framework lies a Hamiltonian formulation of classical mechanics extended to include stochastic effects due to gas collisions and photon recoil. The dynamics is captured by a set of stochastic differential equations that describe both translational and rotational degrees of freedom. These include deterministic terms from conservative potentials (e.g., gradient forces), non-conservative radiation pressure forces (scattering), and stochastic Langevin-type contributions.
We will first present the general classical model and outline how to numerically simulate its dynamics. Then, we will highlight how the classical model connects to its quantum counterpart, and point out the key structural similarities with the quantum Langevin equations derived later. Throughout, we rely on standard tools from classical Hamiltonian mechanics, such as Poisson brackets and the Euler-Lagrange formalism, to maintain a transparent connection between physical intuition and analytical tractability. }
\label{fig:overview_diagram}
\end{figure}

%
In this section, we provide an overview of common theoretical tools employed to describe levitated resonators in the context of roto-translational motion in optomechanics. The ability to model the experimental system precisely is crucial, particularly when investigating extremely small effects, such as those arising from roto-translational motion in optomechanics. These tools enable researchers to analyze and predict the behavior of optically trapped particles, taking into account different particle shapes and focusing on the intricate dynamics of their roto-translational motion. We will begin by introducing the classical model of motion, followed by a discussion of numerical simulation strategies. Finally, we comment on how this classical framework connects to quantum optomechanical descriptions via the quantum Langevin equations, which we derive later.

To guide the reader through the structure of the theoretical framework, we show in Fig.~\ref{fig:overview_diagram}  an overview diagram depicting the components and their key equations and how they flow into the construction of the full roto-translational dynamics.

\subsection{The classical model\label{sec:model}}

\begin{figure}
\includegraphics[width=\textwidth]{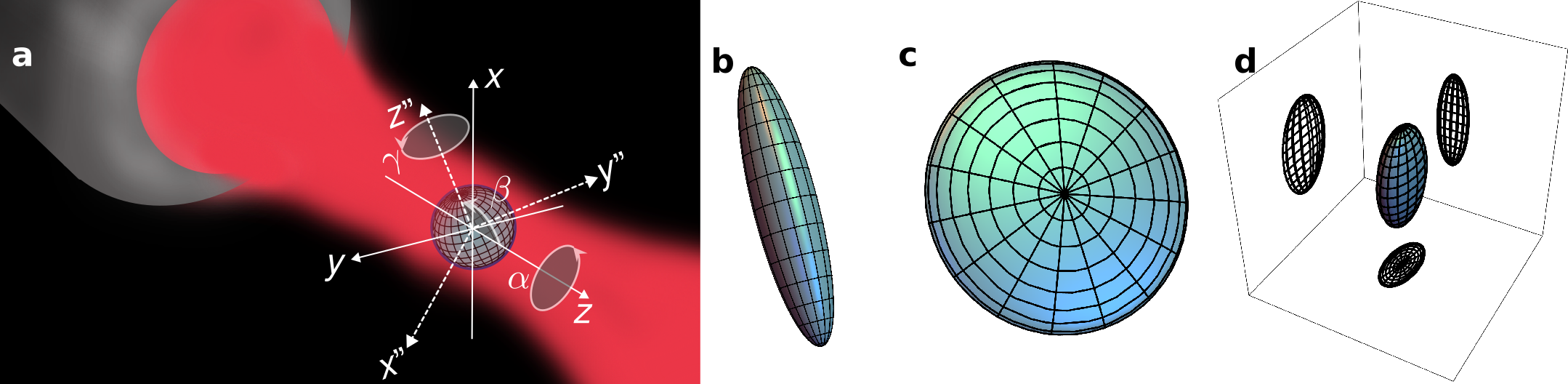}
\caption{
Schematics of the different geometries studied in an optical trap using numerical simulations in this work. 
(a) Illustration of a spherical nanoparticle trapped by a focused laser beam (optical tweezer), showing both the laboratory frame (\(x, y, z\)) and the body frame (\(x'', y'', z''\)) of the object. The three rotational degrees of freedom (Euler angles) are indicated: \(\alpha\) (rotation about the lab-frame \(z\)-axis), \(\beta\) (tilt about the \(y\)-axis), and \(\gamma\) (twist about the body-frame \(z''\)-axis). 
(b) Symmetric prolate ellipsoid, i.e., elongated along one axis, with principal semi-axes \(R_1 = R_2 < R_3\), where \(R_1\) and \(R_2\) correspond to the radii in the transverse directions and \(R_3\) is the longer radius along the symmetry axis. 
(c) Symmetric oblate ellipsoid, i.e., flattened along one axis, with \(R_1 = R_2 > R_3\), where \(R_3\) is the shorter radius along the symmetry axis. 
(d) Asymmetric ellipsoid with three distinct principal semi-axes \(R_1 < R_2 < R_3\), resulting in asymmetry in all three spatial dimensions. The projections onto each coordinate plane highlight the different cross-sections of the object. 
}
\label{fig:exp} 
\end{figure}

We consider the situation shown in Fig.~\ref{fig:exp}\,a where a particle is trapped in an optical tweezer. In Figs.~\ref{fig:exp}\,b-\ref{fig:exp}\,d we show the archetypal nanoparticle shapes that we will explore. Apart from the simple sphere (Fig.~\ref{fig:exp}\,a), we will consider general ellipsoidal particles, which will allow us to explore the main categories of rigid bodies. That is, for cylindrical symmetry, a prolate (Fig.~\ref{fig:exp}\,b) and oblate (Fig.~\ref{fig:exp}\,c) ellipsoid can represent a rod-like and disk-like particle while a non-degenerate ellipsoid represents an asymmetric-top particle (Fig.~\ref{fig:exp}\,d). Since we will limit the analysis to the Rayleigh regime (particle size much smaller than the trapping field wavelength), these four geometries give a full phenomenological spectrum of the observable dynamics.

The starting point of our analysis is the optomechanical interaction Hamiltonian given by~\cite{jackson1998classical,millen2020optomechanics}: 
\begin{equation}
    H_{\text{opt}}=-\frac{1}{2} V\epsilon_0 (\bm{E}_\text{tot}^\top \chi_\text{lab} \bm{E}_\text{tot})\label{eq:opt}
\end{equation}
where $V$ is the particle volume, $\epsilon_0$ is the electric permittivity of free space, $\chi_\text{lab}$ is the susceptibility tensor in the laboratory reference frame, and $\bm{E}_\text{tot}$ is the total electric field. The use of a tensorial susceptibility $\chi_\text{lab}$ is essential for capturing the anisotropic optical response of non-spherical particles. This $3\times3$ matrix describes how the induced dipole moment aligns with the local electric field, taking into account different polarizabilities along the principal axes of the nanoparticle. The total electric field can be decomposed as:
\begin{equation}
    \bm{E}_\text{tot} =\bm{E}_\text{d}+\bm{E}_\text{f},\label{eq:fields}
\end{equation}
where $\bm{E}_\text{d}$ ($\bm{E}_\text{f}$) is the electric field generating the optical trap (the electric field scattered off the nanoparticle). From Eqs.~\eqref{eq:opt} and \eqref{eq:fields} we find a contribution $\propto \bm{E}_\text{d}^2$ and $\propto \bm{E}_\text{d} \bm{E}_\text{f}$: the former term gives rise to the gradient forces/torques, while the latter gives rise to non-conservative terms (deterministic and stochastic radiation pressure forces/torques). These two contributions are treated in detail in Sec.~\ref{sec:gradient} (for the gradient potential) and Secs.~\ref{sec:det}–\ref{sec:stoc} (for the deterministic and stochastic scattering terms, respectively).

The typical experimental implementation exploits high numerical aperture objectives or lenses to obtain a diffraction limited focus. The accurate description of the resulting electric field can be fairly complicated especially if aberrations are significant~\cite{novotny2012principles}. We thus consider a simplified description that retains the main features of the true field and describe it as an stigmatic Gaussian beam, i.e. free of aberrations, with a general polarization state.

\subsubsection{The dynamics\label{sec:Dynamics}}
To specify the position of a nanoparticle we need three numbers $(x,y,z)$, and to specify its orientation we need three angles $(\alpha,\beta,\gamma)$: in total we have six degrees of freedom as seen in Fig.~\ref{fig:exp}a. In this work we will use a Hamiltonian formalism, where for each degree of freedom we have to introduce also the conjugate momentum, and as such, the location of the system in phase space is characterized by twelve numbers. The corresponding dynamics is given by coupled stochastic differential equations:
\begin{equation}
    \dd\boldsymbol{r}= \boldsymbol{\partial_{p}}H_{\text{free}}\dd t\;,
    \label{eq:dr}
\end{equation}
\begin{equation}
    \dd\boldsymbol{p}=  -\boldsymbol{\partial_{r}}H_{\text{gradient}}\dd t+\dd\boldsymbol{p}^{\text{(ds)}}+\dd\boldsymbol{p}{}^{\text{(dc)}}+\dd\boldsymbol{p}{}^{\text{(ss)}}+\dd\boldsymbol{p}{}^{\text{(sc)}}\;,
    \label{eq:dp}
\end{equation}
\begin{equation}
    \dd\boldsymbol{\phi}=  \boldsymbol{\partial_{\pi}}(H_{\text{free}}+H_{\text{gradient}})\dd t\;,
    \label{eq:dphi}
\end{equation}
\begin{equation}
    \dd\boldsymbol{\pi}= -\boldsymbol{\partial_{\phi}}(H_{\text{free}}+H_{\text{gradient}})\dd t+\dd\boldsymbol{\pi}^{\text{(ds)}}+\dd\boldsymbol{\pi}{}^{\text{(dc)}}+\dd\boldsymbol{\pi}{}^{\text{(ss)}}+\dd\boldsymbol{\pi}{}^{\text{(sc)}}\;,
    \label{eq:dpi}
\end{equation}
\noindent where $\boldsymbol{r}=(x,y,z)^{\top}$, $\boldsymbol{p}=(p_{x},p_{y},p_{z})^{\top}$, $\boldsymbol{\phi}=(\alpha,\beta,\gamma)^{\top}$, and $\boldsymbol{\pi}=(\pi_{\alpha},\pi_{\beta},\pi_{\gamma})^{\top}$are the center of mass position, center of mass conjugate momentum, angle, and angle conjugate momentum, respectively. The angles $\alpha$, $\beta$, $\gamma$ denote the Euler angles in the z-y'-z" convention (see Fig.~\ref{fig:exp} a)). $H_{\text{free}}$ and $H_{\text{gradient}}$ denote the free Hamiltonian and the gradient potential respectively; $d\boldsymbol{p}^{\text{(ds)}}$, $d\boldsymbol{{\pi}}^{\text{(ds)}}$ denote the non-conservative deterministic photon scattering (ds) terms, $d\boldsymbol{p}^{\text{(dc)}}$, $d\boldsymbol{{\pi}}^{\text{(dc)}}$ the non-conservative deterministic gas collision (dc) terms, $d\boldsymbol{p}^{\text{(ss)}}$, $d\boldsymbol{{\pi}}^{\text{(ss)}}$ the stochastic photon scattering (ss) terms, and $d\boldsymbol{p}^{\text{(sc)}}$, $d\boldsymbol{{\pi}}^{\text{(sc)}}$ the stochastic gas collision (sc) terms. 

We will briefly discuss how to solve Eqs.~(\ref{eq:dr})-(\ref{eq:dpi}) using numerical methods in Sec.~\ref{subsec:Characterization-of-the}. Loosely speaking, Eqs.~(\ref{eq:dr})-(\ref{eq:dpi}) can be interpreted as prescriptions for the evolution of the system in terms of small increments and a discretized time-step, similarly as one does in the numerical resolution of (non-stochastic) differential equations. For more details about the mathematical formalism, known as Ito calculus, see for example~\cite{evans2012introduction}. 
 
 We now discuss separately the conservative, non-conservative deterministic, and the stochastic terms on the right hand-side of Eqs.~(\ref{eq:dr})-(\ref{eq:dpi}). In particular, we will provide a simple classical derivation of the optomechanical forces and torques starting from the optomechanical interaction Hamiltonian in Eq.~\eqref{eq:opt}. The analysis relies on basic principles of Hamiltonian mechanics, in particular on the Poisson bracket
 \begin{equation}
      \{ A , B \} =\frac{\partial A}{\partial\mathsf{x}} \frac{\partial B}{\partial\mathsf{p}} -\frac{\partial A}{\partial\mathsf{p}} \frac{\partial B}{\partial\mathsf{x}}, \label{eq:Pb}
 \end{equation}
 where $\mathsf{x}$ ($\mathsf{p}$) denotes the generalized position (generalized conjugate momentum). Using Eq.~\eqref{eq:Pb} one can write Hamilton's equations of motion as  $\dot{\mathsf{x}}=\{ \mathsf{x} , H \} $ and $\dot{\mathsf{p}}=\{ \mathsf{p} , H \} $, where $H$ denotes the Hamiltonian~\cite{goldstein2011classical}. 
 
 \subsubsection{Conservative terms}

The dynamics of a nanoparticle can be modelled as a rigid rotor~\cite{goldstein2011classical,arnol2013mathematical} which is characterized by three angles (and their conjugate angle momenta). To describe the orientation of the particle we will adopt the  z-y'-z'' convention illustrated Fig.~\ref{fig:exp}a. The transformation from the particle reference frame (x'',y'',z'' axis) to the laboratory reference frame (x,y,z axis) is given by applying three rotations: first around the z''-axis (angle $\gamma$), then around the y'-axis (angle $\beta$), and finally around the z-axis (angle $\alpha$). Using a matrix representation we can write the transformation explicitly as:
\begin{equation}
    \mathcal{R}\equiv 
\left[\begin{array}{ccc}
\cos(\alpha) & -\sin(\alpha) & 0 \\
\sin(\alpha) & \cos(\alpha) & 0 \\
0 & 0 & 1 \\
\end{array}\right]
\left[\begin{array}{ccc}
\cos(\beta) & 0 & \sin(\beta)  \\
0 & 1 & 0 \\
-\sin(\beta) & 0 & \cos(\beta)  
\end{array}\right]
\left[\begin{array}{ccc}
\cos(\gamma) & -\sin(\gamma) & 0 \\
\sin(\gamma) & \cos(\gamma) & 0 \\
0 & 0 &1
\end{array}\right]. \label{eq:R}
\end{equation}
The matrix $\mathcal{R}$ encodes the orientation of the nanoparticle with respect to the laboratory reference frame. In particular, the nanoparticle is characterized by four tensors: 
\begin{equation}
    I_\text{lab}= \mathcal{R} I \mathcal{R}^\top, \qquad \chi_\text{lab}=\mathcal{R} \chi\mathcal{R}^\top, 
    \qquad 
    F^\text{(t)}_\text{lab}= \mathcal{R} F^\text{(t)} \mathcal{R}^\top, \qquad  F^\text{(r)}_\text{lab}=\mathcal{R} F^\text{(r)}\mathcal{R}^\top, 
    \label{eq:tensors}
\end{equation}
where $I$ is the moment of inertia tensor, $\chi$ is the susceptibility tensor,  $F^\text{(t)}$ is the translational friction tensor, and $F^\text{(r)}$ is the rotational friction tensor. The former two encode inertial and  optical properties, respectively, while the latter two encode the interaction of the nanoparticle with the residual gas particles.
Here we have assumed for simplicity that  the tensors are diagonal in the same frame of reference: $I\equiv \text{diag}(I_1,I_2,I_3)$, $\chi\equiv \text{diag}(\chi_1,\chi_2,\chi_3)$, $F^\text{(t)}\equiv \text{diag}(F^\text{(t)}_1,F^\text{(t)}_2,F^\text{(t)}_3)$, and $F^\text{(r)}\equiv \text{diag}(F^\text{(r)}_1,F^\text{(r)}_2,F^\text{(r)}_3)$. In a more general situation one might need to consider four separate body frames, and hence three additional rotational matrices for each separate body frame, which would significantly complicate the analysis. This assumption is physically reasonable for many common nanoparticle geometries (e.g., spheroids, ellipsoids or rods) where the principal axes of inertia, polarizability and friction tensors naturally align due to the structural symmetry of the particle or the homogeneous material composition.

\subsubsection{Kinetic terms}\label{sec:kinetic}
The kinetic terms of the Hamiltonian can be written explicitly as~\cite{goldstein2011classical,arnol2013mathematical}
\begin{alignat}{1}
H_{\text{free}}=\frac{p_{x}{}^{2}+p_{y}{}^{2}+p_{z}{}^{2}}{2M}+\bigg( & \frac{\mathcal{P}_{1}(\beta,\gamma)}{2I_{1}}+\frac{\mathcal{P}_{2}(\beta,\gamma)}{2I_{2}}+\frac{\pi_{\gamma}{}^{2}}{2I_{3}}\bigg),\label{eq:Hfree}
\end{alignat}

where 
\begin{alignat}{1}
 \mathcal{P}_{1}(\beta,\gamma)& =\csc^{2}(\beta)[\cos(\gamma)(\pi_{\alpha}-\pi_{\gamma}\cos(\beta))-\text{\ensuremath{\pi_{\beta}}}\sin(\beta)\sin(\gamma)]^{2}, \label{eq:P1}\\
\mathcal{P}_{2}(\beta,\gamma) & =\csc^{2}(\beta)[\sin(\gamma)(\text{\ensuremath{\pi_{\alpha}}}-\pi_{\gamma}\cos(\beta))+\pi_{\beta}\sin(\beta)\cos(\gamma)]^{2} \label{eq:P2},
\end{alignat}
$I=(I_{1},I_{2},I_{3})$ is the moment of inertia tensor in the body-frame, $M$ is the total mass of the system, and $\csc(~\cdot~)\equiv 1/\sin (~\cdot~)$. We note that the rotational kinetic terms in Eqs.~\eqref{eq:P1} and \eqref{eq:P2} couple the angular dynamics (unlike the translational dynamics which for a free particle is completely independent along the three axis).
In Sec.~\ref{sec:Numerical-analysis} we will see that the cross-couplings terms (with additional ones arising also from the optomechanical interaction discussed below) gives rise to highly non-trivial rotational dynamics. 

\subsubsection{Gradient potential}
\label{sec:gradient}
 We combine Eqs.~\eqref{eq:opt} and \eqref{eq:fields} and focus on the term $\propto\bm{E}_d^2$. In particular,  the gradient potential is given by:
\begin{equation}
    H_{\text{gradient}}=-\frac{1}{2}V\epsilon_0 \bm{E}_d^\top \chi_\text{lab} \bm{E}_d, \label{eq:gradient}
\end{equation}
i.e., the electric field $\bm{E}_d$ induces the dipole  $\chi_\text{lab} \bm{E}_d$. We consider the electric field of a Gaussian beam given by
\begin{equation}
    \bm{E}_d= i\sqrt{\frac{P}{2\epsilon_0c\sigma_L}} (\bm{\epsilon}_d\, u\, e^{i\omega_L t} -\bm{\epsilon}_d^* u^* e^{-i\omega_L t}), \label{eq:Ed}
\end{equation}
where $P$ is the laser power, $c$ is the speed of light, $\sigma_{L}$ is an effective beam cross-section, $u$ is the mode function,  $\omega_L$ is the angular frequency, and $\boldsymbol{{\epsilon}}_{d}=(b_{x},ib_{y},0)^{\top}$ is the  polarization vector.
Since $b_{x}^{2}+b_{y}^{2}=1$, it is convenient to use $b_{x}=\cos{\psi}$ and $b_{y}=\sin{\psi}$ so that $\psi$ fully parametrizes the field polarization state. In particular, we consider $u$ to be a modified Gaussian mode function:
\begin{equation}
u(\boldsymbol{r})=\frac{w_{0}}{w(z)}\text{exp}\left(-\frac{x^{2}/ a_{1}+y^{2}a_{1}}{w(z)^{2}}\right)e^{-ik_L z},\label{eq:umode}
\end{equation}
 where $\bm{r}=(x,y,z)^\top$, $a_{1}$ quantifies the asymmetry between the $x$ and $y$ axis, $w(z)=w_{0}\sqrt{\left(1+\frac{z^{2}}{z_{R}^{2}}\right)}$ (with $w_{0}$ an effective beam waist), $z_{R}$ is the Rayleigh range, $k_L=\omega_L/c=\frac{2\pi}{\lambda}$, and $\lambda$ is the laser wavelength. The mode function in Eq.~\eqref{eq:umode} represents a first-order approximation and allows to simplify considerably the analytical description. For example, the effective beam cross-section can be computed by integrating the square of the mode function over the transverse plane, i.e.,  
\begin{equation}
\sigma_{L}= \int_{-\infty}^{\infty}\int_{-\infty}^{\infty} \vert u(x,y,0)\vert^2 \,dx\, dy=\pi w_{0}^{2}/2.    
\end{equation}
From Eqs.~\eqref{eq:Ed} and \eqref{eq:umode} we also find the simple relation $P=I_0 \sigma_L$, where $I_0$ is the time-averaged intensity at the beam focus. However, a more refined approximation of the field of a diffraction-limited focus is typically necessary (this is discussed in~\ref{app:gauss}). 

Combining Eqs.~\eqref{eq:gradient}-\eqref{eq:umode} we find after time-averaging the following gradient potential:
\begin{alignat}{1}
H_{\text{gradient}}=&-\frac{VP}{2 c \sigma_{L}}\vert u(\boldsymbol{r})\vert^{2}  \bigg\{b_x^{2}\big[\text{\ensuremath{\chi_{1}}}\big(\cos(\alpha)\cos(\beta)\cos(\gamma)-\sin(\alpha)\sin(\gamma)\big)^{2}\nonumber \\
 & \qquad+\text{\ensuremath{\chi}}_{2}\big(\cos(\alpha)\cos(\beta)\sin(\gamma)+\sin(\alpha)\cos(\gamma)\big)^{2}+\text{\ensuremath{\chi}}_{3}\cos^{2}(\alpha)\sin^{2}(\beta)\big]\nonumber \\
 & \quad+b_y^{2}\big[\text{\ensuremath{\chi}}_{1}\big(\sin(\alpha)\cos(\beta)\cos(\gamma)+\cos(\alpha)\sin(\gamma)\big)^{2}\nonumber \\
 & \qquad+\text{\ensuremath{\chi}}_{2}\big(\cos(\alpha)\cos(\gamma)-\sin(\alpha)\cos(\beta)\sin(\gamma)\big)^{2}+\text{\ensuremath{\chi}}_{3}\sin^{2}(\alpha)\sin^{2}(\beta)\big]\bigg\}.\label{eq:Hgradient}
\end{alignat}
Note that Eq.~\eqref{eq:Hgradient} introduces couplings among translational degrees of freedom (encoded in the factor $\vert u(\boldsymbol{r})\vert^{2}$), the rotational degrees of freedom (encoded in the bracket $\{~\cdot~\}$ spanning four rows), as well as cross-couplings between translational and rotational degrees of freedom (arising from the mixed terms from the former two contributions).

\subsubsection{Non-conservative optical terms} \label{sec:ncopt}

In this section we derive the non-conservative optical terms affecting the motion of the nanoparticle. Non-conservative optical forces and torques arise from Eqs.~\eqref{eq:opt} and \eqref{eq:fields}, in particular, from the cross-coupling term $\propto \bm{E}_f \bm{E}_d$ :
\begin{equation}
    H_{\text{scattering}}=-V\epsilon_0 (\bm{E}_f^\top \chi_\text{lab} \bm{E}_d), \label{eq:scattering}
\end{equation}
where we recall that $\bm{E}_d$ is the electric field generated by the laser, and $\bm{E}_f$ denotes the free electric field. We can interpret Eq.~\eqref{eq:scattering} in the following way: the incoming electric field (factor $\bm{E}_d$) is scattered by the nanoparticle (factor $V\chi_\text{lab}$) into the outgoing electric field (factor $\bm{E}_f$), i.e., $\bm{E}_d$ drives the initially empty modes of $\bm{E}_f$. The outgoing electric field can be decomposed in plane-wave modes~\cite{loudon2000quantum}:
\begin{equation}
    \bm{E}_\text{f}=i\sum_\nu \int \frac{d\bm{k}}{(2\pi)^3} \sqrt{\frac{\hbar \omega_k}{2\epsilon_0}} \left(
    \bm{\epsilon}_{\bm{k},\nu} e^{i \bm{k}\cdot \bm{r}} e^{-i \omega_k t} a_{\bm{k},\nu}
    -\bm{\epsilon}_{\bm{k},\nu}^* e^{-i \bm{k}\cdot \bm{r}} e^{i \omega_k t} a_{\bm{k},\nu}^* \right), \label{eq:Ef}
\end{equation}
where $\bm{k}$ ($\nu$) is the wave-vector (polarization index), $\bm{\epsilon}_{\bm{k},\nu}$ the polarization vector,  $\omega_k= \vert \bm{k} \vert c$ is the angular frequency, and $a_{\bm{k},\nu}$ denotes the complex-valued plane wave modes. The modes satisfy the following property
 \begin{equation}
     \{a_{\bm{k},\nu},a_{\bm{k}',\nu'}^*\}_\text{Pb} = -\frac{i}{\hbar}(2\pi)^3 \delta^{(3)}(\bm{k}-\bm{k}') \delta_{\nu,\nu'},  \label{eq:key}
 \end{equation}
 where $\{~\cdot~,~\cdot~\}_\text{Pb}$ denotes the Poisson bracket. Eq.~\eqref{eq:key} can be used in Eq.~\eqref{eq:Pb} by writing the modes as 
 \begin{equation}
 \label{eq:mode}
     a_{\bm{k},\nu} = (X_{\bm{k},\nu} + i P_{\bm{k},\nu})/\sqrt{2},
 \end{equation}
 where \( X_{\bm{k},\nu} \) and \( P_{\bm{k},\nu} \) correspond to the amplitude and phase quadratures of the field, respectively. 
 
We insert Eqs.~\eqref{eq:Ef} and \eqref{eq:Ed} in Eq.~\eqref{eq:scattering} to obtain the interaction Hamiltonian expressed in terms of the electromagnetic modes:
\begin{equation}
    H_{\text{scattering}} = \sum_\nu \int \frac{d\bm{k}}{(2\pi)^3}  \sqrt{\frac{\hbar \omega_k P V^2}{4 c \sigma_L}}a_{\bm{k},\nu} e^{-i(\omega_k-\omega_L)t} A_{\bm{n},\nu}^* + \text{c.c.},
       \label{eq:magic}
\end{equation}
where c.c. denotes the complex conjugate, and we have introduced the nanoparticle operator
\begin{equation}
 A_{\bm{k},\nu}= \bm{\epsilon}_{\bm{n},\nu}^\top \chi_\text{lab}  \bm{\epsilon}_d^*    u^*  e^{-i \bm{k}\cdot \bm{r}} .\label{eq:nano}
\end{equation}
We recall that $u\equiv u(\bm{r})$, $\chi_\text{lab}\equiv \chi_\text{lab}(\bm{\phi})$, and hence $A_{\bm{k},\nu }\equiv A_{\bm{k},\nu}(\bm{r}, \bm{\phi})$, where  $\bm{r}=(x,y,z)$ and $\bm{\phi}=(\alpha,\beta,\gamma)$. 

The next step is to now eliminate $a_{\bm{k},\nu}$ from Eq.~\eqref{eq:magic} and obtain the resulting dynamics for the nanoparticle characterized by $A_{\bm{n},\nu}$. To this end we compute the time-evolution of the electromagnetic mode:
\begin{equation}
    \dot{a}_{\bm{k},\nu}= \{a_{\bm{k},\nu},H_{\text{scattering}}\}= -i\sqrt{\frac{\omega_k P V^2}{4 \hbar c \sigma_L}}  A_{\bm{k},\nu} e^{i(\omega_k-\omega_L)t},\label{eq:adot}
\end{equation}
where we have used Eqs.~\eqref{eq:key} and \eqref{eq:magic}. We can then readily integrate Eq.~\eqref{eq:adot} to find:
\begin{equation}
    a_{\bm{k},\nu} (t) =   -i\sqrt{\frac{\pi^2\omega_k P V^2}{4 \hbar c \sigma_L}}  A_{\bm{k},\nu} \delta (\omega_k-\omega_L) + a_{\bm{k},\nu} (0),
    \label{eq:at}
\end{equation}
where we have used $\int_0^t e^{i(\omega_k-\omega_L)s} ds \approx \pi \delta (\omega_k-\omega_L)$, i.e., we are making the approximation that the incoming and outgoing field mode have the same frequency.  

We next calculate the equation of motion of the nanoparticle momentum:
\begin{alignat}{1}
    \dot{\mathsf{p}}_j&=\{\mathsf{p}_j,H_{\text{scattering}}\} =-\frac{\partial H_{\text{scattering}}}{\partial \mathsf{x}_j}\nonumber\\
    &=\sum_\nu \int \frac{d\bm{k}}{(2\pi)^3}  \sqrt{\frac{\hbar \omega_k P V^2}{4 c \sigma_L}}a_{\bm{k},\nu} e^{-i(\omega_k-\omega_L)t} \frac{\partial A_{\bm{k},\nu}^*}{\partial \mathsf{x}_j} + \text{c.c.}, \label{eq:pdot}
\end{alignat}
where ${\mathsf{p}}_j=p_x,p_y,p_z, \pi_\alpha, \pi_\beta, \pi_\gamma$ and $\mathsf{x}_j=x,y,z,\alpha, \beta,\gamma$. We now finally insert Eq.~\eqref{eq:at} into \eqref{eq:pdot} to find:
\begin{alignat}{1}
    \dot{\mathsf{p}}_j = &-i\sum_\nu \int \frac{d\bm{k}}{(2\pi)^3}  \frac{\pi \omega_k P V^2}{4 c \sigma_L}  \delta (\omega_k-\omega_L) e^{-i(\omega_k-\omega_L)t} A_{\bm{k},\nu} \frac{\partial A_{\bm{k},\nu}^*}{\partial \mathsf{x}_j}  - \text{c.c.} \nonumber\\
    &+  \sum_\nu \int \frac{d\bm{k}}{(2\pi)^3}  \sqrt{\frac{\hbar \omega_k P V^2}{4 c \sigma_L}} a_{\bm{k},\nu} (0) e^{-i(\omega_k-\omega_L)t} \frac{\partial A_{\bm{k},\nu}^*}{\partial \mathsf{x}_j}  +\text{c.c.}. \label{eq:p2}
\end{alignat}
The right-hand side of Eq.~\eqref{eq:p2} will reduce to the known optical forces and torques affecting the nanoparticle: the contribution arising from the first (second) line is discussed separately in Sec.~\ref{sec:det} (Sec.~\ref{sec:stoc}).

\subsubsection{Deterministic radiation pressure forces and torques} \label{sec:det}
We rewrite the first line Eq.~\eqref{eq:p2} in spherical coordinates:
\begin{equation}
    \dot{\mathsf{p}}_j = -i\sum_\nu \int d\bm{n} \int d\omega_k 
      \frac{\pi \omega_k^3 P V^2}{4 (2\pi)^3 c^4 \sigma_L}  \delta (\omega_k-\omega_L) e^{-i(\omega_k-\omega_L)t} A_{\bm{k},\nu} \frac{\partial A_{\bm{k},\nu}^*}{\partial \mathsf{x}_j}  - \text{c.c.} \label{eq:drpft1}
\end{equation}
where we have changed the integral measure $d\bm{k}=k^2 dk d\bm{n}$, and used the relation $\omega_k=k c$. From Eq.~\eqref{eq:drpft1} we can then integrate in $d\omega_k$ to readily find:
\begin{alignat}{1}
    \dot{\mathsf{p}}_j =& -i\hbar \frac{\gamma_{s}}{2} \sum_\nu \int d\bm{n}  A_{\bm{n},\nu} \partial_j A_{\bm{n},\nu}^* -  \text{c.c.} \nonumber\\
    &= - \hbar \gamma_{s} \sum_\nu \int d\bm{n} ~\text{Im} ( A_{\bm{n},\nu}^* \partial_j A_{\bm{n},\nu}),\label{eq:p3}
\end{alignat}
where we have used $\int_0^\infty d\omega_k \delta (\omega_k-\omega_L) F(\omega_k)=F(\omega_L)$ (with $F$ denoting the integrand), the particle operator $A_{\bm{n},\nu}$ corresponds to $A_{\bm{k},\nu}\vert_{\omega_k=\omega_L}$, and we have used the compact notation for the partial derivative $\partial_j=\frac{\partial}{\partial \mathsf{x}_j}$. In Eq.~\eqref{eq:p3} have also introduced the scattering rate:
\begin{equation}
    \gamma_{s}=\frac{\pi \omega_L^3 P V^2}{ (2\pi)^3 2 c^4 \sigma_L \hbar}=\frac{\tilde{\sigma}_{R}}{\sigma_{L}}\frac{P}{\hbar\omega_{L}}, \label{eq:gammas}
\end{equation}
where $\tilde{\sigma}_{R}=\frac{\pi^{2}V_{0}^{2}}{\lambda^{4}}$ is an effective scattering cross-section, and we recall that the cross-section $\sigma_{L}=\pi w_0^2/2$ characterizes the Gaussian beam ($w_0$ is the beam waist).

By inserting Eq.~\eqref{eq:nano} in Eq.~\eqref{eq:p3} we can find the deterministic radiation pressure forces and torques. To illustrate the procedure, we consider the translational motion along the z-axis (i.e., we set $\mathsf{p}_j=p_z$ and $\mathsf{x}_j=z$):
\begin{alignat}{1}
    \dot{p}_z &= - \hbar \gamma_{s} \sum_\nu \int d\bm{n} ~\text{Im} ( A_{\bm{n},\nu}^* \partial_z A_{\bm{n},\nu})\nonumber\\
    & =- \hbar \gamma_{s} \sum_\nu \int d\bm{n} ~\text{Im} \big( (
    \bm{\epsilon}_{\bm{n},\nu}^\top \chi_\text{lab}  \bm{\epsilon}_d^*    u^*  e^{-i \bm{k}\cdot \bm{r}})^* 
    \partial_z \bm{\epsilon}_{\bm{n},\nu}^\top \chi_\text{lab}  \bm{\epsilon}_d^*    u^*  e^{-i \bm{k}\cdot \bm{r}}\big), \nonumber\\
     & =- \hbar \gamma_{s} \sum_\nu \int d\bm{n} ~\text{Im} \big( i (k_L -k n_3)  \vert u\vert^2
    \vert \bm{\epsilon}_{\bm{n},\nu}^{*\top} \chi_\text{lab}^*  \bm{\epsilon}_d \vert^2  \big). \label{eq:p333}\end{alignat}
where in the last line, we have used the fact that $u\propto e^{-ik_L z}$ (see Eq.~\eqref{eq:umode}). To resolve the summation over the polarizations, we will now use the completeness relation
\begin{equation}
     \sum_\nu (\bm{\epsilon}_{\bm{n},\nu}^{*\top})_i (\bm{\epsilon}_{\bm{n},\nu}^\top )_j = \delta_{i,j}- \bm{n}_i \bm{n}_j, \label{eq:cr}
\end{equation}
where $\bm{n}_i$ is the unit vector along the i-th axis. Using Eq.~\eqref{eq:cr} in  Eq.~\eqref{eq:p333} we then find:
\begin{alignat}{1}
    \dot{p}_z &= - \hbar \gamma_{s} \int d\bm{n} ~\text{Im} \big( i k_L (1 - n_3)  \vert u\vert^2 (\delta_{ij} -n_i n_j) (\chi_\text{lab}^*  \bm{\epsilon}_d )_i (\chi_\text{lab}  \bm{\epsilon}_d^* )_j. \label{eq:p35}
\end{alignat}
The only non-zero term after the integration arises from the even terms $\propto k_L  (\delta_{ij} -n_i n_j)$, while the odd terms $\propto n_i$ and  $\propto n_i n_j n_k$ give a vanishing contribution (which can be concluded from symmetry considerations of the integrand over the unit sphere). The explicit integral can be evaluated by writing $d\bm{n}=\text{sin}(\phi) d\theta d\phi$ and decomposing the unit vector as $\bm{n}=(\text{sin}(\phi)\text{cos}(\theta),\text{sin}(\phi)\text{sin}(\theta),\text{cos}(\phi))$, giving a prefactor $8\pi/3$. Analogous steps can also be used for the other degrees of freedom, resulting in  the following non-zero radiation pressure forces and torques:

\begin{alignat}{1}
dp_{z}^{\text{(ds)}} & =\frac{8\pi\hbar\gamma_{s}}{3}\frac{2\pi}{\lambda}\vert u(\boldsymbol{r})\vert^{2} \left(
\frac{1}{2} \left(b_y^2-b_x^2\right) \sin (2 \alpha ) \cos (\beta ) \sin (2 \gamma ) (\chi_1-\chi_2) (\chi_1+\chi_2)\right.\nonumber\\
   &\qquad\,\,\,\,\,-\frac{1}{16} \cos (2 \gamma ) (\chi_1-\chi_2) (\chi_1+\chi_2)  \left(2 \left(b_y^2-b_x^2\right) \cos (2 \alpha )(\cos (2 \beta )+3)+4 \sin ^2(\beta )\right) \nonumber \\
   &\qquad\,\,\,\,\,-\frac{1}{8}\left(\chi_1^2+\chi_2^2-2 \chi_3^2\right) \left(2 \left(b_x^2-b_y^2\right) \cos (2 \alpha ) \sin ^2(\beta )-\cos (2 \beta )\right) \nonumber\\   
   &\left.\qquad\,\,\,\,\,+\frac{3}{8} \left(\chi_1^2+\chi_2^2\right)+\frac{1}{4} \chi_3^2\right) dt,\label{eq:sforce}\\
d\pi_{\alpha}^{\text{(ds)}} & =\frac{2\pi b_{x}b_{y}\hbar\gamma_{s}}{3}\vert u(\boldsymbol{r})\vert^{2}\bigg(-2\sin^{2}(\beta)\cos(2\gamma)(\text{\ensuremath{\chi_{1}}}-\text{\ensuremath{\chi}}_{2})(\text{\ensuremath{\chi}}_{1}+\text{\ensuremath{\chi}}_{2}-2\text{\ensuremath{\chi}}_{3})\nonumber \\
 & \qquad\qquad\qquad\qquad\qquad\,\,\,\,\,+\cos(2\beta)\left(\text{\ensuremath{\chi}}_{1}^{2}+2\text{\ensuremath{\chi}}_{3}(\text{\ensuremath{\chi}}_{1}+\text{\ensuremath{\chi}}_{2})-4\text{\ensuremath{\chi}}_{1}\text{\ensuremath{\chi}}_{2}+\text{\ensuremath{\chi}}_{2}^{2}-2\text{\ensuremath{\chi}}_{3}^{2}\right)\nonumber \\
 & \qquad\qquad\qquad\qquad\qquad\,\,\,\,\,+3\text{\ensuremath{\chi}}_{1}^{2}-2\text{\ensuremath{\chi}}_{3}(\text{\ensuremath{\chi}}_{1}+\text{\ensuremath{\chi}}_{2})-4\text{\ensuremath{\chi}}_{1}\text{\ensuremath{\chi}}_{2}+3\text{\ensuremath{\chi}}_{2}^{2}+2\text{\ensuremath{\chi}}_{3}^{2}\bigg)dt,\label{eq:ds1}\\
d\pi_{\beta}^{\text{(ds)}} & =\frac{8\pi b_{x}b_{y}\hbar\gamma_{s}}{3}\vert u(\boldsymbol{r})\vert^{2}\sin(\beta)\sin(\gamma)\cos(\gamma)(\text{\ensuremath{\chi}}_{1}-\text{\ensuremath{\chi}}_{2})(\text{\ensuremath{\chi}}_{1}+\text{\ensuremath{\chi}}_{2}-2\text{\ensuremath{\chi}}_{3})dt,\label{eq:ds2}\\
d\pi_{\gamma}^{\text{(ds)}} & =\frac{8\pi b_{x}b_{y}\hbar\gamma_{s}}{3}\vert u(\boldsymbol{r})\vert^{2}\cos(\beta)(\text{\ensuremath{\chi}}_{1}-\text{\ensuremath{\chi}}_{2})^{2}dt,\label{eq:ds3}
\end{alignat}

From Eq.~\eqref{eq:p3}, we can recover the Rayleigh scattering cross-section for a spherical particle. We first note that for a spherically isotropic particle, the susceptibility tensor simplifies to
$\chi_\text{lab}=\chi_0\mathbb{I} $,
where $\mathbb{I}$ is the $3\times3$ identity matrix, $\chi_0\equiv 3 \frac{n^2-1}{n^2+2}$, and $n=\sqrt{\epsilon_r}$ is the refractive index (and $\epsilon_r$ is the relative permittivity (dielectric constant/function)) of the nanoparticle. Since $\dot{\mathsf{p}}_j\propto \vert A_{\bm{k},\nu}\vert^2 \propto \vert(\bm{\epsilon}_{\bm{n},\nu}^\top \bm{\epsilon}_d^*)\vert^2 \propto \text{sin}^2(\phi)$ (with $\phi$ is the polar angle) we find after integrating over the solid angle the factor $8\pi/3$. Combining the prefactors we find the Rayleigh scattering cross section
\begin{equation}
\sigma_{R}\equiv \frac{8\pi}{3} \tilde{\sigma}_{R} \chi_0^2=\frac{24 \pi^3 V^2}{\lambda^4}\frac{(n^2-1)^2}{(n^2+2)^2}. \label{eq:rayleight}
\end{equation}
In this limit Eq.~(\ref{eq:sforce}) reduces to the scattering force of a spherical nanoparticle (i.e., we set $\chi_0\equiv\chi_1=\chi_2=\chi_3$):
\begin{equation}
    \boldsymbol{F}_{s}=(0,0, \frac{\sigma_{R}}{\sigma_L}\frac{P}{c}\vert u(x,y,z)\vert^{2})^{\top},
\end{equation} 
which continuously transfers linear momentum to the nanoparticle (together with the gradient force, they displace the equilibrium position of the nanoparticle in the positive $z$ direction). Similarly, Eqs.~(\ref{eq:ds1})-(\ref{eq:ds3}) correspond to the transfer of angular momentum from the optical field to the particle, albeit only when we have either non-linear polarization (such that the optical field carries spin angular momentum) or for a particle with a susceptibility tensor with a non-degenerate diagonal (i.e., an anisotropic particle). These terms drive the rotational degrees of freedom and induce spinning or librational motion.

\subsubsection{Stochastic radiation pressure force and torque} \label{sec:stoc}
\begin{figure}[ht]
\centering
\begin{tikzpicture}[
    node distance=1cm and 2cm,
    every node/.style={font=\footnotesize, align=center},
    box/.style={draw, rectangle, rounded corners, text width=3.5cm, minimum height=1.2cm, text centered, fill=gray!10},
    arrow/.style={->, thick},
]

\node[box] (vacuum) {Vacuum field \\ fluctuations};
\node[box, right=of vacuum] (modes) {Input field modes \\ $a_{\bm{k},\nu}(0)$ \\ (Eq.~\eqref{eq:mode})};
\node[box, right=of modes] (scatter) {Scattering by \\ nanoparticle \\ (Eq.~\eqref{eq:scattering})};
\node[box, below=of scatter] (field) {Outgoing field \\ $\bm{E}_f$ (Eq.~\eqref{eq:Ef})};
\node[box, left=of field] (force) {Stochastic Radiation pressure \\ force (Eq.~\eqref{eq:p66})};
\node[box, left=of force] (noise) {Stochastic Langevin \\ forces and torques \\ (Eq.~\eqref{eq:Pns}, \eqref{eq:Pins})};

\draw[arrow] (vacuum) -- (modes) node[midway, above, label] {Input noise};
\draw[arrow] (modes) -- (scatter);
\draw[arrow] (scatter) -- (field);
\draw[arrow] (field) -- (force);
\draw[arrow] (force) -- (noise);

\end{tikzpicture}
\caption{Flowchart showing how vacuum field fluctuations and input mode amplitudes ($a_{\bm{k},\nu}(0)$) lead to stochastic radiation pressure forces and torques on the nanoparticle through scattering. Equation references indicate key steps in the derivation.}
\label{fig:noise_flow}
\end{figure}
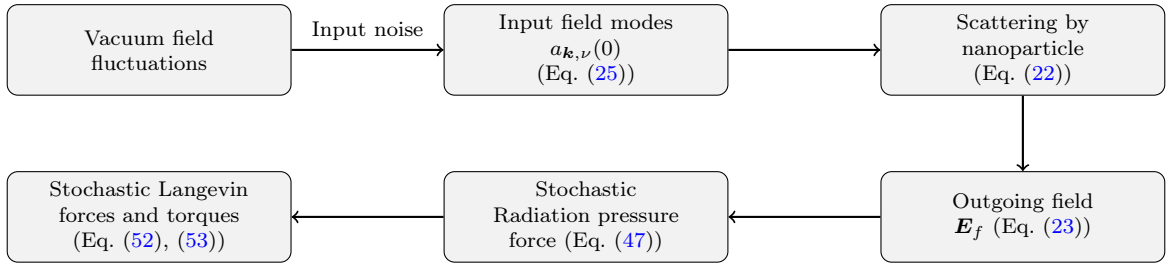

The interaction of the trapped nanoparticle with the optical field leads to scattering of photons, which imparts a random recoil to the particle. The origin of this randomness lies in the vacuum (or input) fluctuations of the electromagnetic field modes, encoded in the initial conditions \( a_{\bm{k},\nu}(0) \). These fluctuations ultimately manifest as stochastic radiation pressure forces and torques acting on the particle. In the following, we derive an effective stochastic description of these forces by systematically tracing how the randomness of the input field modes propagates into Langevin-like noise terms in the equations of motion. This is shown schematically in Fig.~\ref{fig:noise_flow}. To this end, we begin from the scattering Hamiltonian, transform to spherical mode coordinates, and derive the resulting noise terms using standard input-output theory.

We consider the second line of Eq.~\eqref{eq:p2} and rewrite it in spherical coordinates:
\begin{equation}
    \dot{\mathsf{p}}_j = \sum_\nu \int d\bm{n} \int d \omega_k \frac{\omega_k^2}{(2\pi)^3c^3}
\sqrt{\frac{\hbar \omega_k P V^2}{4 c \sigma_L}} a_{\bm{k},\nu} (0) e^{-i(\omega_k-\omega_L)t} \frac{\partial A_{\bm{n},\nu}^*}{\partial \mathsf{x}_j}  +\text{c.c.}, \label{eq:p5}
\end{equation}
where we have used $d\bm{k}=k^2 dk d\bm{n}$, and  $\omega_k=k c$. In addition, we change from plane wave modes $a_{\bm{k},\nu}$ to spherical modes $a_{\omega_k,\bm{n}, \nu}$ which obey the following Poisson bracket:
\begin{equation}
      \{a_{\omega_k,\bm{n}, \nu},a_{\omega_k',\bm{n}', \nu'}^*\}=-\frac{i}{\hbar}\delta(\omega_k-\omega_k') \delta^2 (\bm{n}-\bm{n}')\delta_{\nu,\nu'}. \label{eq:key3}
\end{equation}

\noindent Comparing Eqs.~\eqref{eq:key} and \eqref{eq:key3} we note that the two decompositions (plane-wave versus spherical) differ by a factor $\sqrt{(2\pi)^3 \frac{c^3}{\omega_k^2}}$ (one finds this factor by transforming the three dimensional Dirac delta from Cartesian to spherical coordinates). Taking this latter factor into account we obtain from Eq.~\eqref{eq:p5} the following equation:
\begin{equation}
    \dot{\mathsf{p}}_j = \sum_\nu \int d \omega_k \int d\bm{n} 
    \sqrt{\frac{\hbar \omega_k^3 P V^2}{ (2\pi)^3 4 c^4 \sigma_L}} a_{\omega_k,\bm{n}, \nu} (0) e^{-i(\omega_k-\omega_L)t} \frac{\partial A_{\bm{k},\nu}^*}{\partial \mathsf{x}_j}  +\text{c.c.}. \label{eq:p6}
\end{equation}
We then make the approximation that only the frequencies close to $\omega_L$ will give a non-negligible contribution: specifically, we replace $\omega_k$ with $\omega_L$ in the square root prefactor of Eq.~\eqref{eq:p6} as well as introduce $A_{\bm{n},\nu}=A_{\bm{k},\nu}\vert_{\omega_k=\omega_L}$. Hence from  Eq.~\eqref{eq:p6} we immediately obtain:
\begin{equation}
    \dot{\mathsf{p}}_j = \hbar \sum_\nu \int d\bm{n} 
    \sqrt{\frac{\pi \hbar \omega_L^3 P V^2}{ (2\pi)^3 4 c^4 \sigma_L \hbar}} a^\text{(in)}_{\bm{n},\nu}  \frac{\partial A_{\bm{n},\nu}^*}{\partial \mathsf{x}_j}  +\text{c.c.}. \label{eq:p66}
\end{equation}
We have defined the noise term~\cite{gardiner2004quantum}:
\begin{equation}
    a^\text{(in)}_{\bm{n},\nu} (t) = \frac{1}{\sqrt{\pi}}\int_0^\infty d \omega_k  a_{\omega_k,\bm{n}, \nu} (0) e^{-i (\omega_k-\omega_L) t}.
\end{equation}
which satisfies
\begin{equation}
    \{a^\text{(in)}_{\bm{n},\nu} (t),(a^\text{(in)}_{\bm{n},\nu} (t'))^*\} =-\frac{i}{\hbar}\delta(\bm{n}-\bm{n}') \delta_{\nu,\nu'} \delta (t-t'),
\end{equation}
as can be verified using Eq.~\eqref{eq:key3}. We can then decompose the noise in the real and imaginary part:
\begin{equation}
    a^\text{(in)}_{\bm{n},\nu} (t)= \frac{1}{\sqrt{2 }} \left( X^\text{(in)}_{\bm{n},\nu} (t) + i P^\text{(in)}_{\bm{n},\nu} (t)\right),
\end{equation}

where $X^\text{(in)}_{\bm{n},\nu} (t)$ and $P^\text{(in)}_{\bm{n},\nu} (t)$ can be interpreted as independent  amplitude and phase noise, respectively.
From Eq.~\eqref{eq:p66} we then immediately find:
\begin{alignat}{1}
    \dot{\mathsf{p}}_j &=  \hbar \sqrt{\frac{\gamma_s}{4}} \sum_\nu \int d\bm{n}  
    (X^\text{(in)}_{\bm{n},\nu} \partial_j A_{\bm{n},\nu}^*+i P^\text{(in)}_{\bm{n},\nu} \partial_j A_{\bm{n},\nu}^*) + \text{c.c.} \nonumber\\
    &= -\hbar  \sqrt{\gamma_s} \sum_\nu \int d\bm{n}  
   ~\left(\text{Re}(\partial_j A_{\bm{n},\nu}^*) X^\text{(in)}_{\bm{n},\nu}
   +\text{Im}(\partial_j A_{\bm{n},\nu}^*) P^\text{(in)}_{\bm{n},\nu}\right)
    \label{eq:p15}
\end{alignat}
where $A_{\bm{n},\nu}$ denotes $A_{\bm{k},\nu}\vert_{\omega_k=\omega_L}$, $\partial_j= \frac{\partial}{\partial \mathsf{x}_j}$,  and $P^\text{(in)}_{\bm{n},\nu} = \frac{dU_{\bm{n},\nu}}{dt}$, where $U_{\bm{n},\nu}(t)$ is a Wiener process encoding the cumulative phase noise in the mode labeled by $(\bm{n},\nu)$. This stochastic process is real-valued, continuous, and satisfies $\mathbb{E}[U_{\bm{n},\nu}(t)] = 0$ and $\mathbb{E}[U_{\bm{n},\nu}(t) U_{\bm{n}',\nu'}(t')] = \delta(\bm{n}-\bm{n}') \delta_{\nu,\nu'} \min(t,t')$, such that $P^\text{(in)}_{\bm{n},\nu}$ is its formal time derivative and corresponds to white noise. Similarly, $X^\text{(in)}_{\bm{n},\nu}(t)$ can be identified as the time derivative of an independent Wiener process $V_{\bm{n},\nu}(t)$ encoding amplitude noise, i.e., $X^\text{(in)}_{\bm{n},\nu} = \frac{dV_{\bm{n},\nu}}{dt}$. Finally, inserting Eq.~\eqref{eq:nano} in Eq.~\eqref{eq:p15} we then obtain the stochastic forces and torques:
\begin{alignat}{1}
d\boldsymbol{p}_{k}^{\text{(ss)}} & = -\hbar\sqrt{\gamma_{s}}\sum_{\nu=1}^{2}\int d\bm{n}\,\left(\text{Re}\left(\bm{\epsilon}_{\bm{n},\nu}^{*\top}\chi_\text{lab}\boldsymbol{\epsilon}_{d}\partial_{\boldsymbol{r}_{k}}\left[u e^{ik_L\bm{n}\cdot\boldsymbol{r}}\right]\right) dX^\text{(in)}_{\bm{n},\nu}\right.\nonumber\\
&\qquad\qquad\qquad\qquad\qquad
\left.+\text{Im}\left(\bm{\epsilon}_{\bm{n},\nu}^{*\top}\chi_\text{lab}\boldsymbol{\epsilon}_{d}\partial_{\boldsymbol{r}_{k}}\left[u e^{ik_L\bm{n}\cdot\boldsymbol{r}}\right]\right)dP^\text{(in)}_{\bm{n},\nu} \right),
\label{eq:Pns}\\
d\boldsymbol{{\pi}}_{k}^{\text{(ss)}} & = -\hbar\sqrt{\gamma_{s}}\sum_{\nu=1}^{2}\int d\bm{n}\, \left( \text{Re}\left(\bm{\epsilon}_{\bm{n},\nu}^{*\top}\partial_{\boldsymbol{\phi}_{k}}\left[\chi_\text{lab}\right]\boldsymbol{\epsilon}_{d}u e^{ik_L\bm{n}\cdot\boldsymbol{r}}\right) dX^\text{(in)}_{\bm{n},\nu} \right.\nonumber\\
&\qquad\qquad\qquad\qquad\qquad
\left.+\text{Im}\left(\bm{\epsilon}_{\bm{n},\nu}^{*\top}\partial_{\boldsymbol{\phi}_{k}}\left[\chi_\text{lab}\right]\boldsymbol{\epsilon}_{d}u e^{ik_L\bm{n}\cdot\boldsymbol{r}}\right)dP^\text{(in)}_{\bm{n},\nu} \right)
\label{eq:Pins}
\end{alignat}
where we recall $\chi_\text{lab}=\chi_\text{lab}(\alpha,\beta,\gamma)$ is given in Eq.~\eqref{eq:tensors}, and 
$u=u(x,y,z)$ is defined in Eq.~\eqref{eq:umode}. In Eqs.~\eqref{eq:Pns} and \eqref{eq:Pins} we have also introduced  two families of zero-mean independent Wiener noise increments:
\begin{equation}
 dX^\text{(in)}_{\bm{n},\nu} \equiv X^\text{(in)}_{\bm{n},\nu} dt,\qquad\qquad dP^\text{(in)}_{\bm{n},\nu}  \equiv P^\text{(in)}_{\bm{n},\nu} dt, \label{eq:dXdP}
\end{equation}
which can be readily simulated with a Gaussian random number generator~\cite{evans2012introduction}. The amplitude and phase noises in Eq.~\eqref{eq:dXdP} can be recast also in terms of the mode noise,
\begin{equation}
    \dd a^\text{(in)}_{\bm{n},\nu} = \frac{1}{\sqrt{2}} \left(d X^\text{(in)}_{\bm{n},\nu}+ i \,d P ^\text{(in)}_{\bm{n},\nu} \right), \label{eq:aindef}
\end{equation}
and its complex conjugate $\dd a^\text{(in)*}_{\bm{n},\nu}$.
 
We are left to define the classical expectation values of the input noises $\dd a^\text{(in)}_{\bm{n},\nu}$ and  $\dd a^\text{(in)*}_{\bm{n},\nu}$, i.e., we have to characterize the statistical properties of the two noises. As a guideline we follow the quantum prescription from Eq.~\eqref{eq:noises2QA1} and \eqref{eq:noises2QA2}, formally replacing quantum expectation values, $\langle \, \cdot \, \rangle$, with classical expectation values over different noise realizations, $\mathbb{E}[~\cdot~]$. In particular, we set:
\begin{alignat}{1}
\langle d a^{\text{(in)}}_{\bm{n},\nu} \rangle = 0,\quad 
\langle d a^{\text{(in)}*}_{\bm{n},\nu}\rangle = 0,\quad 
\langle d a^{\text{(in)}}_{\bm{n},\nu} d a^{\text{(in)}}_{\bm{n}',\nu'} \rangle = 0,\quad 
\langle d a^{\text{(in)}*}_{\bm{n},\nu} d a^{\text{(in)}*}_{\bm{n}',\nu'}\rangle = 0, \quad \label{eq:noises2A1}\\
\langle d a^{\text{(in)}*}_{\bm{n},\nu} d a^{\text{(in)}}_{\bm{n}',\nu'} \rangle = \frac{1}{2} \delta(\bm{n}-\bm{n}')\delta_{\nu,\nu'}dt,\quad 
\langle d a ^\text{(in)}_{\bm{n},\nu} d a^{\text{(in)}*}_{\bm{n}',\nu'}\rangle = \frac{1}{2} \delta(\bm{n}-\bm{n}')\delta_{\nu,\nu'}dt. \label{eq:noises2A2}
\end{alignat}
The key difference is that in place of the asymmetric expectation values in Eq.~\eqref{eq:noises2QA2}, which captures the non-commutative nature of quantum operators, we have to take "symmetrized" classical expectation values in Eq.~\eqref{eq:noises2A2}, as classical noises commute. Equivalently, we could have set 
\begin{alignat}{1}
 &\mathbb{E}[dX^\text{(in)}_{\bm{n},\nu}]=0,\quad \mathbb{E}[P^\text{(in)}_{\bm{n},\nu}]=0, \quad 
 \mathbb{E}[dX^\text{(in)}_{\bm{n},\nu} dP^\text{(in)}_{\bm{n},\nu}] =0, \label{eq:noises21} \\
&\mathbb{E}[dX^\text{(in)}_{\bm{n},\nu} dX^\text{(in)}_{\bm{n}',\nu'}] = \frac{1}{2}\delta(\bm{n}-\bm{n}')\delta_{\nu,\nu'}dt,\quad \mathbb{E}[dP^\text{(in)}_{\bm{n},\nu} dP^\text{(in)}_{\bm{n}',\nu'}] = \frac{1}{2}\delta(\bm{n}-\bm{n}')\delta_{\nu,\nu'}dt \label{eq:noises2}
\end{alignat}
as can be verified from Eqs.~\eqref{eq:aindef} and \eqref{eq:noises2A1}-\eqref{eq:noises2A2}. When performing analytical computations we will only use the basic properties of zero-mean independent Wiener noise increments in Eq.~\eqref{eq:noises2A1} and \eqref{eq:noises2A2}.

As a first example we illustrate how to use Eq.~\eqref{eq:Pns} for the simplest case of translational noise affecting a spherical particle trapped in a linearly polarized Gaussian beam. In particular, we will show how to recover the lowest order recoil-heating noise terms along the three-axis presented in~\cite{seberson2020distribution}. We start by recalling that for a spherical particle we have $\chi_\text{lab}=\chi_0 \mathbb{I}$, where  $\mathbb{I}$ is the $3\times3$ identity matrix. From Eq.~\eqref{eq:Pns} we thus find:
\begin{alignat}{1}
d\boldsymbol{p}_{k}^{\text{(ss)}}  = -\hbar\sqrt{\gamma_{s}} \chi_\text{0} \sum_{\nu=1}^{2}\int d\bm{n}\,&\left(\text{Re}\left(\bm{\epsilon}_{\bm{n},\nu}^{*\top}\boldsymbol{\epsilon}_{d}~i k_L (\delta_{3k}- n_k) 
\left(e^{-ik_L(z -\bm{n}\cdot\boldsymbol{r})}\right)\right)dX^\text{(in)}_{\bm{n},\nu} \right. \nonumber\\
 &\quad\left.\text{Im}\left(\bm{\epsilon}_{\bm{n},\nu}^{*\top}\boldsymbol{\epsilon}_{d}~i k_L (\delta_{3k}- n_k) 
\left(e^{-ik_L(z -\bm{n}\cdot\boldsymbol{r})}\right)\right)dP^\text{(in)}_{\bm{n},\nu} \right), 
\label{eq:Pns2}
\end{alignat}
where we have made the approximation $u(\bm{r}) \approx e^{-i k_L z} $ corresponding to a plane wave approximation of the Gaussian beam (see Eq.~\eqref{eq:umode}). To further simplify the analysis we keep only the leading order contribution  $\propto \mathcal{O} (r^0)$  such that Eq.~\eqref{eq:Pns2} reduces to:
\begin{equation}
d\boldsymbol{p}_{k}^{\text{(ss)}}  =-\hbar\sqrt{\gamma_{s}} \chi_\text{0} \sum_{\nu=1}^{2}\int d\bm{n}\,\text{Im}\left(\bm{\epsilon}_{\bm{n},\nu}^{*\top}\boldsymbol{\epsilon}_{d}~i k_L (\delta_{3k}- n_k) \right)dP^\text{(in)}_{\bm{n},\nu},\label{eq:Pns22}
\end{equation}
i.e., we assume the particle is close to the trap center. We note that in Eq.~\eqref{eq:Pns22} only the phase noise $dP^\text{(in)}_{\bm{n},\nu}$ contributes, while the contribution from the amplitude noise $dX^\text{(in)}_{\bm{n},\nu}$ vanishes in the considered limit of small displacements. We then impose an incoming linearly polarized optical field, i.e., $(\boldsymbol{\epsilon}_{d})_j= \delta_{1j}$. Furthermore, we can decompose the outgoing polarization in a basis with real-valued components such that  $\bm{\epsilon}_{\bm{n},\nu}^{*\top}=\bm{\epsilon}_{\bm{n},\nu}^{\top}$. From Eq.~\eqref{eq:Pns22} we thus find:
\begin{equation}
d\boldsymbol{p}_{k}^{\text{(ss)}}  =-\hbar\sqrt{\gamma_{s}} \chi_\text{0} k_L \sum_{\nu=1}^{2}\int d\bm{n}\,
(\bm{\epsilon}_{\bm{n},\nu}^{\top})_1 (\delta_{3k}- n_k) dP^\text{(in)}_{\bm{n},\nu},\label{eq:Pns23}
\end{equation}

We now compute the cross-correlation of the noises:
\begin{equation}
    \bm{\sigma}_{k,k'}^{(tt)}=\mathbb{E}[d\boldsymbol{p}_{k}^{\text{(ss)}} \otimes d\boldsymbol{p}_{k'}^{\text{(ss)}}], \label{eq:sigmatt}
\end{equation}
where $\mathbb{E}$ denotes the stochastic average,  and $\otimes$ is the tensor product.  We insert Eq.~\eqref{eq:Pns2} in \eqref{eq:sigmatt} to find
\begin{equation}
    \bm{\sigma}_{k,k'}^{(tt)}= \gamma_{s} \chi_\text{0}^2 \hbar^2k_L^2
    \sum_{\nu,\nu'=1}^{2}\int d\bm{n}\, \int d\bm{n'}\, 
   (\bm{\epsilon}_{\bm{n},\nu}^{\top})_1 (\bm{\epsilon}_{\bm{n}',\nu'}^{\top})_1 
   (\delta_{3k}- n_k)  (\delta_{3k'}- n'_{k'})
    \mathbb{E}[dP^\text{(in)}_{\bm{n},\nu} dP^\text{(in)}_{\bm{n}',\nu'} ].
    \label{eq:sigmatt2}
\end{equation}
We now recall the property of independent noise increments in Eq.~\eqref{eq:noises2}, i.e., all the modes scattered in different directions or different polarizations are independent. From Eq.~\eqref{eq:sigmatt2} we then immediately find:
\begin{equation}
    \bm{\sigma}_{k,k'}^{(tt)}= \frac{\gamma_{s}}{2} \chi_\text{0}^2 \hbar^2 k_L^2
    \sum_{\nu=1}^{2}\int d\bm{n}\,
    (\bm{\epsilon}_{\bm{n},\nu}^{\top})_1 (\bm{\epsilon}_{\bm{n},\nu}^{\top})_1 
   (\delta_{3k}- n_k)  (\delta_{3k'}- n_{k'}) dt.
    \label{eq:sigmatt24}
\end{equation}
To evaluate the summation over the polarization we use the completeness relation in Eq.~\eqref{eq:cr} to find:
\begin{equation}
    \bm{\sigma}_{k,k'}^{(tt)}= \frac{\gamma_{s}}{2} \chi_\text{0}^2 \hbar^2 k_L^2 \int d\bm{n}\,
     (\delta_{3k}- n_k)  (\delta_{3k'}- n_{k'})  (\delta_{11} -n_1 n_1) dt.
    \label{eq:sigmatt25}
\end{equation}
To perform the integration in Eq.~\eqref{eq:sigmatt25} we use spherical coordinates, where the area element is $d\bm{n}=\text{sin}(\phi) d\theta d\phi$, and we decompose the unit vector as $\bm{n}=(\text{sin}(\phi)\text{cos}(\theta),\text{sin}(\phi)\text{sin}(\theta),\text{cos}(\phi))$, giving non-zero terms $\frac{8\pi}{15}\, \text{diag}(1,2,7)_{k,k'}$. We eventually find:
\begin{equation}
    \bm{\sigma}_{k,k'}^{(tt)}= \frac{4\pi}{15}\gamma_{s} \chi_\text{0}^2  \hbar^2 k_L^2 \,\,\text{diag}(1,2,7)_{k,k'} \,\,dt.
    \label{eq:sigmatt6}
\end{equation}
We then finally use the definition of the scattering rate $\gamma_s$ in Eq.~\eqref{eq:gammas} and the Rayleigh cross-section $\sigma_R$ in Eq.~\eqref{eq:rayleight} to find the final expression:
\begin{equation}
    \bm{\sigma}_{k,k'}^{(tt)}= \frac{1}{10}\frac{\sigma_R}{\sigma_L} \frac{P}{\hbar \omega_L} \hbar^2 k_L^2 \,\,\text{diag}(1,2,7)_{k,k'} \,\,dt.
    \label{eq:sigmatt7}
\end{equation}
We have thus explicitly recovered the results quoted in \cite{seberson2020distribution}.  The recoil heating noise contributions are thus given by 
\begin{equation}
d\boldsymbol{{p}}{}^{\text{(ss)}}_k=  \sqrt{\bm{\sigma}_{k,k'}^{(tt)}} dW_k, \label{eq:dpn}
\end{equation}
where $\bm{\sigma}_{k,k'}^{(tt)}$ denotes the $k,k'$-element in Eq.~\eqref{eq:sigmatt7}, and $dW_k$ ($k=x,y,z$) denote independent Wiener noise terms. 

We can readily generalize the steps from Eqs.~\eqref{eq:Pns2}-\eqref{eq:sigmatt7} to the case of a spherical particle in an elliptically polarized optical field. We set the polarization vector to $\bm{\epsilon}_{\bm{n}}=(b_x,i b_y, 0)^\top$. We then find in place of Eq.~\eqref{eq:Pns22}:
\begin{alignat}{1}
d\boldsymbol{p}_{k}^{\text{(ss)}}  =-\hbar\sqrt{\gamma_{s}} \chi_\text{0} \sum_{\nu=1}^{2}\int d\bm{n}\,&\left(\text{Re}\left(\bm{\epsilon}_{\bm{n},\nu}^{*\top}\boldsymbol{\epsilon}_{d}~i k_L (\delta_{3k}- n_k) \right)dX^\text{(in)}_{\bm{n},\nu}, \right. \nonumber\\
&\left. \quad+\text{Im}\left(\bm{\epsilon}_{\bm{n},\nu}^{*\top}\boldsymbol{\epsilon}_{d}~i k_L (\delta_{3k}- n_k) \right)dP^\text{(in)}_{\bm{n},\nu}\right),\label{eq:Pns22ee}
\end{alignat}
i.e., both the amplitude and phase noise give in this case non-vanishing contributions. Performing then analogous steps to the ones discussed in Eqs.~\eqref{eq:Pns22}-\eqref{eq:sigmatt7} we eventually find:
\begin{equation}
    \bm{\sigma}_{k,k'}^{(tt)}= \frac{1}{10}\frac{\sigma_R}{\sigma_L} \frac{P}{\hbar \omega_L} \hbar^2 k_L^2 \,\,\text{diag}(1+b_y^2,2-b_y^2,7)_{k,k'} \,\,dt,
    \label{eq:sigmatt7e}
\end{equation}
which reduces back to Eq.~\eqref{eq:sigmatt7} when we set $b_y=0$. When we set instead $b_x=b_y=1/\sqrt{2}$, i.e., circular polarization, we find that the noise acting on x and y has the same magnitude $\sqrt{\sigma^\text{(tt)}_{x,x'}}=\sqrt{\sigma^\text{(tt)}_{y,y'}} \propto \sqrt{3/2}$, as expected from symmetry considerations.

As a second illustrative example we will consider an axially symmetric particle (e.g., a nanorod or a disk-shaped nanoparticle)  placed in an optical field with circular polarization. We set $\chi_\text{s}\equiv\chi_1=\chi_2$, $\chi_\text{L}\equiv\chi_3$, and $\bm{\epsilon}_{d}=(1,i, 0)^\top/\sqrt{2}$. To simplify the analysis we will suppose that at the leading order we can neglect cross-couplings between the translational/rotational degrees of freedom. In this limit the result in Eq.~\eqref{eq:sigmatt7e} remains valid for translations (where we set $b_x=b_y=1/\sqrt{2}$), and we are left to analyze the rotational recoil-heating noise. In particular, from Eq.~\eqref{eq:Pins} we have:
\begin{alignat}{1}
d\boldsymbol{\pi}_{k}^{\text{(ss)}} =-\hbar\sqrt{\gamma_{s}}  \sum_{\nu=1}^{2}\int d\bm{n}\,
&\left( \text{Re}\left(\bm{\epsilon}_{\bm{n},\nu}^{*\top} \partial_{\phi_k} [\chi_\text{lab} ]\boldsymbol{\epsilon}_{d} \right)dX^\text{(in)}_{\bm{n},\nu} \right. \nonumber\\
&\left.\quad+\text{Im}\left(\bm{\epsilon}_{\bm{n},\nu}^{*\top} \partial_{\phi_k} [\chi_\text{lab} ]\boldsymbol{\epsilon}_{d} \right)dP^\text{(in)}_{\bm{n},\nu} \right),\label{eq:Rot}
\end{alignat}
where as discussed above we have neglected the couplings to translational degrees of freedom. 
\noindent
We then follow analogous steps from Eqs.~\eqref{eq:Pns23}-\eqref{eq:sigmatt7} to find the rotational noise correlation matrix
$\bm{\sigma}^{\text{(rr)}}_{k,k'}=\mathbb{E}[d\boldsymbol{\pi}_{k}^{\text{(ss)}} \otimes d\boldsymbol{\pi}_{k'}^{\text{(ss)}}]$.
In particular, we find the following correlation matrix components:
\begin{equation}
   \bm{\sigma}^{\text{(rr)}}_{k,k'}  
   =\frac{  \pi\gamma_{s} \hbar^2 }{3}  \vert \chi_L-\chi_s \vert^2 \,\,\text{diag}\left(
 \sin ^2(\beta ) (3-\cos (2 \beta )),2,0\right) dt, \label{eq:sigmarot}
\end{equation}
 and thus the rotational recoil heating noise contributions are given by 
\begin{equation}
d\boldsymbol{{\pi}}{}^{\text{(ss)}}_k=  \sqrt{\bm{\sigma}_{k,k'}^{(rr)}} dW_k, \label{eq:dpnn}
\end{equation}
where $\bm{\sigma}_{k,k'}^{(rr)}$ denotes the $k,k'$-element in Eq.~\eqref{eq:sigmarot}, and $dW_k$ ($k=\alpha,\beta,\gamma$) denote independent Wiener noise terms. As expected, the noise on $d\pi_\gamma$ is absent since the optical field cannot distinguish different $\gamma$ orientations of an axially symmetric particle (such is the case for a nanorod or a disk-shaped particle). We further consider the limit  $\beta\approx\pi/2$, e.g., one can picture a nanorod spinning in the x-y plane (see Fig.~\ref{fig:exp}). In place of Eq.~\eqref{eq:sigmarot} we then find 
\begin{equation}
   \bm{\sigma}^{\text{(rr)}}_{k,k'}  
   =\frac{  \pi\gamma_{s} \hbar^2 }{3}  \vert \chi_L-\chi_s \vert^2 \,\,\text{diag}\left(
 4,2,0\right) dt, \label{eq:sigmarot2}
\end{equation}
which is in accordance with basic symmetry considerations. In particular, in this limit the problem is invariant with respect to rotations in the x-y plane here  corresponding to $\alpha$ rotations. This symmetry is reflected in 
Eq.~\eqref{eq:sigmarot2} which does not have any angular dependency, and thus leads to (constant-prefactor) white-noises affecting the $\alpha$ and $\beta$ motions.

Following a similar procedure as outlined in the above examples we can also obtain the translational and rotational noises for the general case of an arbitrary anisotropic particle with elliptical polarization.  Starting from Eqs.~\eqref{eq:Pns} and \eqref{eq:Pins} one can perform analogous steps as illustrated in Eqs.~\eqref{eq:Pns2}-\eqref{eq:sigmarot2}, where one has to keep $\chi_\text{lab}$, and set the polarization vector to the parametrization $\bm{\epsilon}_{d}=(b_x,i b_y, 0)^\top$. Although the final formulae can become quite long, they can be readily coded for numerically simulating the dynamics of the nanoparticle.

There is however one important question which did not arose in the above examples. Typically one will end up with a non-diagonal noise correlation matrix (in place of the diagonal ones in Eqs.~\eqref{eq:sigmatt7e} or \eqref{eq:sigmarot} discussed above). In order to obtain the noise terms from the correlation matrix one is left to perform a ``square root'' of the correlation matrix. For non-diagonal matrices we have to use the generalization of the ``square root'' known as the Cholesky decomposition~\cite{trefethen2022numerical}. We discuss in \ref{app:Correlated-noises} the general procedure, i.e., how to find from an arbitrary number of noises a simple description with one noise per degree of freedom as we have done in the illustrative examples above. 

However, we will neglect photon-recoil noises when numerically simulating such systems in Sec.~\ref{sec:Numerical-analysis} as the noise from thermal gas collisions dominates in the considered pressure range.


\subsubsection{Non-conservative collisional terms}
\begin{table}[h]
  \label{tab:symbols_theory}
  \setlength{\tabcolsep}{8pt}
  \renewcommand{\arraystretch}{1.3}
  \begin{tabular}{@{}ll@{}}
    \toprule
    \textbf{Symbol} & \textbf{Description} \\
    \midrule
    $\bm{p}$, $\bm{\pi}$ & Linear and angular momenta in the laboratory frame \\
    $\bm{\mathtt{p}}$, $\bm{\mathtt{L}}$ & Linear and angular momenta in the body frame \\
    $F^{\text{(t)}}$, $F^{\text{(r)}}$ & Translational and rotational friction tensors in the body frame (diagonal) \\
    $I$ & Moment of inertia tensor in the body frame (diagonal) \\
    $\mathcal{R}$ & Rotation matrix from body to lab frame \\
    $\mathcal{M}$ & Jacobian relating Euler angle derivatives to angular velocity \\
    $\bm{\omega}$ & Angular velocity in the body frame \\
    $\bm{U}$, $\bm{V}$ & Wiener processes modeling translational and rotational noise in the body frame \\
    $d\bm{U}_\text{lab}$, $d\bm{V}_\text{lab}$ & Corresponding noise terms in the lab frame \\
    $T$ & Temperature of the surrounding gas \\
    $k_B$ & Boltzmann constant\\
    \bottomrule
    \end{tabular}
\caption{Overview of symbols used in the description of non-conservative collisional terms, including their definitions and reference frames.}
\end{table}


In this section we provide a bottom-up model for the non-conservative terms arising from the scattering of residual gas molecules. We recall that the deBroglie wavelength of the gas particle is $\lambda \sim 2\pi\hbar/\sqrt{m_g k_B T}$, where $m_g$ is the mass of a gas particle, and $T$ is the temperature of the gas. For the room-temperature regime, which we will consider when numerically simulating the system in Sec.~\ref{sec:Numerical-analysis}, we find $\lambda \sim 10^{-2}~\text{nm}$ which is much smaller than the length-scale $\sim 100\text{ nm}$ of the typical nanoparticle investigated in levitated optomechanics. As such a gas molecule can resolve the finer details of the nanoparticle shape and we cannot work in the approximation of anisotropic point-particle (as we did  for the optical terms in Sec.~\ref{sec:ncopt}). 

We suppose that we are working in the molecular flow regime~\cite{cercignani1988boltzmann} such that the mean-free path of a gas molecule is much larger than the extension of the nanoparticle. In this regime we can obtain the translational friction tensor $F^\text{(tr)}$ and the rotational friction tensor $F^\text{(rot)}$ using analytical/numerical calculations~\cite{Cavalleri2010Gas,ahn2018optically}. As discussed below Eq.~\eqref{eq:tensors} the two friction tensors are assumed diagonal in the body-frame of the object (i.e., the same reference frame where the moments of inertia tensor $I$ becomes diagonal). The starting point of this analysis is the dynamics in the body frame where it acquires a particularly simple form as dictated by symmetry considerations~\cite{ford1979rotational}.

Let us first discuss translational motion. The dynamics in the body frame takes the form of a simple Langevin equation:
\begin{equation}
\frac{d}{dt}\bm{\mathtt{p}} + \bm{\omega}\times \,\bm{\mathtt{p}} +F^\text{(t)} \,\bm{\mathtt{p}}= \sqrt{2 M k_B  T\,F^\text{(t)}} \,\frac{d\bm{U}}{dt},
\label{eq:plab0}
\end{equation}
where $\bm{\mathtt{p}}$ is the momentum in the body frame, the friction tensor $F^\text{(t)}$ is diagonal, $M$ is the nanoparticle mass, and $d\bm{U}=(dU_1,dU_2,dU_3)$ denote three independent Wiener noise increments. 

The first two terms on the left-hand side of Eq.~\eqref{eq:plab0} can be traced back to the conservation of linear momentum of a free particle in the inertial laboratory frame, i.e., they arise from $\frac{d}{dt} \bm{p}=0$. In particular, the angular frequency $\bm{\omega}$ in the body frame arises from the term 
\begin{equation}
    \mathcal{R}^\top\frac{d}{dt}[\mathcal{R}]\,\,\bm{\mathtt{p}}=\bm{\omega}\times \,\bm{\mathtt{p}}. \label{eq:omegaderiv}
\end{equation}
The third term on the left-hand side and the term on the right-hand side of Eq.~\eqref{eq:plab0} have been added to model damping and noise fluctuations, respectively.

We then recall the relation between the laboratory frame momentum $\bm{p}$ and the body frame momentum $\bm{\mathtt{p}}$ given by:
\begin{equation}
    \bm{\mathtt{p}}=\mathcal{R}^\top\bm{p}. \label{eq:pody}
\end{equation} 
Using Eqs.~\eqref{eq:omegaderiv} and \eqref{eq:pody} we can then transform Eq.~\eqref{eq:plab0} to the laboratory frame:
\begin{equation}
d\bm{p} = -F_\text{lab}^\text{(t)} \,\bm{p} \, dt + \left[ \sqrt{2 M k_B T\,F^\text{(t)}}\right]_\text{lab} \,d\bm{U}_\text{lab},
\label{eq:plab}
\end{equation}
where $\bm{p}$ is the usual momentum in the laboratory reference frame, and $[~\cdot~]_\text{lab}= \mathcal{R}[~\cdot~]\mathcal{R}^\top$ denotes the other quantities in the laboratory frame. The first and second term on the right-hand side of Eq.~\eqref{eq:plab} corresponds to the damping and noise fluctuations, respectively.

We next discuss the rotational motion. We now start from Euler-Langevin equation in the body frame:
\begin{equation}
    \frac{d}{dt} \bm{\mathtt{L}} +\bm{\omega} \times \bm{\mathtt{L}} + F^{\text{(r)}} \bm{\omega}=\sqrt{2k_B T F^{(r)}} \frac{d\bm{V}}{dt} \label{eq:E-L}
\end{equation}
where $\bm{\mathtt{L}}$ is the angular momentum in the body frame, the rotational friction tensor $F^{\text{(r)}}$ is diagonal, and $d\bm{V}=(dV_1,dV_2,dV_3)$ denotes three independent Wiener noise increments. 

In order to find the dynamics for the conjugate angle momenta $\bm{\pi}$ from Eq.~\eqref{eq:E-L} we need to recall some relations. First, we note that the angular momentum is simply a rescaled angular frequency:
\begin{equation}
    \bm{\mathtt{L}}=I \,\bm{\omega}, \label{eq:ang}
\end{equation}
where $I$ is the moment of inertia tensor in the body frame. We then recall the relation between the time-derivatives of the angle vector $\dot{\bm{\phi}}$ and the angular frequency vector $ \bm{\omega}$:
\begin{equation}
    \bm{\omega}=\mathcal{R}^\top\mathcal{M}\,\,\dot{\bm{\phi}}.\label{eq:phidot}
\end{equation}
The transformation matrix $\mathcal{M}$ can be derived from Eq.~\eqref{eq:omegaderiv}, with the components given by:
\begin{equation}
      \mathcal{M}\equiv 
\left[\begin{array}{ccc}
0& -\sin(\alpha) & \cos(\alpha)\sin(\beta)  \\
0 & \cos(\alpha) & \sin(\alpha) \sin(\beta) \\
1 & 0 & \cos(\beta) \\
\end{array}\right].  \label{eq:mmatrix}
\end{equation}
Finally, the relation between the the body-frame angular momenta $\bm{\mathtt{L}}$ and the conjugate angle momenta $\bm{\pi}$ in the laboratory frame is given by:
\begin{equation}
    \bm{\pi} = \mathcal{M}^\top \mathcal{R}\,\,\bm{\mathtt{L}}, \label{eq:Lpi}
\end{equation}
where $\mathcal{R}$ first transform the angular momentum from the body to the laboratory frame, and then  $\mathcal{M}^\top$ transforms it to the conjugate angle momenta.
Using Eqs.~\eqref{eq:ang}-\eqref{eq:Lpi} we can then transform Eq.~\eqref{eq:E-L} to find:
\begin{equation}
   d \bm{\pi} =   -\partial_{\bm{\phi}} H_\text{free} \,dt
   -\mathcal{M}^\top F^\text{(r)}_\text{lab} \,I^{-1}_\text{lab}\, (\mathcal{M}^\top)^{-1}\bm{\pi}  \,dt
      + \mathcal{M}^\top \left[\sqrt{2k_B T F^{(r)}}\right]_\text{lab}
     d\bm{V}_\text{lab}, \label{eq:pilab}
\end{equation}
where  $[~\cdot~]_\text{lab}= \mathcal{R}[~\cdot~]\mathcal{R}^\top$ denotes the quantity in the laboratory frame. The first term on the right-hand side of Eq.~\eqref{eq:pilab} is the Hamiltonian contribution, i.e.,
\begin{equation}
    \partial_{\bm{\phi}} H_\text{free}=
    (\mathcal{M}^\top \frac{d}{dt}(\mathcal{M}^\top)^{-1}) \bm{\pi}, \label{eq:hamil}
\end{equation}
where $H_\text{free}$ has been defined in Eq.~\eqref{eq:Hfree}). The validity of Eq.~\eqref{eq:hamil} can be verified using Eqs.~\eqref{eq:ang}-\eqref{eq:Lpi}, and the differentiation rules of ordinary calculus. The second and third terms on the right-hand side of Eq.~\eqref{eq:pilab} corresponds to the damping and noise terms, respectively.

\subsubsection*{Deterministic damping terms}
In this section we discuss the deterministic damping terms from Eqs.~\eqref{eq:plab} and ~\eqref{eq:pilab}:
\begin{alignat}{1}
d\boldsymbol{p}^{\text{(dc)}} & =  -F_\text{lab}^\text{(t)} \,\bm{p} \, dt ,\label{eq:Pd}\\
d\boldsymbol{{\pi}}^{\text{(dc)}} & =-\mathcal{M}^\top F^\text{(r)}_\text{lab} \,I^{-1}_\text{lab}\, (\mathcal{M}^\top)^{-1}\bm{\pi}  \,dt.\label{eq:Pid}
\end{alignat}

To illustrate how to use Eqs.~\eqref{eq:Pd} and \eqref{eq:Pid} we suppose that the coupling between translational and rotational degrees of freedom in the damping terms is negligible, and derive the damping model corresponding to the classical limit of the dissipative Caldeira-Leggett model \cite{caldeira1983path}. For the translational damping in Eq.~\eqref{eq:Pd} this amounts to assuming the limit of a spherical nanoparticle with a diagonal friction tensor in the body frame. In particular, we set
\begin{equation}
    F^\text{(t)}= \gamma_c \,\mathbb{I}, \label{eq:Ft}
\end{equation}
where $\mathbb{I}$ is the $3\times3$ identity matrix,  and the collisional damping rate is given by~\cite{epstein1924On,Cavalleri2010Gas}
\begin{equation}
    \gamma_c= \frac{4\,\pi m_g R^2 v_t P_g}{3 k_B T m}(1+\pi/8), \label{eq:damping}
\end{equation}
$P_{g}$ ($T$) is the gas pressure (temperature), $R$ ($m$) is the nanoparticle radius (mass), $m_{g}$ ($v_{t}$) is mass (thermal velocity) of a gas particle, and $k_{B}$ is Boltzmann's constant. In this limit we find that Eq.~\eqref{eq:Pd} reduces to
\begin{equation}
    d\boldsymbol{p}^{\text{(dc)}}  =  -\gamma_c \,\bm{p} \, dt, \label{eq:Pd2}
\end{equation}
which is the well-known model of translational damping for an isotropic particle. For the rotational damping we set friction tensor to be proportional to the moment of inertia tensor $I$:
\begin{equation}
    F^\text{(r)}= \gamma_c \,I, \label{eq:Fr}
\end{equation}
where we suppose that the damping coupling $\gamma_c$ is again the one defined in Eq.~\eqref{eq:damping}. In this case we find that Eq.~\eqref{eq:Pid} simplifies to
\begin{equation}
d\boldsymbol{{\pi}}^{\text{(dc)}}  = -\gamma_c \bm{\pi}  \,dt.\label{eq:Pid2}
\end{equation} 
The toy model obtained in Eq.~\eqref{eq:Pd2} and \eqref{eq:Pid2} also arises in the classical limit of the dissipative Caldeira-Leggett model (see \ref{app:Derivation}). More refined models can be developed using analytical/numerical calculations of the friction tensors~\cite{Cavalleri2010Gas,ahn2018optically}.

\subsubsection*{Stochastic collisional terms}
In this section we consider the stochastic collisional terms derived in Eqs.~\eqref{eq:plab} and ~\eqref{eq:pilab}:
\begin{alignat}{1}
d\boldsymbol{p}^{\text{(sc)}} & =\left[ \sqrt{2 M k_B T\,F^\text{(t)}}\right]_\text{lab} \,d\bm{U}_\text{lab},\label{eq:Pnc}\\
d\boldsymbol{{\pi}}^{\text{(sc)}} & =       \mathcal{M}^\top \left[\sqrt{2k_B T F^{(r)}}\right]_\text{lab}
     d\bm{V}_\text{lab}.\label{eq:Pinc}
\end{alignat}
As an illustrative example we continue with the model from the previous section. We insert the friction tensors given in Eqs.~\eqref{eq:Ft} and \eqref{eq:Fr} in Eqs.~\eqref{eq:Pnc} and \eqref{eq:Pinc}, respectively, to find:
\begin{alignat}{1}
d\boldsymbol{p}^{\text{(sc)}} & = \sqrt{2 M k_B T\,\gamma_c} \,\mathcal{R} \,d\bm{U},\label{eq:Pnc3}\\
d\boldsymbol{{\pi}}^{\text{(sc)}} & =    \mathcal{M}^\top \mathcal{R} \sqrt{2k_B T \gamma_c I}
     \,d\bm{V}. \label{eq:Pinc3}
\end{alignat}
We recall that $d\bm{V}=(dV_1,dV_2,dV_3)$ and $d\bm{U}=(dU_1,dU_2,dU_3)$ are independent zero-mean Wiener noise increments which satisfy the properties:
\begin{equation}
\mathbb{E}[dV_k dV_{k'}]=\delta_{k,k'} \,dt,
\quad\mathbb{E}[dU_k dU_{k'}]=\delta_{k,k'}\,dt, 
\quad\mathbb{E}[dV_k dU_k]=0. \label{eq:UV}
\end{equation}
Let us now compute the translational correlation matrix
$\bm{\sigma}^{(tt)}_{k,k'}=\mathbb{E}[d\boldsymbol{p}^{\text{(sc)}}_{k} \otimes d\boldsymbol{p}^{\text{(sc)}}_{k'}]$
and the rotational correlationa matrix
$\bm{\sigma}^{(rr)}_{k,k'}=\mathbb{E}[d\boldsymbol{\pi}^{\text{(sc)}}_{k} \otimes d\boldsymbol{\pi}^{\text{(sc)}}_{k'}]$. In particular, using Eqs.~\eqref{eq:Pnc3}-\eqref{eq:UV} we find:
\begin{equation}
\bm{\sigma}^{(tt)}_{k,k'}
= 2 M k_B T \gamma_c \, \text{diag}(1,1,1)_{k,k'}, \label{eq:stoc1}
\end{equation}
and
\begin{equation}
   \begin{array}{c}
\\
\bm{\sigma}^{(rr)}_{k,k'}\\
 \\
\end{array}
=  2k_B T \gamma_c
\left[
   \begin{array}{ccc}
 \bm{\sigma}^{(rr)}_{1,1} & \sin
   (\beta ) \sin (\gamma ) \cos (\gamma ) (I_2-I_1)   & \text{J3} \cos (\beta ) \\
\bm{\sigma}^{(rr)}_{2,1} & \frac{1}{2} (\cos (2 \gamma )
   (I_2-I_1)+I_1+I_2) & 0 \\
 I_3\cos (\beta ) & 0 & I_3 \\
\end{array}
\right], \label{eq:stoc2}
\end{equation}
where 
\begin{alignat}{1}
    \bm{\sigma}^{(rr)}_{1,1}=&\sin ^2(\beta ) \left(I_1 \cos ^2(\gamma )+I_2 \sin ^2(\gamma )\right)+I_3 \cos ^2(\beta ), \\
    \bm{\sigma}^{(rr)}_{2,1}=&\sin (\beta ) \sin (\gamma ) \cos (\gamma ) (I_2-I_1). 
\end{alignat}
It is instructive to write in place of Eqs.~\eqref{eq:Pnc3} and \eqref{eq:Pinc3} the following stochastic terms:
\begin{alignat}{1}
d\boldsymbol{p}^{\text{(sc)}} & = \sqrt{2 M k_B T\,\gamma_c}\, d\bm{U},\label{eq:Pnc4}\\
d\boldsymbol{{\pi}}^{\text{(sc)}} & = \sqrt{\bm{\sigma}^{(rr)}}
     \,d\bm{V},\label{eq:Pinc4}
\end{alignat}
which reproduce the same noise correlation matrices in Eqs.~\eqref{eq:stoc1} and \eqref{eq:stoc2}.  The form of Eq.~\eqref{eq:Pnc4} is now particularly simple, matching the well-known stochastic terms in the limit of an isotropic particle. For the non-diagonal matrix $\bm{\sigma}^{(rr)}$ the square root is to be interpreted as the matrix obtained from the Cholesky decomposition (see \ref{app:Correlated-noises}), with the explicit formula arising from Eq.~\eqref{eq:Pinc4} too long to be reported here. However, let us consider the special case where $I_L\equiv I_1=I_2$, $I_s\equiv I_3$ and $\beta\approx \pi/2$, e.g., when $I_s\ll I_L$ one can think about a nanorod confined to the x-y plane. We then find from Eq.~\eqref{eq:Pinc4}:
\begin{alignat}{1}
d\boldsymbol{{\pi}}^{\text{(sc)}} & = \sqrt{2k_B T\,\gamma_c} 
\,\,\text{diag}(\sqrt{I_L},\sqrt{I_L},\sqrt{I_s}) \,d\bm{V},\label{eq:Pinc5}
\end{alignat}
which matches the expected form from symmetry considerations. In particular, the rotational noise is smallest along the symmetry axis of the nanoparticle, corresponding to the noise $\propto \sqrt{I_s}$, and degenerate along the two other axis.

\subsection{Simulation}\label{sec:Numerical-analysis}

In this section we consider the model developed in the previous section and explore the typical dynamical features that emerge depending on the particle geometry. We will consider a regime where recoil processes can be neglected and focus on the dynamics driven by stochastic thermal noise in the non-trivial regime where the equation of motion cannot be linearized. For simplicity we will consider different limits for an ellipsoidal nanoparticle for which there are semi-analytical expressions for susceptibility~\cite{bohren_absorption_1998}. This will allow us to fairly accurately represent the phenomenology that emerges in each case. The model developed in the previous section is, of course, valid for any Rayleigh particle, the particle size is considerably smaller than the wavelength of light, ~\cite{rademacher2022measurement}, but typically one has to rely on numerical calculations to obtain the susceptibility which are usually based on the discrete dipole approximation or on a finite-difference time-domain approach.

We will consider an ideal sphere, a rod-like particle (prolate ellipsoid), a disk-like particle (oblate ellipsoid) and an asymmetric top (non degenerate ellipsoid). The use of an ellipsoidal geometry is convenient for two reasons, in the Rayleigh range it represents a reasonable approximation for the light-particle interaction and the susceptibility tensor can be calculated analytically. The tensor elements are given by $\chi_i=(\epsilon_r-1)/[1+(\epsilon_r-1)N_i]$ where $\epsilon_r$ is the relative permittivity of the material and $N_i$ is called depolarization factor and it quantifies the optical anisotropy. It is given by~\cite{schafer2021cooling,rudolph2021theory}:

\begin{equation}\label{eq_anisotropy}
  N_i= \frac{R_1 R_2 R_3}{2} \int_0^
  \infty \frac{ds}{(s+R_i^2)\sqrt{(s+R_x^2)(s+R_y^2)(s+R_z^2)}}.
\end{equation}

In the following, we will use as a guide to the simulations results the analytical estimation of the trap frequencies of all degrees of freedom. It is quite useful, then, to explicitly write simple expression for each. A zero order estimation of the trap frequencies, without any assumption on geometry, can be obtained by Taylor expanding $H_{\text{gradient}}$ in Eq.~\ref{eq:Hgradient}, i.e., $\omega_q^2=(2/m_q) \partial^2 H_{\text{gradient}}/\partial q^2$, calculated at the focus with $q=(x,y,z,\alpha,\beta,\gamma)$ and $m_q=(m,m,m,J_1,J_2,J_3)$. In this way one obtains

\begin{align}\label{eq:trapfreq}
        \omega_q^2=& \frac{I_{0}}{c\,\rho} \bigg\{ 
    \frac{2(\chi_3 \cos^2 \psi+a1^2 \chi_2 \sin^2 \psi)}{a1\, w_{0}^2}, 
    \frac{2(a1^2 \chi_3 \cos^2 \psi+ \chi_2 \sin^2 \psi)}{a1\, w_{0}^2},
    \frac{(\chi_3 \cos^2 \psi+a1^2 \chi_2 \sin^2 \psi)}{z_{R}^2},\nonumber\\
    & \frac{m}{J_1} (\chi_3-\chi_2) \cos 2 \psi,
    \frac{m}{J_2} (\chi_3-\chi_1) \cos^2  \psi,
    \frac{m}{J_3} (\chi_2-\chi_1) \sin^2 \psi
    \bigg\}
\end{align}

\noindent where $I_{0}= 2 P /\sigma_{L}$ is the intensity at the potential center. Eq.~\ref{eq:trapfreq} shows a square root dependence on the trapping field power. However, more general expressions must consider the steady state displacement along the tweezer propagation axis and corrections due to the scattering force (see~\ref{app:gauss} for additional details). Furthermore, the typical thermal variance of a trapped nanoparticle is typically large enough to make the Gaussian profile of the trapping potential evident. Indeed, the spectral line-shapes can easily exhibit distortions due to non-negligible (softening) Duffing terms~\cite{gieseler2013thermal,Zheng2020Robust,zemanek2020using}.

\subsubsection{Setting up the simulation}\label{subsec:Characterization-of-the}

We numerically integrate the particle equations of motion Eqs.~\ref{eq:dr}-\ref{eq:dpi} by implementing a $4^{\text{th}}$ order Runge-Kutta algorithm~\cite{platen2010numerical}. This means solving for twelve coupled equations driven by thermal noise. While the stochastic terms driving linear momenta are uncorrelated, the stochastic torques are correlated. These are constructed by following Eq.~\ref{eq:Pinc} which exploits linear combinations of 9 uncorrelated stochastic terms. Each simulation outputs 12 time traces which are then analyzed in Fourier space by calculating their power spectral density (PSD). For computational efficiency we run $\simeq30$ parallel simulations for each parameter set which are then averaged in a single PSD.

In a general situation the motion of the nanoparticle is affected by several parameters which we separate in different categories. These include:

\textit{Particle parameters}: the mass ($M$), volume ($V$), density ($\rho$), relative dielectric constant ($\epsilon_r$), the moment of inertia ($I_{1}$, $I_{2}$, $I_{3}$) and susceptibility ($\chi_{1}$,$\chi_{2}$,$\chi_{3}$) both along the principal axis in the particle body frame. Notice that for the susceptibility tensor the following hierarchy $\chi_1< \chi_2< \chi_3$ is assumed. 

\textit{Tweezer parameters}: field ellipticity ($b_{x}$,$b_{y}$), the laser wavelength ($\lambda$), the $x-y$ asymmetry of the Gaussian trap $a_{1}$, the Rayleigh range ($z_{R}$), the effective beam waist ($w_{0}$) and laser power at the focus ($P$). 

\textit{Environmental parameters}:  gas pressure ($p$), the gas temperature ($T$), the mass of the gas particles ($m$).

Common parameters that are kept constant in the following sections are summarized in Tab.~\ref{tab1}.

{
\begin{table}[h]
  \centering
  \setlength\extrarowheight{7pt}
\begin{tabular}{|c|c|}
  \hline
  Particle   & \thead{Silicon\\$\rho=2330$\,Kg/m$^3$; $\epsilon_\text{r}=12$;}   \\
  \hline
  Tweezer    & \thead{$\lambda=1550$\,nm; $P=300$\,mW;\\
                $w_{0}=1.06$\,$\mu$m;\,\, $a_{1}=1.126$; \\
                }    \\
  \hline
  Environment& $T=300$\,K; $\gamma_c=\frac{\sqrt{2\pi m_\text{g}} (8+\pi) P \bar{R}^2}{3 m \sqrt{k_B T}}$\\
  \hline
  \end{tabular}
\caption{Summary of common parameters used for the simulations in the following sections. Here, $\bar{R}$ is the radius of a sphere with the same volume of the geometry considered in the simulation. }
  \label{tab1}
\end{table}
}

\subsubsection{Translational Motion}
The first situation we will consider is an isotropic particle, i.e., a sphere of radius $R$ and $\chi_{1}=\chi_{2}=\chi_{3}$, for which the dynamics reduces to only translational non-trivial motion. In this case, the rotational dynamics is \textit{silent} so that it does not couple to the light field and cannot be affected by it (and it is not interferometrically detectable). Then, the angular degrees of freedom simply undergo free diffusion. 

Typically, a nanoparticle trapped in an optical tweezer will exhibit an overdamped dynamics at atmospheric pressure which transitions to the underdamped regime once the pressure reaches low vacuum ($\simeq10$\,mbar)~\cite{li2010measurement}.

\begin{figure}[ht]
\centering \includegraphics[width=\textwidth]{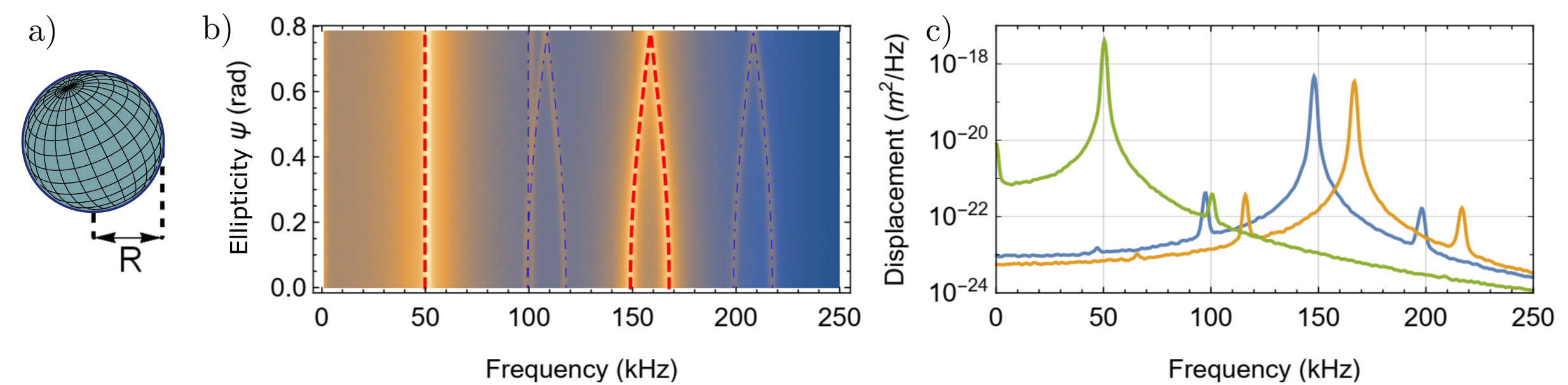} 
\caption{Numerical simulations for a spherical nanoparticle of radius $R=80$\,nm as shown in panel a) at a pressure of $0.5$\,mbar. The total simulated time is $2.5$\,s. b) Center of mass motion as a function of ellipticity parameter $\psi$. The density plot depicts the sum of PSDs represented in log scale with yellow the maximum. Also shown are analytical estimates of the trapping frequencies (Red-dashed) and the main contributions of their non-linear mixing (Blue-dot-dashed). c) PSDs considering linear polarization for the tweezer field along the $x$ (Blue), $y$ (yellow) and $z$ (green) axes. \label{fig:sphere} }
\end{figure}

The center of mass thermal motion of a tweezer held nanoparticle in vacuum is well represented in the literature. However, a parameter less explored is the influence on the dynamics of the elliptical polarization of the trapping field. We show in Fig.~\ref{fig:sphere}\,b) a 2D plot of the PSD of the combined center of mass degrees of freedom as a function of ellipticity. The density plot shows the sum of the $x$, $y$, $z$ PSDs. The peaks corresponding to the harmonic trap frequencies $\omega_q$ are clearly visible and represent the dominant features of the PSDs. These are compared to analytical expectations (red-dashed lines). The non-linearity of the potential is apparent since the second harmonic of $\omega_z$ can clearly be seen. Furthermore, non-linear mixing of the trap frequencies is also present with peaks at $\omega_x \pm \omega_z$ and $\omega_y \pm \omega_z$ (shown as blue dot-dashed lines). As the ellipticity is changed, $z$ remains largely unaffected while the frequency difference $\omega_x -\omega_y$ diminishes until the two modes become degenerate when the polarization is circular. This is the only effect of elliptical polarization for a spherical particle and is simply due to the vanishing asymmetry of the Gaussian mode profile. As we will show in the next section $\psi$ has a more important role when the rotational degrees of freedom come into play.

For completeness, we also show in Fig.~\ref{fig:sphere}\,c), the PSDs for $x$, $y$ and $z$ for the case of linear polarization where the effects of the nonlinear potential can clearly be seen.

\subsubsection{Rod-like particle (prolate ellipsoid)\label{sec:prolate}}

Here we consider a particle with cylindrical symmetry such that $R_{1}=R_{2}<R_{3}$. This is directly mapped to the susceptibility and moments of inertia tensors so that we have $\chi_{1}=\chi_{2}<\chi_{3}$ and $I_{1}=I_{2}>I_{3}$. These assumptions allows to approximate fairly well a rod-like nanoparticle. For this geometry, the rotational dynamics becomes apparent and the particle will tend to align its long axis to the polarization vector. Starting by considering linear polarization ($\psi=0$), it is possible to show that the steady state values for the angular degrees of freedom are, in this case, $(\alpha,\beta)=(0,\pm\pi/2)$. The motion is affected by the gradient force as well as the non-conservative deterministic terms. The cylindrical symmetry of this geometry, implies that one of the angular degrees of freedom, i.e., $\gamma$, does not couple to the optical field so it will undergo free diffusion; $\alpha$ and $\beta$ will librate around the steady state with trap frequencies $\omega_{\alpha}$ and $\omega_{\beta}$  which can be estimated by Taylor expanding the potential, as for the center of mass.

For a rod-like particle, however, the intrinsic non-linearity of the rotational dynamics emerges strongly since the evolution of $\alpha$ and $\beta$ are non-linearly coupled to each other through the free diffusion of $\gamma$. This case has already been treated in the context of optical levitation in Refs.~\cite{seberson2019parametric,bang2020five,trojek2012optical} where it has been shown that this coupling gives rise to precession around the polarization vector. This also means that $\alpha$ and $\beta$ hybridize so that the normal modes of the motion must be redefined. Notice, that non-nonlinearities simply due to the non-harmonicity of the trapping potential are still present and are typically more important than for the center of mass since the potential is significantly shallower.

\begin{figure}[ht]
\centering \includegraphics[width=1\linewidth]{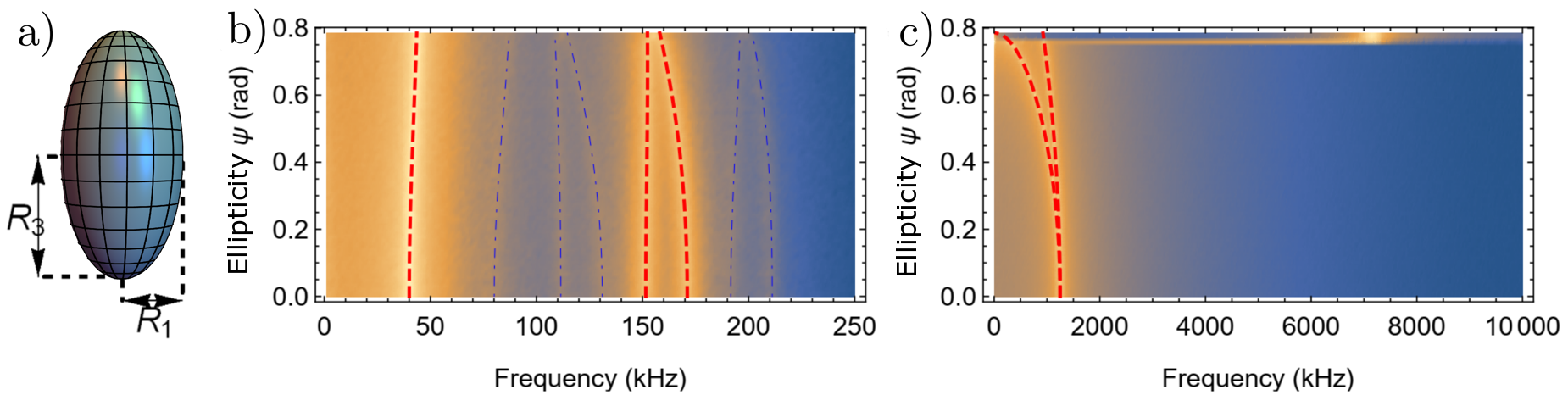} 
\caption{Numerical simulations for a prolate ellipsoidal nanoparticle. a) The geometry considered has $R_3 > R_2 = R_1$  which can represent a rod-like particle. The simulated values are $R_1=75$\,nm and $R_3=2\,R_1$ at a pressure of $5$\,mbar for a total simulated time of $400$\,ms. b) Center of mass motion and c) angular motion as a function of ellipticity parameter $\psi$. The density plots show the sum of the PSDs, for the angular degrees of freedom $\gamma$ is not included since it is undergoing free diffusion. In both 2D plots are shown analytical estimates of the trapping frequencies (Red-dashed) and the main contributions of their non-linear mixing (Blue-dot-dashed).
\label{fig:rodlike}}
\end{figure}

We show in Fig.~\ref{fig:rodlike} the PSDs of the particle degrees of freedom as a function of the field ellipticity parameter $\psi$. The center of mass motion is shown in Fig.~\ref{fig:rodlike}~b) which quite closely resembles Fig.~\ref{fig:sphere} both in terms of the non-linearity exhibited and dependence on $\psi$. Notice, however, that in this case $\omega_x$ and $\omega_y$ do not converge to to their mean when the polarization reaches a circular state, but instead it is shifted towards $\omega_x$. The non diffusing angular degrees of freedom are shown in Fig.~\ref{fig:rodlike}~c), as for the center of mass, the 2D plot shows the sum of the $\alpha$ and $\beta$ PSDs. For linear polarization $\omega_\alpha$ and $\omega_\beta$ are degenerate with a spectral peak strongly broadened by the non linear coupling with $\gamma$. As the ellipticity is increased the degeneracy is lifted with both $\omega_\alpha$ and $\omega_\beta$ shifted at lower frequencies at the same time the torque exerted on $\alpha$ by the non-conservative deterministic terms increases. Once reached a sufficient ellipticity $\alpha$ has a finite probability of escaping the trapping potential and to start spinning. For the simulated dynamics in Fig.~\ref{fig:rodlike}~c) this occurs at $\psi\simeq0.7$\,rad where it can be seen that $\alpha$ spins-up until an equilibrium between the driving torque and friction due to the gas is reached, in this case $\dot{\alpha}/2\pi\simeq7$\,MHz. Notice that the fluctuations of the  frequency of spinning follow a fluctuation-dissipation relation~\cite{vanderLaan2020Optically}. Higher order harmonics and frequency mixing can also be observed.

\subsubsection{Disk-like particle (oblate ellipsoid)\label{sec:disk}}

\begin{figure}[ht]
\centering \includegraphics[width=1\linewidth]{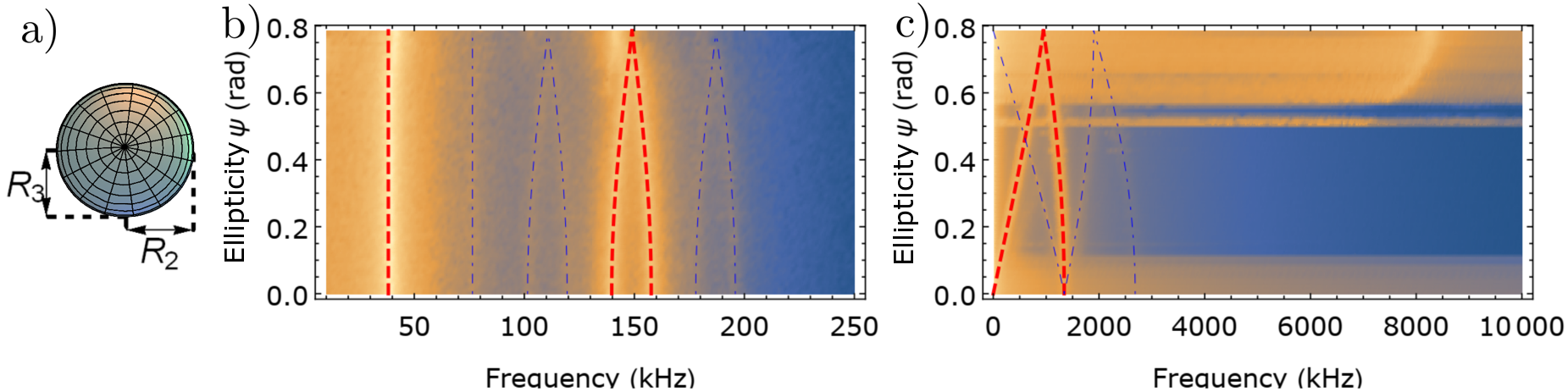} \caption{Numerical simulations for an oblate ellipsoidal nanoparticle. a) The geometry considered has $R_3 = R_2 > R_1$ which can represent a disk-like particle. The simulated values are $R_1=64$\,nm and $R_3=2\,R_1$ at a pressure of $5$\,mbar for a total simulated time of $50$\,ms. b) Center of mass motion and c) angular motion as a function of ellipticity parameter $\psi$. The density plots show the sum of the PSDs. In c) the high value wide-band sections emerge since there is a finite probability of $\gamma$ ($\alpha$) escaping the trapping potential this is highly likely for small (large) values of $\psi$. In both 2D plots are shown analytical estimates of the trapping frequencies (Red-dashed) and the main contributions of their non-linear mixing (Blue-dot-dashed). 
\label{fig:disklike}}
\end{figure}

The second possibility to retain a cylindrical symmetry for the particle but observe a different dynamics is to consider an oblate ellipsoid which can approximate a disk-like particle. In this case, we assume $R_{1}<R_{2}=R_{3}$ so that we have $\chi_{1}<\chi_{2}=\chi_{3}$ and $I_{1}>I_{2}=I_{3}$. As in the previous case, for linear polarization, the axis with the largest polarizability aligns to the field polarization. For an oblate ellipsoid, however, the largest polarizability is degenerate along two axes. As a consequence, the steady state mean value for $\beta=\pi/2$ but both $\alpha$ and $\gamma$ are freely diffusing. As the ellipticity is increased a confining potential starts to emerge for $\gamma$. The potential is shallow at first, so that $\gamma$ can easily escape. At intermediate values, both $\beta$ and $\gamma$ are stably trapped with $\alpha$ still diffusing. This transition is clearly shown in Fig.~\ref{fig:disklike}\,c). The steady state values of $(\beta,\gamma)=(\pm\pi/2,0)$ at this point, indicating that the "large face" of the ellipsoid is orthogonal to the tweezer field propagation direction. As for a rod-like particle, the diffusion of $\alpha$ nonlinearly couples the motion of $\beta$ and $\gamma$ so that the two modes hybridize.

The limiting case of circular polarization is quite different from the rod-like particle scenario. Once the scattering torques overcome the restoring force of the trapping potentials, the motion transitions towards a different steady state with $\alpha$ spinning at high frequency, in this case  $\dot{\alpha}/2\pi\simeq8$\,MHz. In this configuration, the "large face" of the particle is orthogonal to the tweezer polarization plane and spins around the field propagation axis ($z$ axis). 

This drastic change of the steady state has an impact on the center of mass has well. As in the previous case, when the particle is spinning the $x$ and $y$ trap frequencies become degenerate but are shifted to a lower frequency as shown in Fig.~\ref{fig:disklike}\,b). When the particle librates both $x$ and $y$ trap frequencies are proportional the the large particle susceptibility but once the particle starts spinning the frequencies are proportional to the large and small susceptibility averaged over the spinning frequency.

The dynamics of disk-like particles are prone to instability, mainly arising from radiation pressure. Recently, it has been shown, both theoretically~\cite{Seberson2020Stability} and experimentally~\cite{winstone2022optical}, that stable trapping can be more easily achieved in counter propagating setups where the role of radiation pressure is marginal.

\subsubsection{Asymmetric top (non degenerate ellipsoid)\label{sec:top}}

In this section, we consider an asymmetric top particle which means that all three radii are different. We assume $R_{1}<R_{2}<R_{3}$, as a consequence, the same hierarchy will be found in the susceptibility tensor and its inverse for the moments of inertia tensor. For this geometry, all three angular degrees of freedom couple to the trapping field so that the particle orientation can potentially be fixed in the laboratory frame with three distinct trapping frequencies. Nonetheless, the dynamics of the angular degrees of freedom can still appear relatively complex due to the shallow trapping potential.

\begin{figure}[t!]
\centering \includegraphics[width=1\linewidth]{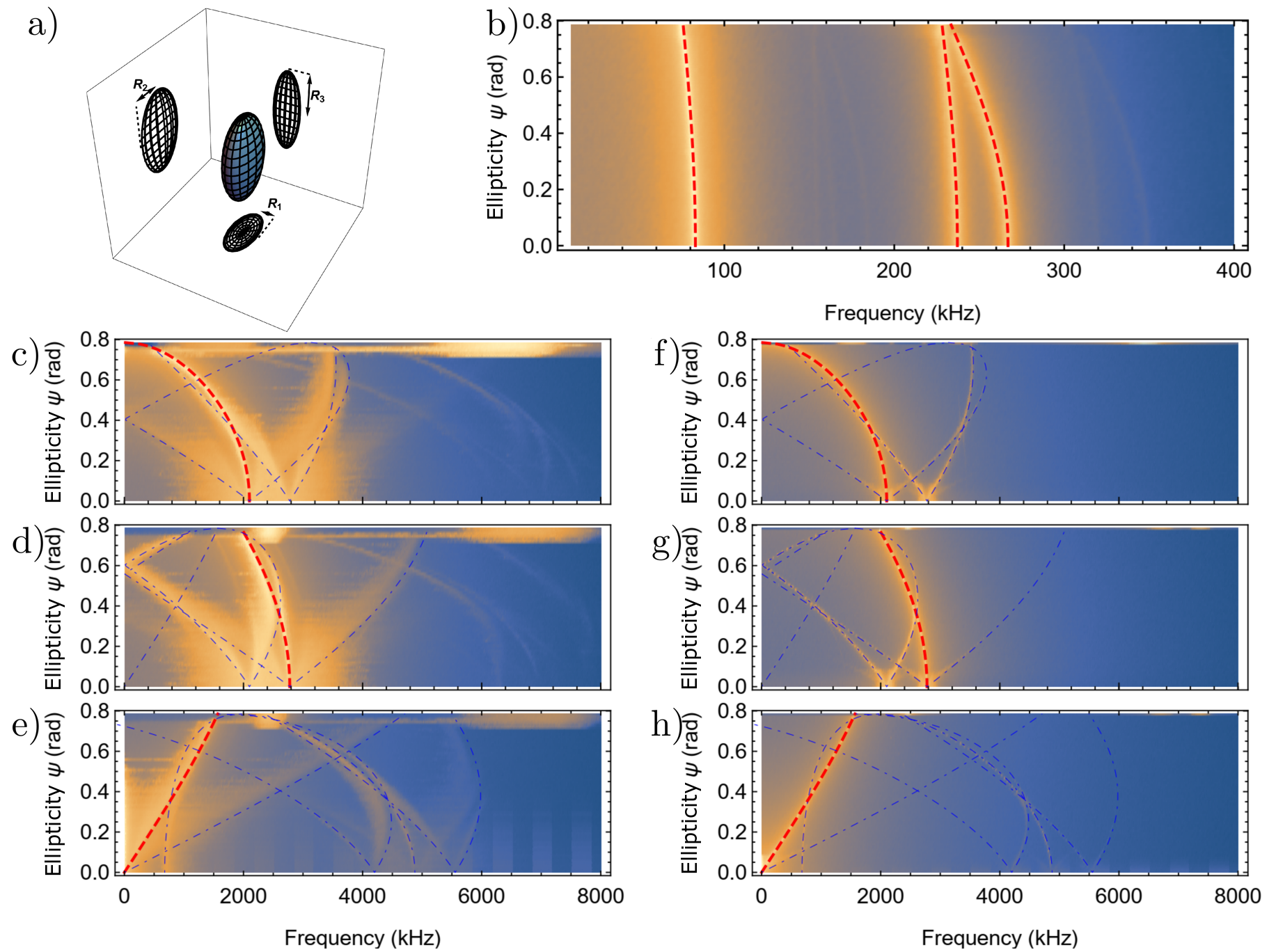}
\caption{Numerical simulation for an asymmetric-top nanoparticle. a) The geometry considered is a shell of thickness $h$ and has external radii $R_3 > R_2 > R_1$.  The simulated values are $R_1=42$\,nm and $R_2=57$\,nm, $R_3=91$\,nm and $h=15$\,nm at a pressure of $0.5$\,mbar for a total simulated time of $780$\,ms. For these simulations the power of the trapping field was fixed to $600$\,mW. b) center of mass motion as a function of ellipticity parameter $\psi$. The density plots show the sum of the PSDs of $x$, $y$ and $z$. Panels c), d) and e) show the PSD of $\alpha$, $\beta$ and $\gamma$ respectively as a function of $\psi$. For this geometry, the angular degrees of freedom are shown separately since the spectra are fairly more complex. To provide a simplified picture of the dynamics, we show in panels f), g) and h) a simulation with identical parameters except for a lower temperature of $T=3$\,K for the thermal bath which greatly reduce the nonlinearity due to shallow potentials. In all 2D plots are shown analytical estimates of the trapping frequencies (Red-dashed) and the main contributions of their non-linear mixing (Blue-dot-dashed). \label{fig:hollow}}
\end{figure}

We show in Fig.~\ref{fig:hollow}\,b) a density plot of the sum of the $x$, $y$ and $z$ PSDs as a function of the tweezer field ellipticity. As for the previous geometries, the general behavior is quite similar to that of a spherical particle with the only exception that the convergence of $\omega_x$ and $\omega_y$ for circular polarization occurs at an even lower frequency.

The dynamics for the angular degrees of freedom is far richer. In this case, we show a 2D plot of the simulated PSDs for $\alpha$, $\beta$ and $\gamma$ individually in Fig.~\ref{fig:hollow}\,c), d) and e) respectively along with analytical estimates of the trap frequencies (red-dashed), obtained as before by Taylor expansion, and main contributions from non-linear mixing (blue-dot-dashed). For a linearly polarized tweezer field, there is still no trapping potential for $\gamma$ so that the behavior is quite similar to that of a prolate ellipsoid discussed in Sec.~\ref{sec:prolate}: $\gamma$ diffuses and $\alpha$ and $\beta$ are non-linearly coupled. The only difference is that $\omega_\alpha$ and $\omega_\beta$ are non degenerate. It is important to notice, that for a linearly polarized field, it is often difficult to distinguish an asymmetric top from a prolate ellipsoid since that requires the frequency difference between $\omega_\alpha$ and $\omega_\beta$ to be larger than the broadening due to the diffusion of $\gamma$.

As the ellipticity increases, a trapping potential for $\gamma$ starts to emerge, but the trap depth is extremely shallow. As a consequence, $\gamma$ is likely to escape the potential well and remains trapped only for short times. This keeps $\gamma$ diffusing for large fractions of the evolution with the associated non-linear coupling and spectral broadening for $\alpha$ and $\beta$ present. For the simulated parameter $\gamma$ becomes stably trapped for an ellipticity $\psi\simeq0.5$ as evidenced by the disappearance of low frequency spectral features in Fig.~\ref{fig:hollow}\,e). Notice that energy conservation implies that as the trap depth for $\gamma$ deepens the potential for $\alpha$ becomes shallower.

The dominant contributions for frequency mixing and mode coupling can be identified as $\omega_\alpha \pm 2\omega_\gamma$ and $\omega_\beta \pm \omega_\gamma$ for $\alpha$ in Fig.~\ref{fig:hollow}\,c); $\omega_\alpha \pm \omega_\gamma$, $\omega_\beta \pm 2\omega_\gamma$ and $\omega_\gamma$ for $\beta$ in Fig.~\ref{fig:hollow}\,d); and finally, $\omega_\beta-\omega_\alpha$, $3\omega_\gamma$, $2\omega_\alpha \pm \omega_\gamma$, $2\omega_\beta \pm \omega_\gamma$ and $\omega_\alpha+\omega_\beta$ for $\gamma$ in Fig.~\ref{fig:hollow}\,e).

When the ellipticity approaches circular polarization, $\omega_\alpha$ tends to zero and the particle starts to spin in the tweezer polarization plane under the influence of the deterministic scattering torque. In this case, however, the spin frequency is highly unharmonic due to the modulation of the scattering torque that emerges, as $\chi_1\neq\chi_2$.

The 2D plots in Fig.~\ref{fig:hollow}\,c), d) and e) are somewhat cumbersome, to gather a clearer picture of the dynamics we show in Fig.~\ref{fig:hollow}\,f), g) and h) a simulation with the same parameters but assuming a lower temperature for the bath, i.e., $T=3$\,K. This allows us to greatly reduce the nonlinearity attributable to the shallow potentials. In particular, since $\gamma$ is now stably trapped even with a small ellipticity ($\psi\simeq0.1$) all spectral features are significantly narrower and more easily identifiable. Another consequence of the smaller thermal excitation is that the particle starts spinning only for a polarization state much closer to circular.

\subsection{Quantum model\label{sec:quantum_model}}
In this section, we look at the quantum counter-part of the classical model derived in Sec.~\ref{sec:model}. We recall that the classical dynamics consisted of the set of Langevin equations (coupled stochastic differential equations), which are  summarized in Eqs.~\eqref{eq:dr}-\eqref{eq:dpi}.  The system is formed by the nanoparticle, which is interacting with the environments consisting of the optical field and of the gas particles in the vacuum chamber. We have treated the nanoparticle as an \emph{open classical system}, effectively reducing the modeling with the two environments to conservative Hamiltonian potentials, as well as to deterministic damping terms and classical noises, which are cooling and heating the nanoparticle, respectively.

To obtain the corresponding quantum equations of motion, now treating the nanoparticle as an \emph{open quantum system}, we can perform the standard quantization procedure, replacing the Poisson brackets with the commutators and classical observables with quantum observables, i.e., 
\begin{equation}
      \left\{ \,\cdot\,,\,\cdot\,\right\}  \rightarrow    -\frac{i}{\hbar}\left[\,\cdot\,,\,\cdot\,\right],\qquad\qquad\qquad O\rightarrow \hat{O}, \label{eq:quantization}
\end{equation}
where $O$ ($\hat{O})$ denotes a classical (quantum) observable. For example, the gradient Hamiltonian in Eq.~\eqref{eq:Hgradient} can be readily transformed to a quantum operator following such a prescription, resulting in the quantum mechanical operator $\hat{H}_\text{gradient}$.

When promoting the free kinetic Hamiltonian to the quantum Hamiltonian operator, the situation is more subtle, and one has to use the Laplace-Beltrami operator~\cite{podolsky1928quantum}:
\begin{equation}
    \hat{H}_\text{free}= \vert  \hat{g} \vert^{-\frac{1}{2}} \hat{\mathsf{p}}_i \vert \hat{g} \vert^{\frac{1}{2}}  \hat{g}^{ij} \hat{\mathsf{p}}_j \label{eq:LB}
\end{equation}
where $\hat{g}_{ij}$ ($\hat{g}^{ij}$) is the metric (inverse metric), and  $\hat{g}$ denote the corresponding metric determinant. The generalized conjugate momentum is simply given by $\hat{\mathsf{p}}_j= -i \hbar \frac{\partial}{\partial \mathsf{x}_j}$, where $\mathsf{x}$ denotes the generalized position (comprising spatial positions and angles of orientation, i.e., $\mathsf{x}_j=x,y,z,\alpha, \beta,\gamma$). The Laplace-Beltrami operator, which generalizes the notion of the Laplace operator from Euclidean space (e.g., working with translational motion in Cartesian coordinates) to non-Euclidean space in differential geometry~\cite{jost2005riemannian}, can be viewed as a method for fixing consistently the operator ordering in the quantum Hamiltonian for any choice of the coordinates.
For translational degrees of freedom described in Cartesian coordinates ($i,j=x,y,z$) the metric elements reduce $\hat{g}_{ij}= \delta_{ij}$, while for the rotational degrees of freedom modeled in the Euler angles illustrated in Fig.~\ref{fig:exp} ($i,j=\alpha,\beta,\gamma$) the metric has the following non-trivial structure
\begin{equation}
\hat{g}_{ij}=  
\begin{bmatrix}
\sin^2(\hat{\beta}) \left(J_1 \cos^2(\hat{\gamma}) + J_2 \sin^2(\hat{\gamma})\right) + J_3 \cos^2(\hat{\beta}) 
& \text{sin} (\hat{\beta}) \sin(\hat{\gamma}) \cos(\hat{\gamma}) (J_2 - J_1) 
& J_3 \text{cos}(\hat{\beta}) \\
\text{sin} (\hat{\beta}) \sin(\hat{\gamma}) \cos(\hat{\gamma}) (J_2 - J_1) & J_1 \sin^2(\hat{\gamma}) + J_2 \cos^2(\hat{\gamma}) & 0 \\
J_3 \cos(\hat{\beta}) & 0 & J_3
\end{bmatrix}. \label{eq:metric}
\end{equation}
More compactly, the metric in Eq.~\eqref{eq:metric} can be also written as $\hat{g}_{ij}=(\hat{\mathcal{M}}^\top  \hat{\mathcal{R}} I \hat{\mathcal{R}}^\top \hat{\mathcal{M}}) _{ij}$, where $\hat{\mathcal{R}}$ and $\hat{\mathcal{M}}$ denote the quantized version of the rotation matrix in Eq.~\eqref{eq:R} and of the transformation matrix in Eq.~\eqref{eq:mmatrix}, respectively,  and $I$ is the moment of intertia tensor in the body frame.

We continue by discussing the modeling of the non-unitary damping terms and input noises (i.e., the quantum version of the deterministic and stochastic non-conservative terms). Such noises are modeled using quantum Wiener processes, which can be seen as a quantum version of the corresponding classical Wiener processes (see for example~\cite{gardiner2004quantum}). Let us look first at the terms related to the interaction of the nanoparticle with the optical field. The deterministic scattering terms in Eqs.~\eqref{eq:sforce}-\eqref{eq:ds3}, corresponding to the scattering force and torques, respectively, are readily transformed into operators $\dd\hat{\boldsymbol{p}}^{\text{(ds)}}$ and $\dd\hat{\boldsymbol{\pi}}^{\text{(ds)}}$ by promoting classical observables to quantum observables as outlined in Eq.~\eqref{eq:quantization}. 

We can proceeded similarly also to obtain the quantum noise terms $\dd\hat{\boldsymbol{p}}{}^{\text{(ss)}}$ and $\dd\hat{\boldsymbol{\pi}}{}^{\text{(ss)}}$, which correspond to the quantum version of the recoil heating terms defined in Eqs.~\eqref{eq:Pns} and \eqref{eq:Pins}, respectively. The quantum noises, unlike their classical counter-part, satisfy the commutation relation~\cite{wiseman2009quantum}:
\begin{equation}
    [d\hat{X}^\text{(in)}_{\bm{n},\nu},d\hat{P}^\text{(in)}_{\bm{n}',\nu'} ]= i \delta(\bm{n}-\bm{n}')\delta_{\nu,\nu'} dt. ~\label{dhatXdhatP}
\end{equation}
Defining the quantum noise for the mode,
\begin{equation}
    d\hat{a}^\text{(in)}_{\bm{n},\nu} = \frac{1}{\sqrt{2}} \left(d\hat{X}^\text{(in)}_{\bm{n},\nu}+ i d\hat{P}^\text{(in)}_{\bm{n},\nu} \right),
\end{equation}
we can also recast Eq.~\eqref{dhatXdhatP} as 
\begin{equation}
    [d\hat{a}^\text{(in)}_{\bm{n},\nu},d\hat{a}^{\text{(in)}\dagger}_{\bm{n}',\nu'}]=  \delta(\bm{n}-\bm{n}')\delta_{\nu,\nu'} dt.
\end{equation}
The quantum noises are characterized by the following properties:  
\begin{alignat}{1}
&\langle d\hat{a}^\text{(in)}_{\bm{n},\nu}\rangle=0,\quad \langle \hat{a}^{\text{(in)}\dagger}_{\bm{n},\nu}\rangle=0 \quad
\langle d\hat{a}^{\text{(in)}}_{\bm{n},\nu} d\hat{a}^{\text{(in)}}_{\bm{n}',\nu'} \rangle = 0,\quad 
\langle d\hat{a}^{\text{(in)}\dagger}_{\bm{n},\nu} d\hat{a}^{\text{(in)}\dagger}_{\bm{n}',\nu'}\rangle = 0 \quad \label{eq:noises2QA1} \\
&\langle d\hat{a}^{\text{(in)}\dagger}_{\bm{n},\nu} d\hat{a}^{\text{(in)}}_{\bm{n}',\nu'} \rangle = 0,\quad 
\langle d\hat{a}^\text{(in)}_{\bm{n},\nu} d\hat{a}^{\text{(in)}\dagger}_{\bm{n}',\nu'}\rangle = \delta(\bm{n}-\bm{n}')\delta_{\nu,\nu'}dt, \label{eq:noises2QA2}
\end{alignat}
or equivalently
\begin{alignat}{1}
 &\langle d\hat{X}^\text{(in)}_{\bm{n},\nu}\rangle=0,\quad \langle \hat{P}^\text{(in)}_{\bm{n},\nu}\rangle=0, \quad 
 \langle d\hat{X}^\text{(in)}_{\bm{n},\nu} d\hat{P}^\text{(in)}_{\bm{n},\nu} \rangle =  \frac{i}{2}\delta(\bm{n}-\bm{n}')\delta_{\nu,\nu'}dt , \label{eq:noises2Q1} \\
&\langle d\hat{X}^\text{(in)}_{\bm{n},\nu} d\hat{X}^\text{(in)}_{\bm{n}',\nu'} \rangle = \frac{1}{2}\delta(\bm{n}-\bm{n}')\delta_{\nu,\nu'}dt,\quad 
\langle d\hat{P}^\text{(in)}_{\bm{n},\nu} d\hat{P}^\text{(in)}_{\bm{n}',\nu'}\rangle = \frac{1}{2}\delta(\bm{n}-\bm{n}')\delta_{\nu,\nu'}dt, \label{eq:noises2Q2}
\end{alignat}
where $\langle~\cdot~\rangle$ denotes the quantum average. One can think about Eqs.~\eqref{eq:noises2QA1}-\eqref{eq:noises2Q2} as the quantum counter-part or generalization of the classical expectation values in Eqs.~\eqref{eq:noises2A1}-\eqref{eq:noises2}.

We are left to discuss the quantum counter-part of the damping terms and of the stochastic noises related to gas collisions. For simplicity we will focus on the commonly used dissipative Caldeira-Leggett model~\cite{caldeira1983path}, which has been introduced in the classical form in Eqs.~\eqref{eq:Ft}-\eqref{eq:Pinc5}. We first note that the deterministic damping terms in Eqs.~\eqref{eq:Pd} and\eqref{eq:Pid} can be readily transformed into quantum operators $\dd\hat{\boldsymbol{p}}{}^{\text{(dc)}}$ and $\dd\hat{\boldsymbol{\pi}}{}^{\text{(dc)}}$, respectively.
The quantum operators $\dd\hat{\boldsymbol{p}}{}^{\text{(sc)}}$ and $\dd\hat{\boldsymbol{\pi}}{}^{\text{(sc)}}$ can be formally obtained from Eqs.~\eqref{eq:Pnc3} and \eqref{eq:Pinc3}, respectively, where we need to introduce the quantum noises $d\hat{V}$ and $d\hat{U}$ in place of the corresponding classical noises. The quantum noises are characterized by the expectation values
\begin{equation}
\langle d\hat{V}_k d\hat{V}_{k'} \rangle = \delta_{k,k'} \,dt, \quad 
\langle d\hat{U}_k d\hat{U}_{k'} \rangle = \delta_{k,k'}\,dt,  \quad
\langle d\hat{V}_k d\hat{U}_k \rangle=0,  \label{eq:UVQ}
\end{equation}
where we have implicitly assumed that the environmental temperature $T$ is large (e.g., at room temperature). Specifically, when $k_B T \gg \hbar \omega$, where $\omega$ is the frequency of a mechanical degree of freedom, we can neglect the asymmetry of the expectations values of the quantum noise as $k_B T/\hbar\omega+1 \approx k_B T/\hbar\omega$ (in contrast to the case of the optical quantum noise discussed in Eqs.~\eqref{eq:noises2QA1}-\eqref{eq:noises2Q2})\cite{wiseman2009quantum}. In short, the quantum features encoded in the expectation values become important for low occupancy thermal states, while for environments at high temperature the difference between the classical and quantum noises become less pronounced.

Putting it all together we find in place of Eqs.~\eqref{eq:dr}-\eqref{eq:dpi}, the following coupled quantum Langevin equations:
\begin{equation}
    \dd\hat{\boldsymbol{r}}= -\frac{i}{\hbar} \left[ \hat{\boldsymbol{r}}, \hat{H}_{\text{free}} \right]\dd t\;,
    \label{eq:drQ}
\end{equation}
\begin{equation}
    \dd\hat{\boldsymbol{p}}=  -\frac{i}{\hbar} \left[ \hat{\boldsymbol{p}}, \hat{H}_{\text{gradient}}\right] \dd t+\dd\hat{\boldsymbol{p}}^{\text{(ds)}}+\dd\hat{\boldsymbol{p}}{}^{\text{(dc)}}+\dd\hat{\boldsymbol{p}}{}^{\text{(ss)}}+\dd\hat{\boldsymbol{p}}{}^{\text{(sc)}}\;,
    \label{eq:dpQ}
\end{equation}
\begin{equation}
    \dd\hat{\boldsymbol{\phi}}=  -\frac{i}{\hbar}  \left[ \hat{\boldsymbol{\phi}},\hat{H}_{\text{free}}+\hat{H}_{\text{gradient}} \right] \dd t\;,
    \label{eq:dphiQ}
\end{equation}
\begin{equation}
    \dd\hat{\boldsymbol{\pi}}= -\frac{i}{\hbar}  \left[\hat{\boldsymbol{\pi}},\hat{H}_{\text{free}}+\hat{H} _{\text{gradient}} \right] \dd t+\dd\hat{\boldsymbol{\pi}}^{\text{(ds)}}+\dd\hat{\boldsymbol{\pi}}{}^{\text{(dc)}}+\dd\hat{\boldsymbol{\pi}}{}^{\text{(ss)}}+\dd\hat{\boldsymbol{\pi}}{}^{\text{(sc)}}\;.
    \label{eq:dpiQ}
\end{equation}

For completeness we also briefly discuss the Stochastic Master Equation (SME) approach~\cite{belavkin1999measurement,jacobs2014quantum,wiseman2009quantum}, which is complementary to the quantum Langevin equations in Eqs.~\eqref{eq:drQ}-\eqref{eq:dpiQ}. In the SME approach, the state of the system is encoded in the conditional statistical operator, $\rho_c$, which evolves according the following equation (written in It\^{o} stochastic form):
\begin{alignat}{1}
& d\hat{\rho}_{c}=	-\frac{i}{\hbar}[\hat{H}_{\text{free}}+\hat{H}_{\text{gradient}},\hat{\rho}_{c}]dt \nonumber\\
&	+\gamma_{s}\sum_{\nu}\int d\bm{n}\mathcal{D}[\hat{A}^\dagger_{\bm{n},\nu}]\hat{\rho}_{c}dt+
\sqrt{\gamma_{s}}\sum_{\nu=1}^{2}\int d\bm{n}\mathcal{H}[\hat{A}^\dagger_{\bm{n},\nu}]\hat{\rho}_{c}dU_{\bm{n},\nu} \nonumber\\
&	+\gamma_{c}\sum_{j=1}^{3}\mathcal{D}[\hat{\bm{L}}_{j}]\hat{\rho}_{c}dt
+\sqrt{\gamma_{c}}\sum_{j=1}^{3}\mathcal{H}[\hat{\boldsymbol{L}}_{j}]\hat{\rho}_{c}dV_{j}
+\gamma_{C}\sum_{\zeta,j=1}^{3}\mathcal{D}[\hat{\boldsymbol{C}}_{\zeta,,j}]\hat{\rho}_{c}dt+\sqrt{\gamma_{c}}\sum_{\zeta,j=1}^{3}\mathcal{H}[\hat{\boldsymbol{C}}_{\zeta,j}]\hat{\rho}_{c}dZ_{\zeta,j}, \label{eq:conditional}
\end{alignat}    
where $U_{\bm{n},\nu}$, $V_{j}$, and $Z_{\zeta,j}$ are taken to be independent zero mean Wiener processes. By taking the average over the noise realizations, $\mathbb{E}[\,\cdot\,]$,  we obtain the dynamics for the unconditional state usually denoted $\hat{\rho}$, i.e., $\hat{\rho}=\mathbb{E}[\hat{\rho_c}]$.
The first line of Eq.~\eqref{eq:conditional} contains the free and the gradient Hamiltonians, $\hat{H}_\text{free}$ and $\hat{H}_\text{grad}$, respectively, which have been discussed above in Eqs.~\eqref{eq:quantization}-\eqref{eq:metric}. 
The second line encodes the non-unitary contributions related to the scattering of the incoming photons from the laser off the nanoparticle. The optomechanical interaction of linear rotors with linearly polarized light has been discussed in~\cite{Stickler2016Rotranslational,stickler2016spatio}, the case of arbitrary rotors with unpolarized light in~\cite{papendell2017quantum}, and the case of arbitrary rotors with elliptical polarization followed in this review in~\cite{torovs2018detection}. The operator $\hat{A}_{\bm{n},\nu}$, can be obtained by quantizing the corresponding classical observable $A_{\bm{k},\nu}\vert_{\omega_k=\omega_L}$, where $A_{\bm{k},\nu}$ has been defined in Eq.~\eqref{eq:nano}, and $\omega_L$ is the angular frequency of the incoming photons. The Rayleigh scattering rate, $\gamma_s$, which controls the scattering process, matches the classical expression given in Eq.~\eqref{eq:gammas}.
The third line of Eq.~\eqref{eq:conditional} corresponds to the interaction between the nanoparticle with the gas particles within the dissipative Caldeira-Leggett model~\cite{caldeira1983path} (continuing with the example from Eqs.~\eqref{eq:Ft}-\eqref{eq:Pinc5}). Rotational diffusion without friction was discussed in \cite{papendell2017quantum}, while the rotational dissipators were first obtained in~\cite{stickler2017rotational}, and then recovered in \cite{torovs2018detection} in the Hamiltonian form used in this review. The overall process is controlled by the collisional damping rate, $\gamma_c$, which can be modeled as in Eq.~\eqref{eq:damping}, and the operators characterizing the net effect of gas collisions are given by:
\begin{equation}
\hat{L}_{j}	=\frac{i\sqrt{2Mk_{b}T}}{\hbar} (\hat{\boldsymbol{r}}+\frac{i\hbar}{2Mk_{b}T}\hat{\boldsymbol{p}}) \cdot\boldsymbol{e}_{j}, \quad
\hat{\boldsymbol{C}}_{\zeta,j}	=\frac{i\sqrt{2k_{b}T\tilde{D}_{\zeta}}}{\hbar}(\hat{\mathcal{R}}\mathbf{e}_{\zeta}
-\frac{i\hbar}{2k_{b}T}\hat{\mathcal{R}}G_{\zeta}I^{-1}\hat{\mathcal{R}}^{\top}
(\hat{\mathcal{M}}^{\top}\hat{\mathcal{R}})^{-1}\hat{\bm{\pi}})\cdot\boldsymbol{e}_{j}, \label{eq:moperators}
\end{equation}
for translational and rotational degrees of freedom, respectively, where $M$ is the nanoparticle mass, $k_B$ is Boltzmann's constant, $T$ is the temperature of the gas, $\hat{\bm{r}}$ ($\hat{\bm{p}}$) is the position (conjugate momentum) operator,  $I$ is the momentum of inertia tensor,  $\tilde{D}_{\zeta}=\frac{1}{2}\text{tr}(I)-I_{\zeta}$, $\boldsymbol{e}_{j}$ is the unit vector along the $j$ axis,
$G_\zeta$ is the generator of rotations about the $\zeta$-axis, $\hat{\mathcal{R}}$ and $\hat{\mathcal{M}}$ correspond to the quantized rotation matrix from Eq.~\eqref{eq:R} and transformation matrix from Eq.~\eqref{eq:mmatrix}), respectively, and $\hat{\bm{\pi}}$ is the conjugate angle momentum operator.
 
In Eq.~\eqref{eq:conditional} we have also adapted the following notation for the dissipators, $\mathcal{D}[\,\cdot\,]$, and the noise terms, $\mathcal{H}[\,\cdot\,]$, adapted from~\cite{wiseman1993interpretation}:
\begin{equation}
\mathcal{D}[\hat{K}]\,\cdot\,	=\hat{K}\,\cdot\,\hat{K}^{\dagger}-\frac{1}{2}\left\{ \hat{K}^{\dagger}\hat{K},\,\cdot\,\right\} ,
\end{equation}
\begin{equation}
\mathcal{H}[\hat{K}]\,\cdot\,	=\hat{K}\,\cdot\,+\,\cdot\,\hat{K}^{\dagger}-\text{tr}[\hat{K}\,\cdot\,+\,\cdot\,\hat{K}^{\dagger}]\,\cdot\,,
\end{equation}
where $\hat{K}$ denotes the operator controlling the non-unitary dynamics. For the interested reader we illustrate how one can recover the classical results derived in Sec.~\ref{sec:model} starting from the quantum model given by the SME in Eq.~\eqref{eq:conditional} in \ref{app:Derivation}.

\subsection{Comparison of trapping modalities}
\label{sec:comparison}

The general Hamiltonian structure developed here applies not only to optical traps but also to other levitation schemes such as magnetic or Paul traps. While our focus in this work is on optically levitated particles, it is useful to highlight how different trap types modulate the model.

\paragraph{Optical traps (focus of this work)} 
As detailed in Sec.~\ref{sec:model}, all the terms introduced in Eqs.~(\ref{eq:opt})–(\ref{eq:dpi}) are relevant. In particular, the optical gradient potential (conservative), radiation pressure (non-conservative), and stochastic recoil forces all arise from the electric field interaction with a dielectric nanoparticle. As such, the full tensorial structure of the susceptibility $\chi$ plays a central role. Active feedback and coherent driving are excluded from the current formulation, but can be incorporated in future extensions.

\paragraph{Magnetic traps} 
Magnetic traps confine particles via field gradients acting on their magnetic dipole moment. The conservative part of the Hamiltonian includes a magnetic potential term $H_{\text{mag}} = -\bm{\mu}\cdot\bm{B}$, where $\bm{\mu}$ is the magnetic moment and $\bm{B}$ the magnetic field. In this case, $\chi$ (electric susceptibility) and optical scattering terms are absent, and the gradient term derives from the field $\bm{B}$ instead of $\bm{E}_d$. The absence of recoil heating can potentially offer a clear advantage when compared to optical traps; however, it is often the case that additional dissipation channels are present. The most common is dissipation through the generation of eddy currents but, depending on the system, other processes can occur~\cite{vinante2020ultralow}.

\paragraph{Paul traps} 
Paul traps use a combination of oscillating and static electric fields to produce a confining potential. The effective potential arises from time-averaging the rapidly oscillating drive fields (pseudo-potential approximation). The susceptibility tensor $\chi$ is again inactive, and damping/noise terms primarily come from gas collisions. Like for magnetic traps, non-conservative optical forces are absent. However, an additional complication arises from the possible (likely) non-uniform charged distribution which implies the presence of multipoles. These can interact directly with oscillating electric fields and significantly impact particle rotational dynamics~\cite{perdriat2024Rotational}. 

This modular perspective allows one to turn on/off specific contributions in the equations of motion depending on the trapping modality.

\section{Control and manipulation of roto-translational motion \label{sec:applications}}

This section provides an overview of the applications of roto-translational levitated optomechanics realized to date. We focus on critical aspects of experimental achievements and promising proposals, discussing the current state-of-the-art and outlining the limitations and challenges associated with each platform. Specifically, we examine the control and manipulation of roto-translational motion, particle characterization via rotational dynamics, force and torque sensing, pressure sensing and viscosity measurements, and inertial and gravitational sensing. These applications highlight how the unique capabilities of roto-translational optomechanics can be leveraged to advance both fundamental research and practical technologies. Hence, we mainly focus on the control and applications of rotational degrees of freedom, which offer distinctive advantages over purely translational dynamics.

\subsection{Particle characterization via rotational motion}\label{sec:part_characterization}

Characterization of single nanoparticles such as viruses, aerosols, and colloidal particles are of particular interest for medical applications, material science, and atmospheric physics \cite{Ying2024Single,pan2019collection,mitra2010nano,wang2022online,zhu2011single,Asghari2021Fast,Priyadarshi2022A,nyandey2024convolutional,Kaye2000Simultaneous,Chen2020Light,Dick2007Multiangle}. Of particular note is the characterization via non-intrusive optical methods where the particle is isolated from the detrimental effects of substrates, solvents and other particles and can be held for extended periods of time where the particle's interaction with the light and surrounding medium can be observed~\cite{wu2017stable,kohler2021tracking,li2021fast,Rybin2023Novel,Kohli2021Measuring,Chen2025Shape,Li2024Morphological}.

As discussed throughout this paper, particles exhibit rotational behavior under two circumstances when levitated in polarized light: \textit{i)} the particle has natural birefringence due to the material properties; \textit{ii)} it has shaped-induced birefringence due to the anisotropic shape of the particle. The refractive index for a birefringent particle is different in two axes. This means the angular momenta exchanged from the polarized light to the particle varies for different axes and so exerts a net force. The rotational motion due to the anisotropic shape can be used to characterize the shape. 

One of the first indications in an experiment that the levitated particle is asymmetrical in shape is how the gas particles interact with the object, particularly in the few millibar regime. For a spherical particle, the way the gas particles scatter from the surface of the object is isotropic, which means the gas damping experienced is the same in all directions. However, for an asymmetrical particle, the gas dampings will reflect the degree by which the symmetry is broken. For example, if we consider only the CoM, a cylindrical symmetry implies two different damping coefficients, one of which will be doubly degenerate. While, for an asymmetric top all three would be different. The same can be shown for the rotational degrees of freedom. For any convex geometry, the expected dampings can be calculated following~\cite{Cavalleri2010Gas,epstein1924On}, on the other hand, a concave geometry requires a more refined approach to take into account multiple collisions by a single gas molecule. For example, in~\cite{ahn2018optically} a Monte Carlo method was used to estimate the gas dampings of a nanodumbbell CoM. Regardless of the method used, the damping coefficients calculated are clearly dependent on a number of particle and environmental parameters. However, their ratio is not and can provide information on the geometrical aspect ratio of the particles.  

This approach can be implemented experimentally by measuring the line widths of the translational motion PSD. When comparing two directions, a linewidth ratio of 1 indicates that the two directions have the same cross-section. And, in general, the higher the linewidth ratio, the higher the aspect ratio of the particle. However, two prerequisites are necessary: \emph{i)} the particle needs to be well aligned since a spinning or tumbling dynamics will average out any differences. \textit{ii)} The Duffing broadening needs to be negligible or traced out~\cite{zemanek2020using,pontin2020ultranarrow,Zheng2020room,Amer2023Simulations}.

The approach was first experimentally demonstrated by Hoang \textit{et al.}~\cite{hoang2016torsional} which measured the damping ratio for several nanodiamond exhibiting a spectral peak potentially compatible with librational motion (as opposed to rotation). By measuring an average linewidth ratio $\gamma_\text{x}/\gamma_\text{y}=0.8$ demonstrated, along with other indicators, that the motion observed was indeed librational. Other notable examples are found in~\cite{bang2020five,vanderLaan2021subkelvin} where a linewidth ratio in the range $1.2-1.3$ was used to confirm trapping of a dumbbell, and in~\cite{gao2024feedback} where three different dampings for the CoM motion $(\gamma_{\text{x}}:\gamma_{\text{z}}:\gamma_{\text{y}}=1.7:1.3:1)$ were measured in conjunction with three librational frequencies indicating a trapped asymmetric top particle. Finally, in~\cite{pontin2023simultaneous} a ratio of $1.2$ was used, along with measured trap frequencies, to model the particle shape as a prolate ellipsoid. We point out, however, that the typical accuracy experimentally achievable is not sufficient by itself to distinguish between \textit{similar} geometries. For example, the difference in linewidths ratio between a dumbbell and a prolate ellipsoid is typically too small when compared to experimental uncertainties. Nonetheless, the method allows to identify the rotor type (see Sec.~\ref{sec:Numerical-analysis}) and consequently the expected dynamics.

\begin{figure}[t!]
\centering \includegraphics[width=\textwidth]{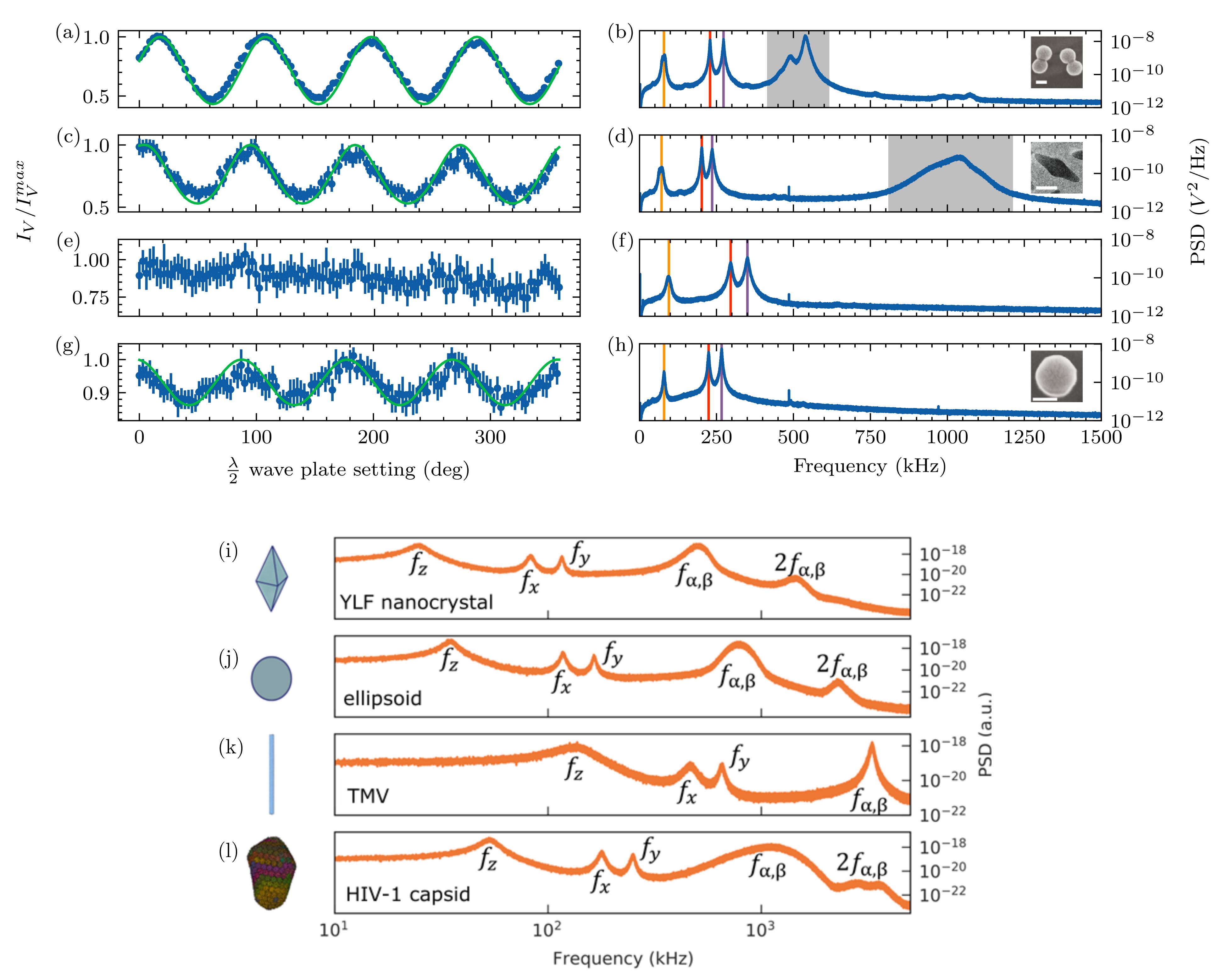}
\caption{\textbf{Effects of particle morphology on light scattering and rotational dynamics of optically levitated nanoparticles.} (\textbf{a--h}) Adapted from Rademacher \textit{et al.},~\cite{rademacher2022measurement}. (\textbf{a, c, e, g}) Normalized vertically polarized light intensity $I_V/I_V^{\text{max}}$ as a function of $\lambda/2$ waveplate angle, showing angularly resolved Rayleigh scattering patterns for different nanoparticle shapes: a silica nanodumbbell (a), an octahedral YLF nanocrystal (c), a liquid methanol nanodroplet (e), and a slightly aspherical silica nanosphere (g). Distinct modulations in scattering intensity arise from shape-induced optical anisotropy and alignment in the optical trap.  (\textbf{b, d, f, h}) Power spectral densities (PSDs) of the particle motion measured via forward scattered light, corresponding to the same particles as in (a, c, e, g), respectively. Insets show SEM images of the trapped particles. The PSDs reveal librational peaks in (b) and (d), absent in the nearly spherical particles of (f) and (h), confirming their symmetry. The shaded grey regions in (b, d) mark the broad librational frequency range coupled via the $\gamma$ rotational motion. (\textbf{i--l}) Adapted from Gosling \textit{et al.},~\cite{gosling2024levitodynamic}.  Simulated levitodynamic spectra of four morphologically distinct particles: (\textbf{i}) a YLF nanocrystal, (\textbf{j}) a silica ellipsoid, (\textbf{k}) a tobacco mosaic virus (TMV), and (\textbf{l}) an HIV-1 capsid. The power spectral densities illustrate distinct peaks associated with translational modes ($f_x$, $f_y$, $f_z$) and librational modes ($f_{\alpha,\beta}$, $2f_{\alpha,\beta}$). These spectra serve as unique optical fingerprints that reflect differences in particle mass distribution, optical susceptibility, and geometry, offering a powerful tool for single-particle identification. Together, these experimental and simulated datasets demonstrate how anisotropic light scattering (a–h) and rotational spectral signatures (i–l) provide complementary, morphology-sensitive characterization methods for individual levitated nanoparticles.}
\label{fig:characterisation}
\end{figure}

When an anisotropic particle is levitated in a linearly polarized beam, the particle will align with the polarization axis. If this polarization axis is rotated so to does the particle. This can be utilized to determine the morphology of the particle. Using a second laser beam as a plane wave on the particle, the Rayleigh scattering pattern of the particle can be mapped by rotating the particle in this beam. The amount of light scattered depends on the asymmetry of the particle. This technique was applied to a variety of particle geometries from perfect spherical nanodroplets to octahedral nanocrystals~\cite{rademacher2022measurement,Schellenberg2023Mass,Li2024Structure} and is shown in Fig.~\ref{fig:characterisation} with the ability to resolve shape differences down to a few nanometers. 

The librational motion of levitated nanoparticles is dependent on it's shape and size \cite{rashid2018precession,hoang2016torsional}. This can be utilized to characterize the nanoparticle in a technique called Levitodynamic spectroscopy \cite{gosling2024levitodynamic}.  The librational frequency, linewidth and ratio of the translational linewidths all depend upon the shape and size of the nanoparticle. By combining these various measures of the nanoparticle, a fuller characterization of the nanoparticle can be achieved. This has been demonstrated with two sets of colloidally grown YLF nanocrystals of different sizes trapped in a single beam of optical tweezers. Here very distinctive behavior is observed between the two sets and even within the sets with size differences as small as a few nanometers could be resolved. In addition, the position of the librational frequencies of a nanoparticle levitated in an optical lattice has been shown to determine the radii of an ellipsoid with a precision of 0.08\% \cite{kamba2023nanoscale}. Determination via the rotational behavior of a levitated objects enables the measurement of the morphology of a single nanoparticle, isolated from backgrounds such as suspending solvents or substrates.

Material properties of levitated particles such as the ultimate tensile strength can be determined from rotational motion. When particles are rotated at high speeds (>GHz rotation frequencies), they experience strong centrifugal forces that rip apart the particle. The frequency at which the nanoparticle is pulled apart gives information about the stress the material can withstand~\cite{ahn2018optically}.

\subsection{Roto-translational cooling}\label{controlANDmanipualtion}
Cooling both translational and rotational degrees of freedom of levitated nanoparticles is a key prerequisite for accessing the quantum regime of motion and for enabling high-precision control of all six mechanical degrees of freedom. Over the past few years, significant experimental progress has been made in implementing active and passive cooling schemes that target not only the center-of-mass motion but also the complex librational and rotational dynamics. These efforts have led to the demonstration of effective feedback control, the development of advanced detection strategies, and the use of coherent light–matter interactions to simultaneously address multiple modes. This section provides an overview of the leading approaches for cooling and driving rotational motion—parametric feedback cooling, cold damping, and coherent scattering—and outlines the major achievements and remaining challenges on the path toward full quantum control of roto-translational dynamics.

\subsubsection{Feedback cooling}
\begin{figure}[t!]
\centering \includegraphics[width=0.9\linewidth]{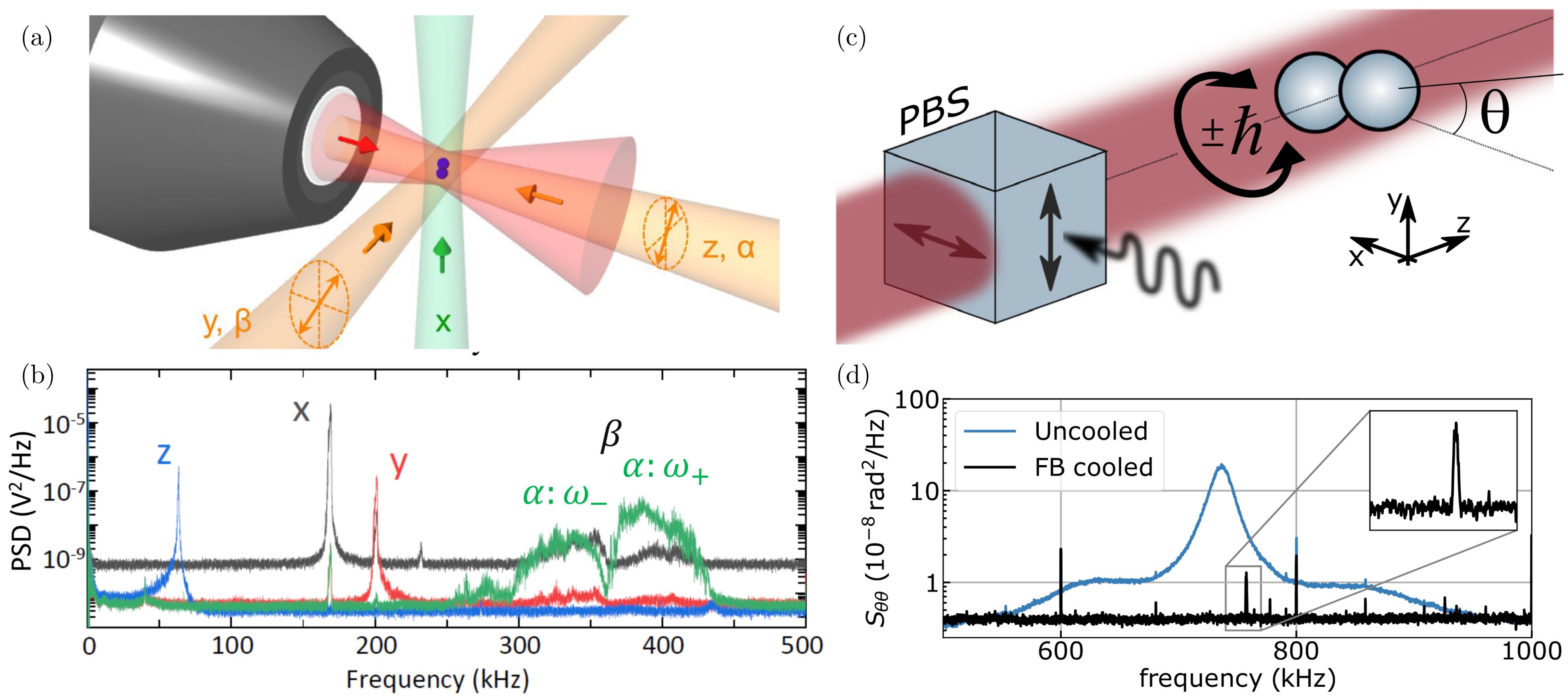}
\caption{
\textbf{Experimental configurations and rotational motion spectra of optically levitated nanodumbbells.} 
(\textbf{a--b}) Adapted from Bang \textit{et al.},~\cite{bang2020five}. 
(\textbf{a}) Schematic of a five-dimensional cooling configuration. A silica nanodumbbell is levitated using a tightly focused 1064 nm linearly polarized laser (red beam), with the particle’s long axis aligning to the polarization axis. Additional cooling beams (green and orange) are used to actively cool the three center-of-mass translational degrees of freedom ($x$, $y$, $z$) and two librational degrees of freedom ($\alpha$, $\beta$). The third rotational mode ($\gamma$), corresponding to free rotation about the dumbbell’s long axis, is not directly coupled to the optical potential. 
(\textbf{b}) Power spectral density (PSD) showing translational peaks at frequencies corresponding to the $x$, $y$, and $z$ motions, and broader torsional resonances labeled $\beta$ and $\alpha:\omega_\pm$, where $\omega_+$ and $\omega_-$ are hybridized librational modes due to coupling with the free spin $\gamma$. All five motional degrees are visible in the spectral response, illustrating successful simultaneous cooling and readout.
(\textbf{c--d}) Adapted from van der Laan \textit{et al.},~\cite{vanderLaan2021subkelvin}. 
(\textbf{c}) Schematic representation of radiation torque shot noise acting on a nanodumbbell. Fluctuations in angular momentum transfer from vacuum fluctuations of the optical field impart a random torque onto the librator. The detection scheme involves a polarizing beam splitter (PBS) and balanced photodetection to read out the libration angle $\theta$. 
(\textbf{d}) Angular PSD $S_{\theta\theta}$ of the nanodumbbell libration measured at two different conditions: without feedback cooling (blue) and with feedback cooling applied (black). Under feedback cooling, the librational motion is confined to a narrow spectral peak at $\sim$750 kHz, with an effective temperature of 240 mK. The residual spectral density reveals the quantum backaction noise floor, enabling experimental access to radiation torque shot noise, a hallmark of rotational quantum optomechanics.}
\label{fig:colddamping} 
\end{figure}

Cold damping, introduced in Sec.~\ref{sec:colddamping}, was first demonstrated on the rotational DoFs in Ref.~\cite{bang2020five}. A silica nanodumbbell was trapped in a linearly polarized optical tweezer. The linear optical torques, necessary for the cooling scheme, were introduced by auxiliary optical beams as shown in Fig.~\ref{fig:colddamping}\,a). The polarization of each beam was tilted with respect to the polarization axis of the trapping laser to exert a torque on the two librational modes (see Eqs.~\ref{eq:ds1}-\ref{eq:ds3}). By modulating the power using acousto-optic modulators a torque proportional to angular velocity could be obtained. The experiment showed cooling of the librational modes, $\alpha$ and $\beta$, to effective temperatures $\simeq10$\,K. It also highlighted the nonlinear coupling between these two DoFs becomes a dominant feature in the case of a cylindrically symmetric nanoparticle, as discussed in  Sec.~\ref{sec:prolate}. The resulting mode hybridization can be seen in Fig.~\ref{fig:colddamping}~b).

More recently, feedback cooling of all three angular DoFs has been demonstrated in Ref.~\cite{kamba2023nanoscale}. In this case, the particle was confined in an optical standing wave trap where the CoM motion was cooled very effectively near ground state with a $n_z=0.69\pm0.18$ (standing wave axis) and $n_x=n_y=6\pm1$ (transverse directions). The strong CoM cooling revealed the presence of spectral lines which were identified as belonging to the librational motion (and their non-linear mixing) indicating a confinement of the particle orientation. The three librational trap frequencies occurred in a range between $10$ and $70$\,kHz. Such low frequencies imply a nearly spherical particle for which the optical trap can only provide a shallow and non-linear potential. This also translates to a low detection efficiency. Nonetheless, the authors showed feedback cooling of the three librations down to temperatures of the order of few tens of mK. To achieve this, the feedback torque was obtained by applying electric fields and exploiting the residual dipole moment of the particle after neutralization. As can be expected, the final effective temperatures were limited by the detection noise floor.  

In a later development of the same experiment, the possibility to implement a 6D feedback cooling scheme, allowed the authors to explore the interplay between CoM motion and librations. By performing a time-of-flight experiment, they were able to reconstruct the velocity distribution of the particle being released from the trap in which the motion was precooled to equivalent occupation numbers as in Ref.~\cite{kamba2023nanoscale}.  The measurement was performed twice with and without cooling of the librational modes and revealed a significantly wider velocity distribution in absence of the librational cooling. This result highlighted the relevance of the rotational DoFs even for experiments that focus on the CoM motion.

Feedback cooling of the librational motion has also been achieved via the applications of non-optical forces particularly in systems which do not employ an optical tweezer. Non-spherical particles levitated in magnetic traps undergo librational motion. In Ref.~\cite{Timberlake2024Linear}, a piezoelectric actuator was used to modulate the position of the trap to obtain a linear feedback force. The superconducting magnetic trap was housed in a ultracryogenic environment at a temperature of $410$\,mK, thus allowing the particle motion detection with a SQUID. In this experiment cooling of a single librational mode was demonstrated down to an effective temperature of $830\pm200$ mK. This value, higher than the environment thermodynamic temperature, reflects the typical difficulties of thermalizing a low frequency oscillator levitated object in a noisy cryogenic environment.

An interesting variation on the application of feedback cooling has been shown in Ref.~\cite{blakemore2022librational}. Here, a large microsphere levitated by a gravito-optical trap was set to spin using a rotating electric field driving the particle permanent electric dipole moment. The implemented feedback scheme then cooled the librational motion around the rotating frame to temperatures of the order of $1$\,K.

\subsubsection{Parametric feedback cooling}

Similarly to translational motion, measurement-based parametric feedback cooling of an optically levitated nanodumbbell has also been demonstrated in Ref.~\cite{vanderLaan2021subkelvin}. A pictorial representation of the experiment is shown in Fig.~\ref{fig:colddamping}~c), the particle was trapped in a linearly polarized light indicating a silent $\gamma$ motion and a coupled and degenerate libration for $\alpha$ and $\beta$ (see Sec.~\ref{sec:prolate}) at a frequency of $750$~kHz. The parametric feedback scheme was implement by modulating the trap intensity and achieved a final effective temperature of 240 mK limited by imprecision noise of the forward detection scheme and by a heating rate potentially compatible with torque shot noise. 

This technique was advanced further by implementing an efficient backward-scattering detection scheme that significantly improved the signal-to-noise ratio in monitoring librational modes~\cite{gao2024feedback}. Here, the particle was an asymmetric top, likely a nanocluster, with three non-degenerate librational frequencies (see Sec.~\ref{sec:top}) which allowed a parametric feedback approach (PLL variant) to be applied to all three librational degrees of freedom. The final effective temperatures measured were $(1.34\pm0.14,15\pm2,4.1\pm0.5)$\,mK for $\alpha,\beta$ and $\gamma$ respectively with the lowest corresponding to an occupation number of $n=84\pm9$ phonons thus close to the regime where sideband asymmetry can provide a calibration free estimate of $n$. The largest contribution to this improved final temperature is the significantly higher detection efficiency $\eta$ obtained by back scattered detection. For the coldest librational mode $\alpha$ this was estimated to be $\eta=0.5\%$ with potential improvements  still achievable by optimizing the local oscillator mode overlap. However, the authors argue that with all things being equal, a cold damping approach would have outperformed the parametric scheme potentially allowing a final occupation of $n\simeq7$ to be reached, which would be significantly closer to the ground state.  

\subsubsection{Coherent scattering}\label{sec:CS_chapter4}

\begin{figure}[t!]
\centering \includegraphics[width=0.9\linewidth]{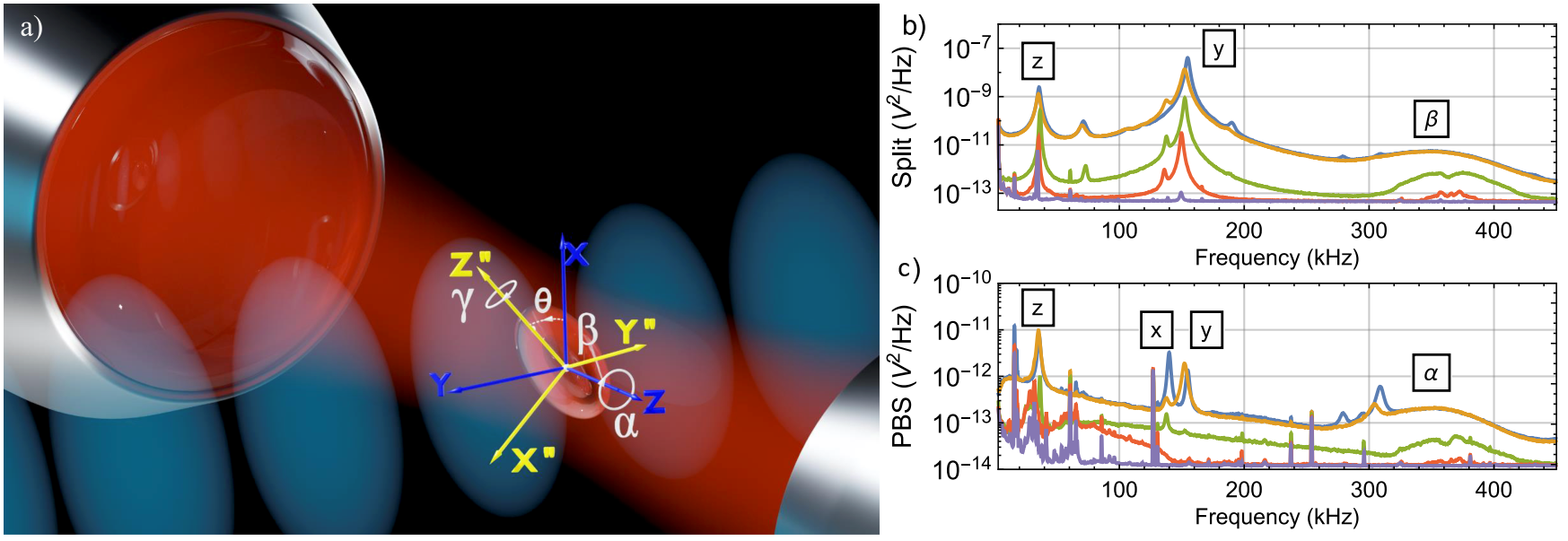}
\caption{
\textbf{Coherent scattering cooling of an ellipsoidal nanoparticle} 
Adapted from Pontin \textit{et al.},~\cite{pontin2023simultaneous}.  
(\textbf{a}) Pictorial view of the experiment and definition of the reference frames. The transformation of coordinates from body to laboratory frame is defined through three Euler angles in the $z-y'-z''$ convention. The tweezer field propagates along the $z$ direction while the cavity axis lies along $y$. (\textbf{b--c}) Motional PSDs of the particle motion as a function of pressure which ranges from $2.5$\,mbar down to $1\times10^{-6}$\,mbar. Both measurements are obtained with a forward scattering homodyne approach (see Sec.~\ref{sec:detection}). The \textit{Split} detection in (\textbf{b}) exploits D-mirrors and provides information on the CoM motion but is also sensitive to $\beta$ while the \textit{PBS} detection in (\textbf{c}) is sensitive to polarization changes and it mostly provides information on the motion of $\alpha$. } 
\label{fig:6D} 
\end{figure}

Another significant advance was presented in Ref.~\cite{pontin2023simultaneous} where a coherent scattering approach was exploited to simultaneously cool all 6 DoFs of a nonspherical nanoparticle. As in a standard CS scheme, the particle was levitated in an optical tweezer and placed at the center of a high finesse optical cavity. A pictorial view of the experiment is shown in Fig.~\ref{fig:6D}~a). In this case, however, the tweezer polarization state was set to be elliptical to a degree that allowed to lift the librational frequencies degeneracy without causing it to spin. The particle shape was modeled as a prolate ellipsoid with good cylindrical symmetry (see Sec.~\ref{sec:prolate}) exploiting the center of mass linewidths ratios (Sec.~\ref{sec:part_characterization}) and the 5 identified trap frequencies which were $(35.7,141.4,152.5)$\,kHz for the CoM, $(z,x,y)$ respectively and $(357,377)$\,kHz for the hybridized $\alpha$ and $\beta$ modes. As introduced in Sec.~\ref{sec:coherent}, the cooling rates cannot be optimal for all DoFs and a compromise needs to be struck. To this end, the particle was placed away from a cavity node at $\phi=0.97$ and with the main axis of the tweezer polarization at an angle $\theta=45\degree $ and the cavity detuning of $\Delta=-362$\,kHz, chosen to enhance cooling of the librations which were in the resolved sideband limit. 

Experimental spectra are shown for different pressures in Fig.~\ref{fig:6D}~b),~c), at the lowest, the effective temperatures were estimated to be $(T_x,T_y,T_z,T_\alpha,T_\beta)=(126\pm17\,\mu$K$, 212\pm16\,\mu$K$, 6\pm2$\,mK$, 6.3\pm1.4$\,mK$, 4.4\pm0.5$\,mK). Interestingly, data presented showed two distinctive features. First, at intermediate pressures the non-linear coupling of the rotational DoFs where highlighted in a striking way, showing hybridization of the $\alpha$ and $\beta$ motion with time dependent trap frequencies occurring through the coupling with the freely diffusing $\gamma$ motion~\cite{seberson2019parametric}. Second, at the lowest pressure a dynamical transition occurred. The weak coupling to the cavity of $\gamma$ was finally sufficient to extract energy from the motion allowing $\gamma$ to be stably trapped in a shallow optical potential (depth $\sim20$\,K). As a consequence, the trap frequencies of $\alpha$ and $\beta$ stabilized as well, bringing the system closer to the deep trapping regime, where the librations behave as independent oscillators. This transition marked a significant shift in the dynamical regime, allowing the conclusion that all 6 DoFs were simultaneously cooled by the CS interaction with a final temperature for gamma estimated to be $\approx1$\,K, though not directly measured.

Nonetheless, the results presented in Ref.~\cite{pontin2023simultaneous} marked the first demonstration of 6D cooling, followed shortly after by similar results achieved through active feedback~\cite{kamba2023nanoscale}. It also highlighted the difficulties in achieving efficient cooling on all DoFs. Indeed, recent results exploited a CS interaction with an asymmetric top particle. In this case, however, the interaction was optimized to cool a single librational DoF and demonstrated ground state cooling~\cite{dania2024highpurity}. This will be discussed more in depth in Sec.~\ref{sec:ground}.

\subsection{Sensing and Measurement \label{sec:sensing}}
The ability to accurately measure forces, torques, and particle characteristics at the nanoscale is crucial for advancing fundamental research and developing new technologies in levitated optomechanics. By exploiting the precise control and isolation of levitated particles, a range of sensing techniques can be employed to measure environmental parameters, and to probe fundamental physical interactions. These approaches leverage both translational and rotational dynamics, offering unparalleled sensitivity in detecting minute forces and torques. This section explores various methods of sensing, from gravitational sensing to high-precision torque and pressure sensing.

\subsubsection{Force and Torque sensing\label{sec:Force}}

Levitated particles provide excellent sensitivity for the detection of accelerations, forces and torques. There is now quite an extensive body of literature that focuses on exploiting the center-of-mass motion for sensing applications. We point interested readers to recent reviews~\cite{gonzalez2021levitodynamics,moore2021searching,jin2024towards} and provide here only a brief overview. We will then focus on torque sensing.

Better acceleration sensitivity is usually achieved with larger particles, making gravito-optical (Sec.~\ref{sec:optical}), magnetic (Sec.~\ref{sec:magnetictraps}) and electrodynamic (Sec.~\ref{sec:electrodynamictraps}) traps best suited for this task~\cite{Timberlake2019Acceleration,vinante2020ultralow,lewandowski2021high,hofer2022high,Vinante2022Levitated,latorre2023superconducting,Post2010Inertial}.  An acceleration sensitivity of $95\pm41$\,ng$/\sqrt{Hz}$ has been experimentally demonstrated~\cite{Monteiro2020Force}  with theoretical predictions indicating possible improvements of several orders of magnitude, down to $3\times10^{-15}$g$/\sqrt{\text{Hz}}$ for a permanent magnet levitated in a $300$\,mK cryogenic environment~\cite{Timberlake2019Acceleration}.

On the other hand, for force sensing applications it is usually advantageous to operate at the higher trap frequencies that can be achieved with optical tweezers. The force sensitivity limited by thermal noise is given by $S_{ff}^{1/2}=\sqrt{4k_B TM\gamma_c}$ for a $1$\,s measurement time. If the external force to be measured is stochastic or a coherent drive the sensitivity then improves with averaging time~\cite{Gosling2024Sensing}. A force detection resolution of $166\pm55 \times 10^{-24}$\,N has been recently demonstrated~\cite{Liang2023Yoctonewton} for a measurement time of $\simeq2700$\,s exploiting a silica nanoparticle levitated in an optical tweezer at a pressure of $8\times 10^{-9}$\,mbar.
Such sensitivities can be exploited for a number of applications ranging from environment characterization~\cite{Millen2014Nanoscale,hebestreit2018measuring}, non-interferometric tests of collapse models~\cite{carlesso2022present} and physics beyond the standard model~\cite{moore2021searching}. A complementary regime is that of impulsive forces for which a good momentum resolution is necessary at short timescales. In this context, levitated nanoparticle can potentially resolve individual collisions with background gas, with potential application to pressure sensing (see Sec.~\ref{sec:pressuresensing}), and has recently allowed the detection of an individual nuclear $\alpha$-decay from a radioactive element embedded on a levitated particle~\cite{wang2024Mechanical}.

\begin{figure}[t!]
\centering \includegraphics[width=\linewidth]{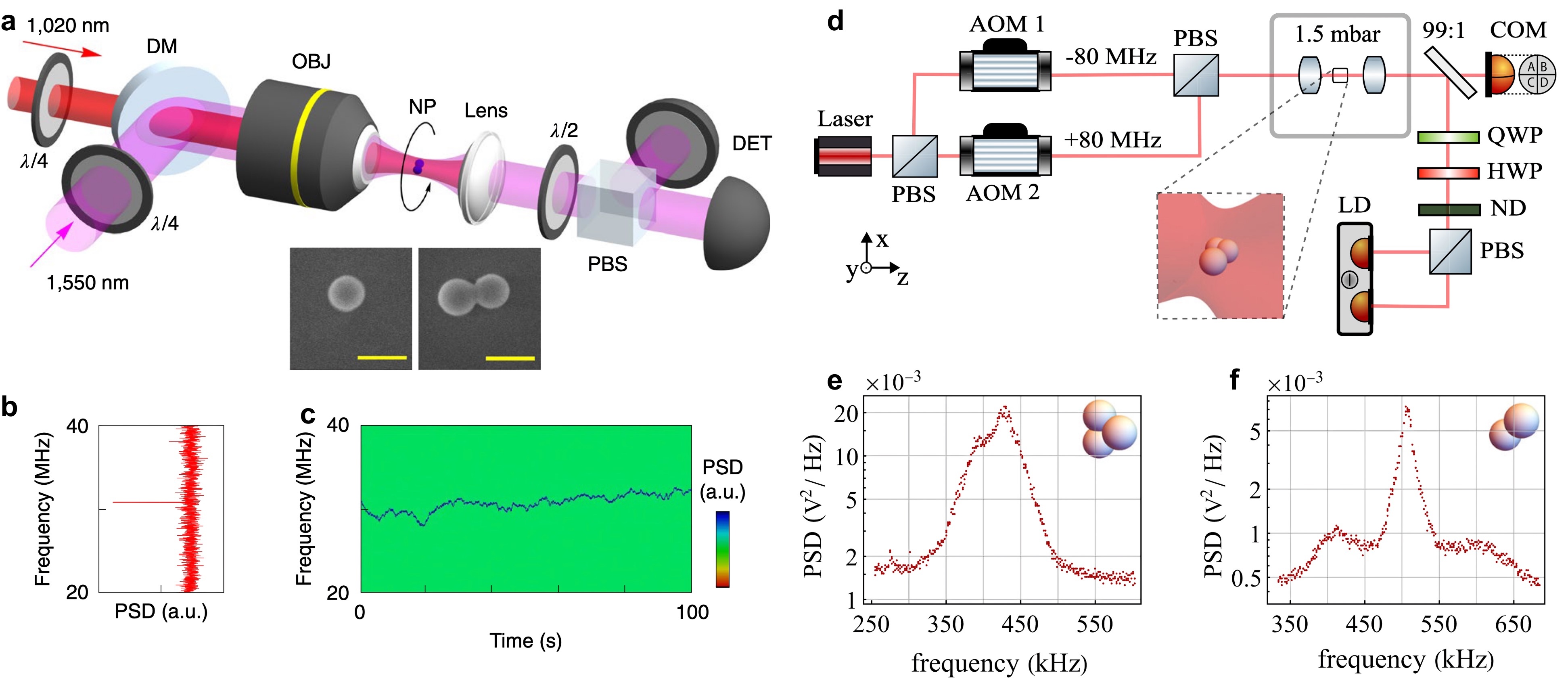}
\caption{
\textbf{Force and torque sensing examples using optically levitated nanoparticles.} 
(\textbf{a--c}) Adapted from Fig.~1 of Ahn \textit{et al.}~\cite{ahn2020ultrasensitive}. 
(\textbf{a}) Experimental schematic of the optical torque detection system: a silica nanoparticle (NP) is optically levitated in vacuum using a tightly focused 1{,}550~nm laser. An auxiliary 1{,}020~nm laser applies an external torque. The polarization of both beams is adjusted with waveplates ($\lambda$/4 and $\lambda$/2), and motion-induced polarization changes are detected with a balanced photodetector (DET) after a polarizing beam splitter (PBS). Inset: SEM images of a silica nanosphere and a dumbbell-shaped nanoparticle (scale bar: 200~nm). 
(\textbf{b}) Power spectral density (PSD) of the rotational motion of a nanodumbbell shows a peak at twice the rotation frequency due to optical birefringence. 
(\textbf{c}) Spectrogram of the rotational PSD over 100~s, highlighting frequency stability and drift under vacuum. 
(\textbf{d--f}) Adapted from Fig.~2 of Zielińska \textit{et al.}~\cite{zielinska2023controlling}. 
(\textbf{d}) Schematic of the optical trapping setup with tunable degree of polarization (DOP): orthogonally polarized components of a 1{,}550~nm laser are frequency-shifted by $\pm80$~MHz using acousto-optic modulators (AOMs), recombined on a PBS, and focused into a vacuum chamber with a high-NA lens. The scattered light is analyzed via homodyne detection to extract libration signals. 
(\textbf{e}) PSD of a “cluster” particle (fully anisotropic) in a linearly polarized trap shows two distinct libration peaks associated with orthogonal angular degrees of freedom. 
(\textbf{f}) PSD of a dumbbell-shaped particle reveals a sharp central libration peak with thermally broadened sidebands, indicative of spinning-induced coupling between libration modes. 
Together, these panels illustrate complementary approaches to torque sensing and rotational dynamics control: GHz rotation detection and sub-attonewton-meter sensitivity~\cite{ahn2020ultrasensitive} versus polarization-controlled libration tuning and Brownian rotation spectroscopy~\cite{zielinska2023controlling}.
}
\label{fig:torque} 
\end{figure}

Torque sensing with anisotropic particles, offers the possibility of going beyond the dynamics of an harmonic oscillator. Indeed, a particularly interesting regime is that of a spinning nanoparticle, otherwise called a nanorotor. As we have shown in Sec.~\ref{sec:det}, in an optical tweezer a particle with sufficient asymmetry, either in the geometry or in the susceptibility, will experience a constant torque which can cause it to spin in the tweezer polarization plane. Examples of experimental demonstrations are~\cite{ahn2018optically} for a nanodumbbell and~\cite{rashid2018precession} for an asymmetric top. An example of the experimental approach is shown in Fig. ~\ref{fig:torque}~a), adapted from~\cite{ahn2020ultrasensitive}.  In the spinning regime, a clear picture of the dynamics can be gathered with a simplified model. The evolution of the rotation frequency $\Omega_{\text{rot}}$ can be expressed as
\begin{equation}
    I \Dot{\Omega}_{\text{rot}}(t)+\gamma_c I  \Omega_{\text{rot}}= \tau^{(ds)}+\tau^{(sc)}+\tau^{(ss)}+\tau_{\text{ext}},\label{eq:torquesensing}
\end{equation}
  
\noindent where we have introduced the torque $\tau^{(i)}=d\pi^{(i)}/dt$. As shown in Sec.~\ref{sec:theory}, $\tau^{(ds)}$ is the deterministic optical torque (see Eqs.~\eqref{eq:ds1}-\eqref{eq:ds3}), $\tau^{(sc)}$ is the stochastic torque due to thermal noise (see Eq.~\eqref{eq:Pinc4}), $\tau^{(ss)}$ is the stochastic torque due to photon scattering (see Eqs.~\eqref{eq:sigmarot}-\eqref{eq:dpnn}), while $\tau_{\text{ext}}$ represents any additional external torques. Eq.~\eqref{eq:torquesensing} allows us to highlight two important factors. First, in absence of external torques, the rotation frequency will spin up until equilibrium is reached between the deterministic torque $\tau^{(ds)}$  and the friction torque which implies $\Omega_\text{rot}=\tau^{(ds)}/(\gamma_c I)$; once the steady state has been reached, fluctuations of $\Omega_{\text{rot}}$ will have a standard deviation $\sigma_{\Omega_{\text{rot}}}=\sqrt{k_B T/I}$ following a fluctuation-dissipation process associated with the thermal torque~\cite{vanderLaan2020Optically}. This dynamics is shown in Fig.~\ref{fig:torque}~b)-c). Second, the effect of external torques is that of changing the rotation frequency. Thus, there is a clear advantage in terms of torque sensitivity compared to the librational scenario, with the sensitivity limited by the frequency stability of $\Omega_{\text{rot}}$ which is determined by thermal torque $\tau^
{(sc)}$  and eventually photon scattering $\tau^
{(ss)}$. As an example, the projected sensitivity for the two scenarios with the same parameter was reported to be (Ref.~\cite{rashid2018precession}) $1\times10^{-29}$\,Nm$/\sqrt{\text{Hz}}$ for the librational case and $4\times10^{-31}$\,Nm$/\sqrt{\text{Hz}}$ for the same particle set to spin. For both cases equal contributions from thermal and recoil torque noise were considered.

The following works have experimentally demonstrated this approach, by spinning silica nanodumbbells that are almost spherical particles with sufficient anisotropy to allow manipulation and detection of the rotational degrees of freedom. It has been shown that these particles can be spun up to frequencies $\simeq1 - 6$\,GHz~\cite{ahn2018optically,reimann2018ghz,jin20216}. The best sensitivity demonstrated, for an almost spherical particle, was $\simeq4\times10^{-27}$\,Nm$/\sqrt{\text{Hz}}$ at a pressure of $1.7\times10^{-5}$\,mbar thus still relatively far from the recoil limit~\cite{ahn2020ultrasensitive}.

Approaching the recoil limit may allow the measurement of Casimir torques~\cite{Parsegian1972Dielectric, Somers2018Measurement} and the elusive vacuum friction~\cite{Pendry1997Shearing,Xu2020Enhancement}. The typical scenario where the Casimir force emerges considers two neutral plates at small distance from each other. Vacuum fluctuations will then give rise to an attractive force between them~\cite{Casimir1948on}.  However, if the plates are optically anisotropic, i.e., birefringent, a torque will be present as well~\cite{Munday2005Torque}. Thus, an anisotropic particle librating near a birefringent surface will experience a static Casimir torque. On the other hand, a fast-rotating neutral particle will experience a frictional torque due to the conversion of quantum and thermal vacuum fluctuations into radiation emission. This vacuum friction is expected to be highly enhanced if the particle is rotating near a surface instead of in free space,  since a surface will add a large local density of electromagnetic states~\cite{Joulain2003Definition,ahn2020ultrasensitive}. For both experiments, the particle needs to be levitated near a surface, which also highlights the necessity of managing other effects. Among these are external stray fields, patch potentials and permanent dipoles. Initial results in this direction are promising, with a demonstrated torque sensitivity of $(5\pm 1)\times 10^{-26}$\,NmHz$^{-1/2}$ at a pressure of $8.1\times 10^{-5}$\,mbar for a nanodumbbell trapped $430$\,nm away from a sapphire surface~\cite{ju2023near}. While this sensitivity is not yet sufficient to measure the standard Casimir torque, the author show that the torque can be significantly enhanced by a gold nanograting up to an estimated value of $\sim 10^{-24}$\,Nm for optimized geometric parameters. The additional challenge arise from the static nature of the latter torque which implies a more complex experimental approach based on a release and recapture protocol~\cite{Xu2017Detecting}. Using different materials such as Yttrium Iron Garnet can also offers great enhancements of vacuum friction~\cite{Khosravi2024Giant}.

Different rotational dynamics can emerge if one considers alternative configurations for the optical trap. As a first example, by controlling the degree of polarization of the trapping field it is possible to obtain a free rotor state where two rotational DoFs are harmonically trapped on the polarization plane but the last can undergo a free evolution. In other words, the orientation of the particle in the polarization plane evolves only under the action of external torques without any confining potentials akin to an optically suspended gyroscope. This has been recently demonstrated by creating an optical tweezer overlapping two fields linearly polarized along orthogonal directions~\cite{zielinska2023controlling}. A simplified optical layout is shown in Fig.~\ref{fig:torque}~d). In this case, the particle longest axis will align itself along the polarization axis of the beam of highest power. However, for equal power, the trap is effectively unpolarized, in the sense that there is no longer a preferred axis but only a preferred plane, so that the particle orientation undergoes free diffusion driven by thermal noise. However, the experimental results shown in~\cite{zielinska2023controlling} have highlighted how sensitive is the dynamics to small imperfections of the trapping potential which can introduce a residual torque. Interestingly, this leads to very different behaviors for symmetric and asymmetric-top particles. An example of typical experiment PSD for these two geometries is show in Fig.~\ref{fig:torque}~e)-f). 

Another notable scenario is the use of structured light for the optical trap. In this case, torques applied to the particle by the trapping field can differ significantly from the typical scenario. In this context, Hu \textit{et al.}~\cite{hu2023structured} have used a levitated silicon nanorod to probe the transverse orbital angular momentum that appears on an interfering counter-propagating trap when an offset is introduced between the two axis of the beams. This offset generates optical vortices with their wavefront spiraling around a phase singularity. The associated torque can cause the nanorod to spin outside the polarization plane, a stark contrast with the behavior in a standard tweezer. Interestingly, if the counter-propagating beams are not interfering, the particle trajectory can evolve along complex orbits~\cite{Raj2022Orbital,Chen2016Characteristics}. Trapping of spherical particles with a Laguerre-Gaussian beam with a topological charge $\leq 10$ has been demonstrated~\cite{Mazilu2016Orbital} but has yet to be extended to different particle geometries.

\subsubsection{Pressure sensing}~\label{sec:pressuresensing}
Collisions that affect both the translational and rotational motion are readily observed at modest pressures in the underdamped regime through a change in the line shape of the power spectral density. However, at lower pressures, the effects on lineshape are minimal, or would take long measurement runs to attain accurate lineshapes \cite{pontin2020ultranarrow}. Pressure-induced changes of the levitated particles ring-down time of the center-of-mass motion after impulsive excitation offer another means to measure pressure \cite{Dania2024Ultrahigh}. Using this method, pressures down to the $10^{-11}$ mbar range were measured.  The drag induced by collisions on induced rotational motion also offers another route to pressure measurement.  As the drag force increases with rotational speed, measurements even at high vacuum appear feasible with this technique~\cite{jin2024towards}, where rotational speeds in the GHz range can be reached \cite{ahn2018optically}.  Initial experiments demonstrated the capability for this type of sensing using levitated silicon nanorods. Here, rotation was accomplished via transfer of spin angular momentum from the trapping beam, where the frequency of the rotor was locked to an external clock using feedback control of the levitated lasers polarization. In this work, it was shown that the phase between the drive that controls the rotor's frequency and the silicon nanorods was sensitive to pressure via the drag. The near real-time phase measurements of the relative phase indicated pressure measurement were achievable with a precision of 0.3\%~\cite{kuhn2017optically}. Importantly, phase lag measured over short time circumvents the need for the longer measurement times of a PSD linewidth. Blakemore \textit{et al.} developed a spinning-rotor vacuum gauge that measured torsional drag based on optically levitated and spinning microsphere~\cite{blakemore2020absolute}. Here pressures in the $10^{-3}-10^{-6}$ mbar could be determined over a 10\,s timescales with 7\% precision. The lower range could be significantly increased by using higher rotation rates that utilized in the 10 kHz range.  More recently \cite{Barker2024collision}, has proposed that the detection of the rate of single collision events between gas particles and a levitated sphere, could be used for pressure measurement even at very high vacuum. While this work focused on measurement of momentum kicks to the particle, rotational kicks via the same collisions would also be induced and represent a possible future direction for this type of sensing.

\subsubsection{Inertial and gravitational sensing}
\label{sec:inertial_sensing}
The force and acceleration sensitivity of levitated systems makes them potentially attractive as compact inertial sensors. For example, there is currently a need for accurate and compact accelerometers and gyroscopes in navigation, particularly in situations where global satellite navigation systems (GNSS) are not available. Sensitive measurements of acceleration can be used for monitoring seismic noise and for measuring variations in gravitational fields. When compared to typical accelerometers constructed from clamped mechanical systems, which have multiple mechanical modes, levitated oscillators have only three simplifying analysis. In addition clamped systems are subject to non-linearities which can be significantly reduced in levitated systems. However, as acceleration sensitivity scales with mass, the much larger oscillator masses available in magnetic levitation make this the most promising platform for development of accelerometers. This levitation platform has already been used for sensitive measurements of gravitational forces~\cite{fuchs2024measuring}. 

The ability to control rotation suggests that the rotational degrees of freedom inherent in levitation can be used for the construction of gyroscopes. While typically larger moments of inertia enhance performance, the ability to rotate nanoparticles to high angular frequencies can also be used to compensate for this with GHz rotational frequencies demonstrated for nanoparticles~\cite{ahn2018optically}. Controlled rotation of both the short axis and long axis of near elliptical nanoparticles has been demonstrated \cite{zielinska2024longaxis}. Optical centrifuges, which have been used extensively for accelerating molecules, can be utilized for levitated nanoparticles~\cite{xiong2025optical}. The controlled rotations could reach rates in excess of 100 MHz with the rotational frequencies set by the rapidly rotating linearly polarized light rather than the gas damping. This opens the possibility towards quantised rotational motion and a quantum gyroscope.

In addition, a gyroscope constructed from an optically trapped micron sized vaterite microparticle has already been demonstrated with bias stability of the order of 0.08 $^{\circ}
$/s \cite{Zeng2024optically}. While this is not state-of-the-art, additional improvements through cooling and control promise to push this to significantly lower levels  with values of $10^{-9}~^{\circ}$/hr predicted by the authors.  Magnetically levitated ferromagnetic gyroscopes have been proposed to be sensitive to exotic spin-dependent interactions, with the potential to test fundamental physics~\cite{fadeev2021ferromagnetic}.

\subsubsection{Magnetometry and Electrometry}
The undamped nature of levitated motion and its concomitant sensitivity to small forces and torques allows levitated systems to be used for mechanical sensing of magnetic and electric fields~\cite{ni2025microscopic}. This is particularly sensitive for AC fields when oscillating at or near the mechanical frequency of the oscillator. Recently, Ahrens \textit{et al.} demonstrated measurement of magnetic fields below the energy resolution limit $E_R$, where $E_R > \hbar$ \cite{Ahrens2025Levitated}. Here an energy resolution that was two orders of magnitude below this limit was demonstrated using hard ferromagnets levitated in a superconducting magnetic field, corresponding to a sensitivity of 20 fT/$\sqrt{\text{Hz}}$. This sensitivity can be ascribed to the macroscopic number of spins that are locked together in the ferromagnet that rotate together. This was measured using detection of the induced magnetic torque, whose resolution was limited by the thermally driven torque. This paper estimates that a modified levitated system is capable of delivering sensitivity that is 4 orders of magnitude lower than the energy resolution limit.
Additionally, weak AC electric fields can also be measured when charged objects are levitated. In an optical trap, detection of weak electric forces has been demonstrated with sensitivities exceeding 1 $\mu$V cm$^{-1}$ Hz$^{-1/2}$ at 1.4 $\times$ $10^{-7}$ mbar with a linearity over 91 dB. \cite{Zhu2023nanoscale}. The ability to tune the mechanical frequencies in some levitated systems, coupled with the fact that the different modes are sensitive to electric fields in 3D allows the creation of a polarization sensitive tunable antenna for low-frequency radio waves \cite{Fu2025optically}. By utilizing correlations induced by electrical fields acting on the levitated system's motion, the direction of the electrical fields can also be resolved~\cite{Gosling2024Sensing}.

\subsection{Roto-translational Quantum Regime}
\label{sec:roto_quantum}
Accessing the quantum regime in rotational optomechanics is an essential goal of ongoing research. Clamped optomechanical systems have achieved ground state cooling~\cite{chan2011laser} and strong coupling~\cite{teufel2011sideband} in cryogenic environments more than a decade ago. Similar results have been demonstrated in recent years in optically levitated systems at room temperature. 
In particular, the cooling of individual translational and orientational degrees of freedom to the motional ground state. Such states have high purity and provide the starting point for future explorations of quantum phenomena and for sensing in the quantum regime. Thus, we will begin this chapter summarizing, in Sec.~\ref{sec:ground}, the main experiments which demonstrated such a result for both CoM and rotational DoFs. 

Levitated optomechanical experiments have also demonstrated the strong-coupling regime and beyond along with quantum-coherent coupling. In Sec.~\ref{sec:strong}, we will review these different regimes, their relevance, and discuss recent experiments that demonstrate their emergence. 

Looking further ahead, the main challenge remains to move past the achievement of ground state cooling and implement more complex experimental protocols with the underlying objective to move towards the generation of non-Gaussian states. In this context, experimental progress has been made focusing on the CoM motion while experiments with rotations are right at the threshold where this exploration becomes possible. On the theoretical side, schemes have been proposed for both CoM and rotations. In Sec.~\ref{sec:Non-classical}, we will summarize the experimental results in this context and provide an overview of theoretical proposals.

Finally, experiments with hybrid systems consisting of nanoparticles coupled to embedded degrees of freedom or external auxiliary modes provide an alternative route towards the quantum regime. Among these the most studied system are NV centers which will be discussed in Sec.~\ref{sec:NV}.

\subsubsection{Ground state cooling}\label{sec:ground}
Ground state cooling is often seen as the starting point for the exploration of quantum phenomena in optomechanical systems (see below Secs.~\ref{sec:strong}-\ref{sec:Non-classical}). To introduce the key concepts related to ground state cooling we first focus on a generic harmonic oscillator with mechanical frequency $\omega$ \cite{bowen2015quantum}. We assume that the harmonic oscillator is in a thermal state:
\begin{equation} \label{rhosimple}
    \hat{\rho} =\sum_{n=0}^{\infty} P_n \vert n \rangle \langle n\vert,
\end{equation}
where $P_n=(1-\xi) \xi^n$ can be interpreted as the probability of being in the number state $\vert n \rangle$, $\xi=\exp (-\frac{\hbar \omega n}{k_B T})$, and $T$ denotes the temperature. In the context of roto-translational levitated optomechanics such a harmonic oscillator can  be used to model a single motional degree of freedom (for the full theoretical description see Sec.~\ref{sec:theory}, while for the discussion about cooling the internal, i.e., bulk, temperature of the nanoparticle see Sec.~\ref{sec:optical}). We can identify two regimes depending on temperature of the system: $\hbar \omega   \gg k_B T$ (the high temperature regime) and $\hbar \omega  \ll k_B T$ (the low temperature regime). In the high temperature regime we find that the mean occupation number $\bar{n}$ is approximately the energy of the system divided by the energy of a single excitation, $\bar{n}\approx\frac{k_B T}{\hbar \omega}$.  In the low temperature regime, the mean occupation becomes $\bar{n}\approx \exp(-\frac{\hbar \omega}{k_B T}) \ll 1$, with the probability, $P(0)= 1- \exp(-\frac{\hbar \omega}{k_B T})$, of being in the ground state $\vert 0 \rangle$ close to unity. Using this observation, we will define that ground state cooling has been achieved experimentally when the probability of being in the ground state is at least $50\%$, i.e., $P_0\geq \frac{1}{2}$. The average phonon number $\bar{n}$ is however a concept that relies on possibility of isolating a single mechanical degree of freedom from the other ones appearing in the description of the system. An alternative figure of merit, which offers more versatility, is given by the state purity, which for a thermal states reduces to the simple formula $\mu = (2\bar{n} + 1)^{-1}$~\cite{paris2003purity}. State purity can be used to quantify the quantumness of a systems even when the light and matter degrees of freedom can no longer be separated (see the ultrastrong coupling regime in  Sec.~\ref{sec:strong}) or when we are dealing with general higher-dimensional mechanical oscillators~\cite{borkje2023quantum}.

\begin{figure}[t!]
\centering \includegraphics[width=0.9\linewidth]{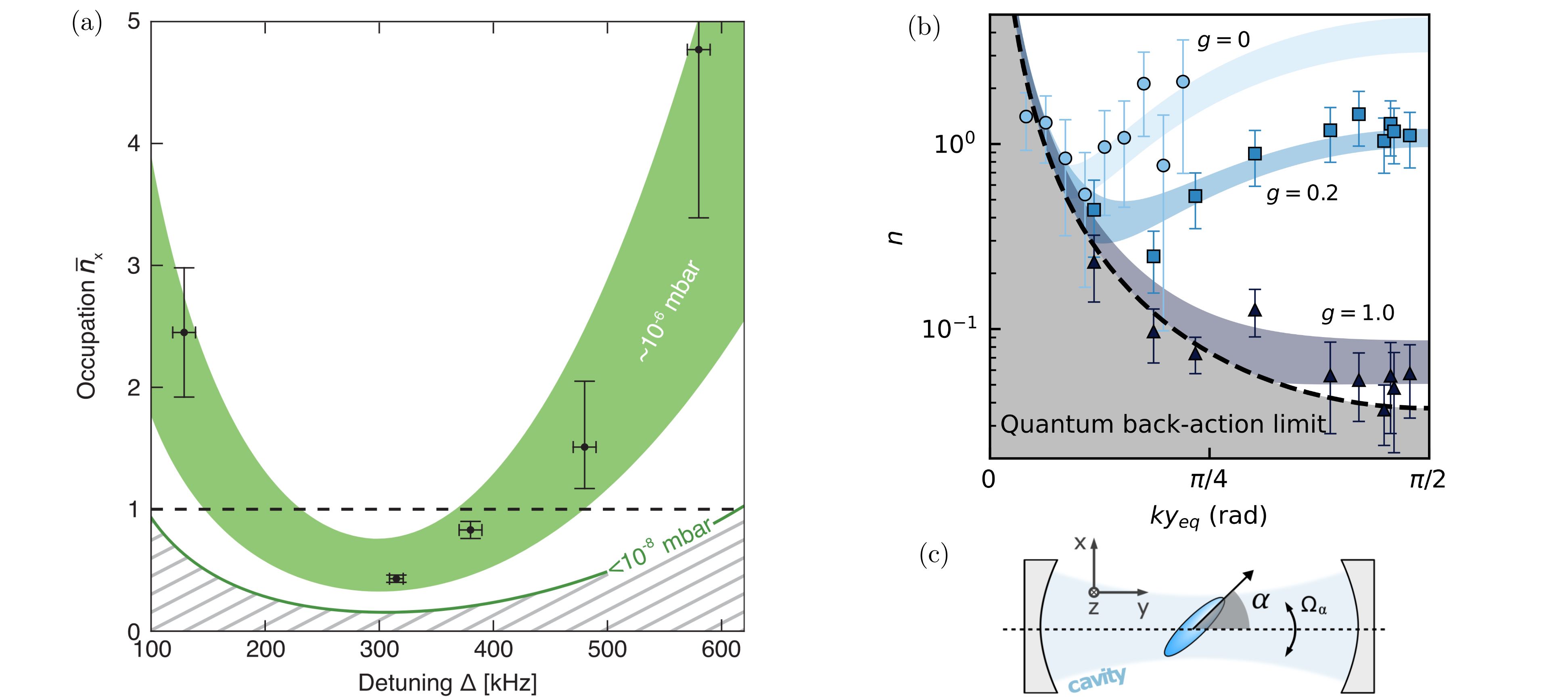}
\caption{
\textbf{Ground-state cooling of translational and rotational degrees of freedom in levitated nanoparticles.}
(\textbf{a}) Adapted from Delić \textit{et al.}~\cite{delic2020cooling}.
Mean phonon occupation $\bar{n}_x$ of the translational mode along the cavity axis as a function of cavity detuning $\Delta$. The lowest occupation $\bar{n}_x = 0.43 \pm 0.03$ is reached near the optimal detuning $\Delta \approx 2\pi \times 315$ kHz. The shaded green region indicates the theoretically expected range under experimental conditions ($\sim10^{-6}$ mbar), and the lower green curve shows the limit at pressures $<10^{-8}$ mbar. An schematic of this scheme is shown earlier in the review (see Fig.~\ref{fig:figureExpSchematic}a).
(\textbf{b,c}) Adapted from Dania \textit{et al.}~\cite{dania2024highpurity}.
(\textbf{b}) Phonon occupation number $n$ of the librational degree of freedom $\alpha$ as a function of equilibrium cavity position $k y_{\mathrm{eq}}$, for three different levels of laser phase noise cancellation gain $g = 0$, $0.2$, and $1.0$. Active suppression of phase noise allows the system to approach the quantum back-action limit (dashed line), reaching $\bar{n}_\alpha = 0.04 \pm 0.01$, corresponding to a 92\% state purity.
(\textbf{c}) Schematic of the experimental geometry for librational ground-state cooling. An anisotropic nanoparticle librates about the tweezer polarization axis with angular frequency $\Omega_\alpha$. The cavity is aligned along the $y$-axis (orthogonal to the tweezer propagation $z$-axis), and the librational coupling is optimized near the cavity anti-node at $\phi = 0$.
}
\label{fig:groundstate} 
\end{figure}

The experimental demonstration of cooling a motional degree of freedom to the quantum ground state was first demonstrated in the coherent scattering setup using a Silica nanoparticle and linear polarization~\cite{delic2020cooling} (see Sec.~\ref{sec:coherent} for a description of the coherent scattering setup). Here we recall that the ideal configuration for cooling the motion along the cavity axis would be near a cavity node ($\phi=\pi/2$) and with the particle scattering in the cavity ($\theta=\pi/2$)~\cite{delic2019cavity,windey2019cavitybased}. In this latter situation the coupling of the mechanical motion along the cavity axis with the cavity mode is maximized, and the motion of the mechanical mode along the cavity axis effectively decouples from the other mechanical modes. The description in this configuration thus reduces to the usual equations of motion of cavity optomechanics with one mechanical and one optical mode, where the optomechanical coupling encodes the interaction in the coherent scattering setup~\cite{gonzalez2019theory}. However, relatively small deviations from the ideal position can lead to effective cross-couplings between the mechanical modes and hybridization, which has to be taken into account for accurate thermometry~\cite{toros2020quantum}. In~\cite{delic2020cooling} the translational motion along the cavity symmetry x-axis was performed at the pressure $10^{-6}$ mbar starting at room temperature with the nominal diameter of the nanoparticle $d=143\pm4$ nm. The optical wavelength was $1064$ nm, and the power in the focus was estimated to be $400$ mW. The cavity had a finesse was $73000$, a length of $1.07$ cm, and a waist of $41.1\;\mu m$. The chosen configuration was close to the optimal with the paper reporting cooling of both the $x$ and $y$ mechanical modes. Thermometry has been performed exploiting small movements away from the cavity node by measuring the Stokes and anti-Stokes sidebands~\cite{leibfried2003quantum,Marquardt2007Quantum,wilson2007theory,genes2008ground}. The reported mean occupation number was $\bar{n}_x=0.43 \pm 0.03$ with the mechanical frequency $\omega_x=2\pi \times 305$ kHz. The measured $\bar{n}_x$ is shown in Fig.~\ref{fig:groundstate} as a function of cavity detuning. Importantly, the experiment~\cite{delic2020cooling} was a major milestone as it pushed levitated optomechanics for the first time into the quantum regime.

Cooling of single translational degrees of freedom has been also achieved in tweezer setups with active feedback schemes. In~\cite{magrini2021realtime,tebbenjohanns2021quantum} the motion of a charged nanoparticle was measured optically and then controlled using a pair of electrodes (see Sec.~\ref{sec:colddamping} for an introduction to this method, which is a variant of the cold damping schemes). The basic idea is to apply a force in the opposite direction of the particle's velocity exploiting the Coulomb interaction~\cite{millen2015cavity}.  The particle's motion was estimated in real-time using the scattered optical field fed into an in-loop homodyne detection~\cite{mancini1998optomechanical}.  The electrodes cooled the translational mode of the nanoparticle along the tweezer propagation z-axis. Specifically, the voltage was applied to the holder of the collection lens located in front of the grounded tweezer objective~\cite{frimmer2017controlling}. Thermometry, certifying ground-state cooling, was performed by measuring the ratio of the Stokes and anti-Stokes scattering rates with a separate out-of-loop heterodyne detection~\cite{tebbenjohanns2020motional,clerk2010introduction,safavi2012observation}. Each of the two experiments~\cite{magrini2021realtime,tebbenjohanns2021quantum} also had distinguishing features, which are important in their own right. 
In \cite{magrini2021realtime} they have used real-time Kalman filtering~\cite{kalman1960new} for the optimal estimation of the motional state of the nanoparticle of diameter $d=143$ nm~\cite{setter2018real,liao2018fpga}, followed by the application of the optimal control~\cite{kalman1960contributions} of the voltage on the two electrodes~\cite{tebbenjohanns2020motional,kamba2021recoil}. The trapping tweezer had linear polarization, numerical aperture $0.95$, optical wavelength $1064$ nm, and the power at the tweezer focus was estimated as $300$ mW. Starting from room temperature and pressure $9.2\times 10^{-9}$ mbar, they have reported the final mean occupation number $\bar{n}_z=0.56 \pm 0.02$ with the mechanical frequency $\omega_z=2\pi \times 104$ kHz. 
In~\cite{tebbenjohanns2021quantum} they have performed the experiment in a cryogenic environment at the temperature $60$ K with the diameter of the particle estimated to be $ 106 \pm 5 $ nm.  Using a digital filter that electronically processes the optical signal in real-time the nanoparticle's velocity was estimated and then used to control the voltage on the electrodes. The tweezer with linear polarization was focused an using aspheric trapping lens with numerical aperture $0.75$, the optical wavelength was $1550$ nm, and the estimated laser power was $1200$ mW. The reported final occupation number was $\bar{n}_z=0.65 \pm 0.04$ at the mechanical frequency  $\omega_z=2\pi \times 77.6$ kHz and at the pressure $3\times10^{-9}$ mbar. 

Ground-state cooling via active feedback was also realized in an optical lattice formed by retro-reflecting the laser~\cite{kamba2022optical}. In such a setup the frequency of the $z$ mechanical mode, corresponding to the motion along the symmetry axis of the optical lattice, becomes larger than the frequencies of the modes associated with the two transverse directions. The $z$ mode has thus the highest frequency, as well as the highest coupling to the optical field, which  makes it apt for ground-state cooling. By introducing sidebands to the main frequency of the trapping laser one generates displaced optical lattices, each exerting a gradient force along the optical lattice, but in opposite directions. By modulating the relative amplitude of the two sidebands it is thus possible to generate a fully controllable force on the nanoparticle.  In~\cite{kamba2022optical} they reported the implementation of optical cold damping using such sideband modulation, 
resulting in the final occupation number $n_z=0.69$ (see also related experiments, where the optical cold damping technique was used~\cite{kamba2023Revealing, vijayan2023scalable}). Such scheme has been also used to cool simultaneously the other motional degrees of freedom of a neutral anisotropic nanoparticle~\cite{kamba2023nanoscale}.  The two transverse translational degrees of freedom were cooled using parametric feedback cooling by modulating the intensity of the trapping laser. The librations were cooled by changing the voltage on nearby electrodes thus controlling the orientation of the electric dipole moment, and hence of the nanoparticle~\cite{Afek2021Control}. 

The ground-state cooling of a rotational degree of an anisotropic silica nanoparticle was reported in the coherent scattering setup~\cite{dania2024highpurity}. The rotational degree $\alpha$ was optically trapped in a harmonic potential, resulting in librational motion (see Fig.~\ref{fig:rotations} for an illustration). As libration can be mathematically described using the same mathematical formulae as in the case of harmonic motion, the general considerations regarding ground-state cooling in the paragraph enveloping Eq.~\eqref{rhosimple} are thus applicable also in this case. Specifically, the rotational degree $\alpha$ was librating about the polarization direction in the tweezer focal plane (i.e., the y-z plane, where the tweezer propagation direction and its linear polarization were used to define the z-axis and y-axis, respectively). The polarization axis was aligned with the cavity symmetry axis such that most of the radiation was scattered out of the cavity, i.e., $\theta=0$, in contrast to the setting, $\theta=\pi/2$ chosen for the cooling of translational motion when the tweezer field mostly scatters into the cavity~\cite{delic2020cooling}. The reference frame definition and the experimental configuration is shown in Fig.~\ref{fig:groundstate}~c). Importantly, unlike for the ground-state cooling of a single translational motion, when optimal cooling is achieved near the cavity-node, in~\cite{dania2024highpurity} the nanoparticle libration was optimally cooled near the cavity anti-node ($k y_{eq}=\pi/2$). The reason is that the coupling of $\alpha$ with the cavity mode $a$ gets maximized in this configuration~\cite{rudolph2021theory,schafer2021cooling,pontin2023simultaneous}. While the radiation torque shot noise is largely unaffected by the position in the cavity (its main contribution stems from the tweezer optical field)~\cite{vanderLaan2021subkelvin}, the laser phase noise is amplified by the strength of the cavity field and is maximal at the cavity antinode~\cite{delic2020cooling,meyer2019resolved}. To mitigate this latter potential source of heating, the phase noise was measured using an Mach-Zehnder interferometer and then actively canceled via a feedback loop using an electro-optic phase modulator~\cite{parniak2021high}. The high impact of phase noise, and its suppression, on the final occupation is shown in Fig.~\ref{fig:groundstate}~b) as a function of position in the cavity standing wave. The heterodyne detection was implemented in free-space using the optical field scattered by the nanoparticle out of the cavity in the opposite direction of the tweezer propagation direction similarly as in~\cite{magrini2021realtime,tebbenjohanns2021quantum} (see the discussion in previous paragraph regarding the out-of-the-loop heterodyne detection). The reported mean occupation number was $\bar{n}=0.04\pm0.01$ at the mechanical frequency $\omega_\alpha= 1.08$ MHz, which was achieved starting from room temperature and at the pressure $p=5\times10^{-9}$ mbar. The corresponding state purity is $92\%$, which is competitive with the purities that can be achieved with tethered mechanical oscillators in cryogenic environments~\cite{qiu2020laser,youssefi2023squeezed}. Importantly, \cite{dania2024highpurity} this established a route for high-purity quantum optomechanics at room temperature.

There have also been experiments  in the coherent scattering setup, where multiple degrees of freedom were cooled to the ground state simultaneously. The first simultaneous cooling of two translational degrees of freedom using linearly polarized light was reported in~\cite{piotrowski2023simultaneous}.  For consistency, we here adopt the same notation from the previous paragraphs following~\cite{delic2020cooling,dania2024highpurity} with $y$ aligned with the tweezer polarization axis (in ~\cite{piotrowski2023simultaneous} the $x$ axis is aligned with the tweezer polarization). The ideal configuration for two-dimensional cooling of the translational degrees of freedom $x$, $y$ is near a cavity node ($\phi=\pi/2$), where the translational optomechanical couplings are maximized similarly as discussed above for the cooling of single degree of freedom along the cavity axis. However, to simultaneously cool both the $x$ and $y$ mechanical degrees of freedom we need to tilt the polarization axis to the optimal value $\theta=\pi/4$, resulting in both the $x$ and $y$ axis at an angle $\pi/4$ with respect to the cavity symmetry axis. In such a "x-y symmetrized" configuration it can be shown that there is a "Goldilocks" zone for the particle size. Specifically, optimal cooling of the two translational degrees of freedom can be realized when the particle is not too small (large) in order to avoid low cooperativity (the emergence of dark modes)~\cite{toros2021coherent-scattering}. Decoupled dark modes, i.e., mechanical degrees decoupled from the cavity field, inhibit the cooling performance similarly in other optomechanical systems~\cite{shkarin2014optically}. In~\cite{piotrowski2023simultaneous} the Silica nanoparticle of nominal diameter $d=143\pm6$ nm was trapped at the pressure $5\pm4 \times 10^{-9}~\text{mbar}$ using the optical field with wavelength $1550\text{nm}$. Two-dimensional ground-state cooling was demonstrated near the optimal setting, with the reported final occupation numbers $0.83$, $0.81$ at the mechanical frequencies $\omega_x=224~\text{kHz}$ and $\omega_y=268~\text{kHz}$, respectively. The final occupation numbers were certified using sideband thermometry~\cite{leibfried2003quantum,Marquardt2007Quantum,wilson2007theory,genes2008ground}. By changing the tilt of the polarization axis to $\theta=\pi/2$ the cooling performance is improved (worsened) for the mode along the cavity axis (axis orthogonal to the cavity symmetry axis): resulting in one-dimensional ground state cooling with occupation number $\sim0.46$ (high phonon occupancy $\sim14$). Notably,~\cite{piotrowski2023simultaneous} indicates that full 3D ground-state cooling could be within reach of future experiments, which will be able to have full control of nananoparticles in the quantum regime, a step in this direction was accomplished~\cite{pontin2023simultaneous}, where the simultaneous cooling of all six degrees of freedom was performed. This was discussed in Sec.~\ref{sec:CS_chapter4}.

Reaching the ground state simultaneously for all the motional degrees of freedom would represent a major milestone of the field which will be relevant for both applications and fundamental research. Theoretical analysis suggests that this might be achievable by exploiting coherent scattering~\cite{rudolph2021theory}. However, the required parameters are quite demanding so combining different techniques to address different degrees of freedom might offer an easier path (see for example discussion above on~\cite{kamba2023nanoscale}). Indeed, there is a complementarity of experimental techniques since the DoFs efficiently cooled by feedback schemes are typically not well managed by coherent scattering and vice versa.


\subsubsection{Strong coupling regime and beyond\label{sec:strong}}

 Most works in levitated optomechanics can be well described in the weak coupling regime, where the optomechanical coupling, $g$, is smaller than the mechanical frequency, $\omega$, i.e., $g\ll\omega_\text{m}$ as well as smaller than the dissipation rates, i.e.,  $\omega$, i.e., $g\ll\kappa,\gamma$. In such a regime we can apply the rotating wave approximation leading neglecting counter propagating terms (in the context of cavity quantum electrodynamics we recall that this approximation leads to the famous Jaynes-Cummings Hamiltonian~\cite{jaynes2005comparison}).
 
 In the context of cavity optomechanics, the strong coupling regime is generally achieved when $g \gg \kappa,\gamma$, where $\kappa$ and $\gamma$ denote the cavity decay rate and mechanical damping rate, respectively, i.e., the interaction between the optics and mechanics occurs on faster time scale than the dissipation. 

 In the sideband resolved regime, where $\omega_\text{m} > \kappa/4$, the condition for the strong coupling regime is equivalent to  $4 g >\kappa$ (assuming $\gamma$ is negligible compared to $\kappa$ as is often the case by working at low pressures). In the strong coupling regime, where coherent effects of the interaction begin to dominate over the decay rates, we can observe normal-mode splitting and an avoided crossing in the spectra.  Loosely speaking, before the photons leak out of the cavity, or the mechanical damping becomes relevant, the nanoparticle interacts coherently with the optical photon multiple times. The ultrastrong is defined when the coupling reaches the condition $g \gtrsim 0.1 \omega_\text{m}$ (i.e., $4 g \gtrsim 0.1 \kappa$)~\cite{forn2019ultrastrong}, and the deep strong coupling regime is achieved at  $g \gtrsim \omega_\text{m}$ (i.e., $4 g \gtrsim  \kappa$)~\cite{mueller2020deep}. 
 
 The quantum-coherent coupling occurs when the coupling $4g$ exceeds the optical and mechanical decoherence rates, $\Gamma_\text{o}=\max(N_\text{o},1) \kappa$ and $\Gamma_\text{m}=\max (N_\text{m},1) \gamma$, respectively, where $N_o$ ($N_m$) is the occupation number of the optical (mechanical) thermal bath. Loosely speaking one can think that for very low bath occupation numbers the decoherence rate saturates to the value of the bare dissipation rates dictated by quantum vacuum fluctuations (in this context to the rates $\gamma$ or $\kappa$). For example, in the case of the intracavity field generated by a laser, the occupation number $N_\text{o}$ is typically negligible (much less than $1$), while the occupation of the thermal bath affecting the mechanics can be estimated as $N_\text{m} \approx \frac{k_B T}{\hbar \omega_\text{m}}$. In this regime, the requirement for the quantum-coherent coupling reduces to the condition $4g >\Gamma_\text{m},\kappa$ corresponding to cooperativities $C=4g^2/(\kappa \Gamma_\text{m})$ larger than unity.

Using the coherent scattering setup the field of levitated optomechanics started exploring the regimes beyond the weak coupling for one~\cite{delosRosSommer2021strong, dare2024ultrastrong} and two-dimensional~\cite{ranfagni2021vectorial} translational motions. To ease the comparison with other experiments we here continue to use the notation introduced earlier: see Sec.~\ref{sec:coherent} for the angles $\theta$ and $\phi$ (i.e., the tilt of the polarization with respect to the cavity symmetry axis and the displacement of the particle from the cavity anti-node, respectively) and Sec.~\ref{sec:ground} for the axes and couplings. 
The strong coupling regime between a mechanical and optical degree was first reported in~\cite{delosRosSommer2021strong}. A nanoparticle of radius $90$ nm was trapped using an optical tweezer with wavelength $1064$ nm and power $150$ mW (focused using a numerical aperture $0.8$). The polarization angle was set to $\theta\approx\pi/2$ (i.e., scattering in the cavity), resulting in coupling, $g_x$, between the cavity mode and the mechanical motion along the cavity axis. The maximum optomechanical coupling $g_x=2.3\kappa$ was achieved when the particle was placed at $\phi=\pi/2$ (that is, at the cavity node), and the mechanical frequency was $\omega_x=2\pi \times 172$ kHz. They reported the observation of normal mode splitting and the associated avoided crossing with the maximum difference between the hybridized mechanical frequencies reaching the value $2g_x=2\pi \times 46$ kHz. The gas chamber was at room temperature and the pressure was kept at $1.4$ mbar and the cavity decay rate was $\kappa=2\pi \times 10$ kHz, which limited the maximum co-operativity to $8\times 10^{-6}$.
In~\cite{dare2024ultrastrong} they have reported reaching the ultrastrong coupling regime using the coherent scattering setup.  The tweezer beam was generated using a laser of wavelength $1064$ nm with the power at the focus set to $400$ mW (focused with a numerical aperture $0.8$), resulting in a harmonic trap along the x-axis with frequency $\omega_x=190$ kHz. To maximize the coupling to the mechanical mode oscillating along the cavity axis, the linear polarization of the tweezer was oriented orthogonally to the cavity symmetry axis (i.e., $\theta=\pi/2$) and the particle was placed at the cavity node (i.e., $\phi=\pi/2$). The nanoparticle of nominal radius $105$ nm was trapped at the pressure $4$ mbar and the cavity linewidth was $\kappa=2\pi \times 193$ kHz. They reported achieving the coupling $g_x=0.55\omega_x$ with the system still operating in the linear regime without any non-negligible non-linearities stabilizing the motion of the system. Such a system is inherently unstable and could find applications for strong mechanical squeezing~\cite{kustura2022mechanical} as well as increased force sensitivity and entanglement generation~\cite{weiss2021large,cosco2021enhanced}.

In~\cite{ranfagni2021vectorial} they reached for the first time the quantum-coherent as well as the strong coupling regime, where a nanoparticle of nominal diameter $170$ nm was trapped at room temperature. The tweezer that formed the trap operated at wavelength $1064$ nm with the power set to $300$ mW (with more details of the setup given in~\cite{calamai2021transfer}). The polarization angle was set to $\theta=72^{\circ}$ resulting in the coupling of the $x$ and $y$ modes to the optical field with rates $g_x=g \sin^2\theta $ and $g_y= g \cos\theta \sin\theta$, respectively, where $g=2\pi \times 30$ kHz is achieved by placing the particle close to the node of the cavity corresponding to $\phi=\pi/2$. In this configuration the reported mechanical frequencies were $\omega_x=2\pi \times 123$ kHz and $\omega_y=2\pi \times 118$ kHz such that the strong coupling condition has been satisfied for both modes $x$ and $y$. The pressure in the nitrogen atmosphere gas chamber was controlled down to values $3\times 10^{-7}$ mbar such that the mechanical damping, $\Gamma_\text{m}$ was dominated by the recoil heating rate of $8.9$ \text{kHz} and $4.6$ \text{kHz} for the $x$ and $y$ mode, respectively. Using the reported value of the cavity decay rate $\kappa= 2\pi\times 57$ kHz we find the co-operativites,  $\sim  0.15$ and $\sim  1.8$, respectively. The optical field formed hybrid optomechanial modes with two mechanical modes in the transverse plane of the tweezer, giving the hybrid modes a distinct vectorial character, i.e., vectorial polaritons.

\begin{figure}[h!]
\centering
\includegraphics[width=\textwidth]{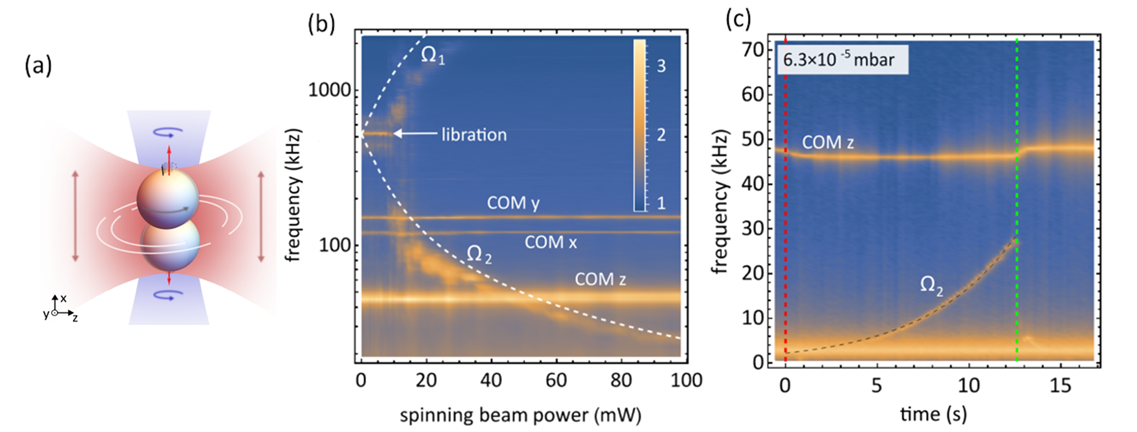}
\caption{\textbf{Long-axis spinning of a levitated nanodumbbell, adapted from}~\cite{zielinska2024longaxis}. (\textbf{a}) A dumbbell is trapped at the focus of a linearly polarized optical tweezer ($\lambda=1550$\,nm). An auxiliary circularly polarized beam ($\lambda=1064$\,nm) is used to apply a controlled torque in the y-z plane. (\textbf{b}) Coupling between two initially degenerate librational frequencies introduced by a rotation around the long axis of the particle. The torque is controlled through the auxiliary (spinning) beam power. As the steady state rotational frequency increase the degeneracy is lifted and the librational frequencies split in $\Omega_{1/2}$. (\textbf{c}) Ring-down measurement of the spinning frequency. At t=0 the spinning beam is switched off. As the spinning frequency decreases, the lower branch of $\Omega_{1/2}$ shows an increased frequency. The dashed green line marks the time at which the spinning beam power is switched back on.
 }
\label{fig:longaxis} 
\end{figure}

The deep strong coupling regime in the context of rotational levitated optomechanics has been reported in~\cite{zielinska2024longaxis}. A nanodumbbell, composed of two spherical silica nanoparticles with nominal diameter $143$ nm, was trapped in a linearly polarized tweezer beam with wavelength $1550$ nm and power $700$ mW (numerical aperture $0.8$). Two orientational degrees of freedom are hence trapped in librational motion about the direction of polarization (the x-axis), with the bare libration frequency denoted by $\Omega_0$. The third rotational degree of freedom, $\psi$, corresponding to rotations about the symmetry axis of the nanodumbbell, was controlled with an auxiliary laser propagating along the x direction. A pictorial view of the approach is shown in Fig.~\ref{fig:longaxis}~(a). Using circularly polarized light in the y-z plane for the auxiliary beam, the nanodumbbell was controllably spun up  to spinning rates exceeding $1$ GHz. This secondary beam was operating at the wavelength $1064$ nm with tunable power up to $120$ mW (focused with a numerical aperture 0.3). The fast spinning induces a mechanical coupling, $g$, between the two librational modes, as a consequence of the intrinsic dynamics of rotors (see the kinetic Hamiltonian in Sec.~\ref{sec:kinetic}). In contrast to optomechanical couplings, where the interaction is between the optical and mechanical degrees of freedom, here the optical field was used to control the value of the mechanical coupling $g$ between two mechanical degrees of freedom. Specifically the coupling is given by $g=(I_3/2I_1)\dot{\psi}_0$, where $I_2$ ($I_3$) is the moment of inertia along the short (long) axis, and $\dot{\psi}_0$ denotes the frequency of rotation along the symmetry axis. Assuming a length-to-diameter ratio of $1.8$ for the nanodumbell, they estimated that the ratio of the two moments of inertia was $I_3/I_1=0.6$ and that the spinning frequencies reached values up to $\dot{\psi}_0=2\pi \times 1.20$ GHz using ringdown measurements. The dumbbell’s spinning rate up to values $30$ MHz was also estimated from the hybrid mode frequencies of the two librations,  $\Omega_{1/2}=\sqrt{\Omega_0^2+g^2}\pm g$, after which the signal-to-noise ratio precludes the application of this method (electronic noise becomes important for the hybrid mode at lower frequency precession mode $\Omega_2$ and the stabilizing effect related to large angular momenta reduces the overall amplitude of the signal). A measurement of the librational splitting is shown in Fig.~\ref{fig:longaxis}\,(b). A better estimate of the spinning rate is obtained using a ring-down method. By switching off the auxiliary laser, and performing a fit to the time-dependent curve of the emerging low frequency precession mode $\Omega_2\equiv \Omega_2(t)$ in the pressure range $10^{-3}$ mbar to $10^{-6}$ mbar. This is shown in Fig.~\ref{fig:longaxis}\,(c). Using this configuration the work has reported reaching the deep strong regime~\cite{frisk2019ultrastrong}, with the ratio $g/\Omega_0$ reaching values close to 3, making it competitive with trapped atoms~\cite{koch2023quantum}.

\subsubsection{Non-classical states and entanglement\label{sec:Non-classical}}
 
\begin{figure}[h!]
\centering
\includegraphics[width=\textwidth]{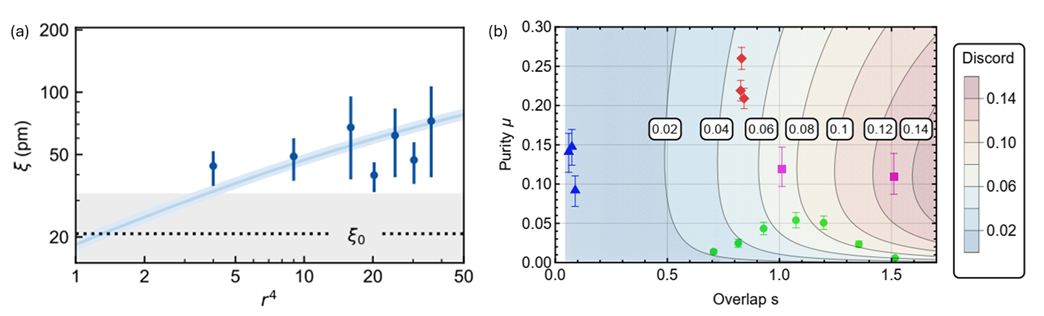}
\caption{\textbf{ Non-classicality beyond ground state cooling.} (\textbf{a}) Adapted from~\cite{rossi2024quantum}. Inferred coherence length as a function of trap frequency ration $r$ used pulse modulation protocol~\cite{janszky1992strong,rashid2016experimental}. Here, experimental data is compared with model prediction (blue solid line) and initial coherence length $\xi_0\approx21$\,pm (dashed line). Data shows an increased coherence length to $\xi=(73\pm34)$\,pm. (\textbf{b}) Two-dimensional state purity as a function of spectral overlap as defined in the text, adapted from~\cite{deplano2025high}. Data points from different experiments (green~\cite{ranfagni2022twodimensional}, magenta squares~\cite{ranfagni2023spectral}, blue triangles~\cite{piotrowski2023simultaneous}, red squares~\cite{deplano2025high}) are shown overlaid with a contour plot of achievable quantum discord. The experiment highlighted the difficulty of simultaneously achieving high purity and large Discord through a CS interaction. The best demonstrated combination of this two parameters consisted in a purity $\mu=0.209$ and a Discord $D\sim0.0482\pm0.0012$.
}
\label{fig:nonclassical} 
\end{figure}

Achieving ground-state cooling and the strong coupling regime and beyond opens the door to preparing and detecting quantum superpositions and entangled motional states (see Secs.~\ref{sec:ground} and \ref{sec:strong}).
A key step towards interferometric schemes was achieved in \cite{rossi2024quantum}, where they reported for the first time a quantum delocalization of single translational degree of freedom of a levitated nanoparticle beyond the spread of the zero-point-motion. The experimental configuration consisted of an tweezer generated by an optical field with wavelength $1550$ nm and linear polarization along the x-axis and numerical aperture 0.8 (here we continue to use the axis notation introduced in Fig.~\ref{fig:exp} (a)). The particle was trapped in a cryogenic environment kept at a temperature of $7$ K with the gas pressure below $10^{-9}$ mbar, making decoherence due to gas collisions negligible, leaving photon recoil as the main source of decoherence. They focused on the z-motion (motion along the tweezer propagation direction), which was trapped at the mechanical frequency $56.5$ kHz, while the total decoherence rate was estimated at $23.7$ kHz. They performed a three step protocol with the sequence consisting of: feedback cooling to the ground state using cold-damping (see Sec.~\ref{sec:colddamping}), pulse modulation of the trapping potential to delocalize, i.e., squeeze, the z-motion~\cite{janszky1992strong}, and time-trace measurements in combination with a retrodiction filter to infer the generated state~\cite{rossi2019observing,lammers2024quantum}. They reported achieving a threefold expansion of the spatial coherence length to $73$ pm with minimal added noise, this is shown in Fig.~\ref{fig:nonclassical}\,(a). This demonstrates that the quantum state of a levitated particle can be coherently spread beyond the ground-state width – a clear signature of non-classical motion - with potential applications for quantum-enhanced force sensing~\cite{burd2019quantum}. 

An important step towards realizing continuous variable entanglement between two mechanical modes was achieved in~\cite{deplano2025high}. Exploiting the coherent scattering setup they reported cooling two mechanical modes of a nanoparticle with nominal radius of $100$ nm in the regime of strong coupling and large spectral overlap. The particle was trapped using a tweezer with wavelength $1064$ nm and with linear polarization oriented at angle $\theta=\pi/4$, i.e., the angle between the cavity symmetry axis and the polarization direction, and placed at the cavity node corresponding to the angle $\phi=\pi/2$ (see Sec.~\ref{sec:coherent}). The power at the focus was set to $250$ mW, resulting in mechanical frequencies $\omega_x=121.1$ kHz and $\omega_y=108.5$ kHz for the $x$ and $y$ mechanical motion in the polarization plane, respectively. The optical cavity had a linewidth of $\kappa=2\pi \times 57$ kHz and the gas chamber was operating at $3\times 10^{-8}$ mbar. They report achieving the maximum purity of the total x and y state to be $\mu=0.209$ with a spectral overlap $s\sim2  (g_x^2+g_y^2)/\kappa \delta \sim 0.84$ ($g_x$ and $g_y$ denote the cavity optomechanical couplings, and $\delta=\omega_x-\omega_y$). They also reported the maximum achieved quantum discord of $D\sim0.0482\pm0.0012$, with the positive value indicating the quantumness of the correlations~\cite{ollivier2001quantum}. This is shown in Fig.~\ref{fig:nonclassical}\,(b), where the results are compared with previous experiments.  Increasing further the spectral overlap hinders cooling, i.e., it becomes less efficient, and thus lowers the achievable state purity, while ideally one would like to increase both the total $x,y$ state purity and have a large spectral overlap in order to generate and verify mechanical entanglement. For comparison, the work in~\cite{ranfagni2023spectral} achieved a spectral overlap of $\sim1.5$, a total $x,y$ state purity $\sim0.11$ and quantum discord of $\sim0.13$. To circumvent this issue, and to demonstrate entangled states~\cite{vitali2007stationary,genes2008simultaneous}, there are proposal involving additional electromagnetic modes~\cite{hartmann2008steady,li2015generation}, cavities~\cite{mazzola2011activating} and modifications with pulsed schemes~\cite{rakhubovsky2020detecting}. Generating entangled states of two or more mechanical modes will pave the way towards innovative quantum information schemes~\cite{Weedbrook2012Gaussian} and testing the limits of quantum mechanics~\cite{marshall2003towards,bassi2013models,gasbarri2021testing}.

On the theoretical side, several proposals have outlined how to generate and verify even more exotic states, such as spatial quantum superpositions (“Schrödinger cat” states) of levitated objects \cite{millen2020quantum, stickler2018probing,romero2011large,romero2011quantum,bateman2014near,geraci2015sensing,wan2016free,pino2018chip,toros2021creating}. These schemes often require the nanoparticle to be in its ground state first, combined with extremely low decoherence environments and precise control. Furthermore, two motional modes of a single levitated nanoparticle have been simultaneously cooled to their joint ground state \cite{piotrowski2023simultaneous}, setting the stage for creating entangled or multimode non-classical states of its motion.  Experiments have also shown how to control two particles in adjacent sites \cite{rieser2022tunable} with proposals suggesting nanoparticles arrays where the quantum resource of entanglement could be used for quantum-enhanced sensing and quantum information processing~\cite{hammerer2010quantum}.  
%
Rotational motion of levitated nanoparticles offers a complementary avenue for non-classical state preparation, with unique features absent in translational oscillations. A free rotor has a quantised angular momentum spectrum rather than a discrete harmonic spectrum, and even an angle-confined “librating” particle exhibits nonlinear dynamics that could enable novel quantum interference effects~\cite{stickler2021quantum}. 
Experiments have made rapid strides in this direction by achieving extreme control over rotations. All six ro-translational degrees of freedom have now been cooled and controlled in tandem, using both active feedback and cavity optomechanics approaches \cite{pontin2023simultaneous, kamba2023nanoscale}. This comprehensive control of a nanoparticle orientation and position at the quantum level is an essential step toward preparing joint rotational-translational quantum states and leveraging them for quantum-enhanced sensing or fundamental tests. Furthermore, optically levitated nanorotors have been spun up to GHz rotation frequencies in vacuum \cite{reimann2018ghz, ahn2018optically}, demonstrating stability and control in the high-speed classical regime. More recently, torsional (librational) modes have been feedback-cooled to sub-Kelvin effective temperatures, corresponding to mean quantum occupation numbers well below 100 quanta \cite{vanderLaan2021subkelvin, blakemore2022librational, gao2024feedback}. A quantum ground state of librational motion (i.e. a single-digit phonon occupation of an angular mode) has very recently been reached experimentally~\cite{dania2024highpurity} and shows that rotational degrees of freedom can be cooled and manipulated to the quantum regime (see Secs.~\ref{sec:ground} and \ref{sec:strong} for an overview of experimental progress).

Theoretical proposals suggest several pathways to fully non-classical rotational states. One approach is to exploit internal spin to control orientation: for instance, a single embedded NV center in a levitated nanodiamond can be strongly coupled to the particle’s rotational motion, allowing preparation of a quantum superposition of orientations \cite{rusconi2022spin}. Aligning the magnetic field with the NV center enables ultrastrong coupling between the NV spin and the diamond’s rotation. This strong interaction allows precise single-spin control over the particle’s three-dimensional orientation and supports protocols to create and detect quantum superpositions of particle orientations. This plays a crucial role when it comes to alignment control in spatial superpositions with NV diamond as outlined in~\cite{Japha2023Quantum}. 
Other schemes use purely optical means – such as elliptically polarized light scattering – to cool and coherently control angular motion \cite{schafer2021cooling}, potentially enabling a quantum ground-state rotor. A recent review provides a comprehensive overview of these prospects for quantum rotations of nanoparticles \cite{stickler2021quantum}. Achieving quantum-controlled rotation would not only expand the toolkit of non-classical states (e.g. allowing superpositions of angular momentum or orientation) but also enable new tests of quantum physics, such as probing rotational decoherence and symmetry-breaking in macro-scale quantum objects.

Additionally, the current state of the art for matter-wave interferometry has been achieved with macromolecules with mass $\sim 10^{-23}\,\text{kg}$ and thus the preparation of non-classical motional states of nanoparticles with masses in the range $\sim 10^{-19}\,\text{kg}$ to $\sim 10^{-17}\,\text{kg}$ would represent a major milestone. Non-classical states of motion would enable sensing based on quantum phases with applications ranging from gravimetry~\cite{rademacher2020quantum} to testing the large scale limit of quantum mechanics~\cite{bassi2013models}.
The orientational degrees of freedom of a nanoparticle are expected to lead to distinct quantum phenomena in a regime previously unexplored by molecular systems. Among these are orientational quantum revivals~\cite{stickler2018probing, schrinski2022interferometric} and tunneling between different orientational states~\cite{ma2020quantum}, also known as the quantum tennis racket effect. 

Finally, various strategies for generating and detecting non-classical correlations, such as entanglement, across diverse physical systems have been theoretically modeled. Starting with one of the foundational publications  \cite{chang2010cavity} which explore cavity optomechanics as a means to induce entanglement between mechanical and optical modes. \cite{ralph2016coupling} proposes coupling superconducting qubits to optical photons to facilitate long-distance quantum communication. \cite{bose2017spin} suggest using spin systems to test the quantum nature of gravity by observing entanglement generated through gravitational interactions. \cite{rudolph2020entangling} discusses methods to entangle levitated systems using the coherent scattering scheme. \cite{rakhubovsky2020detecting} focuses on detecting entanglement in levitated systems using statistical correlations. \cite{weiss2021large} investigates large-scale entanglement using optimal control. \cite{cosco2021enhanced} enhances entanglement detection techniques using a periodic driving scheme. \cite{brando2021coherent} examines coherent scattering and its role in entanglement distribution. Collectively, these studies advance our understanding of entanglement generation and detection, paving the way for future quantum technologies using rotational degrees of freedom in levitated systems.

\subsubsection{NV centers}
\label{sec:NV}

Levitation of nitrogen vacancy (NV) centers embedded in nanodiamond has been demonstrated in optical~\cite{rahman2016burning}, electrical~\cite{perdriat2022Angle}, and magnetic levitation~\cite{gieseler2020single}, as initially discussed in Sec.~\ref{sec:Levitation}. The interactions between the rotational modes and either single or an ensemble of NV spins allows for a range of different interactions in levitated optomechanics. For example, the coupling between levitated ferromagnetic micro-particles in a Paul trap with nearby NV centers in a nanodiaond fixed to a surface was studied experimentally~\cite{huillery2020spin}. They demonstrated that under modest magnetic fields, the induced librational frequencies can reach several hundred kilohertz with quality factors approaching $10^4$. This work suggested that strong coupling was feasible since the mechanical state lifetime was longer than the NV spin coherence lifetime.  Strong coupling between a single NV spins and the rotational mode of nanodiamonds levitated in an ion trap has also been proposed~\cite{ma2017proposal,delord2017strong}. In addition, the torsional/librational motion of levitated nanodiamond arrays could also be used to mediate the coherent coupling between embedded NV centers~\cite{zhang2025scalable}.

Spin–mechanical coupling has been exploited for cooling the torsional motion of electrically microdiamonds with embedded NV centers~\cite{delord2020spin}. Here a large ensemble of NV centers was driven into an excited magnetic state via microwave excitation where a spin-dependent torque was generated. In this process diamond aligns (or anti-aligns) with the applied magnetic field. As the spin lifetime is comparable to the angular oscillation period, the delay between the NV magnetization and the B-field induced angular oscillation creates a velocity dependent torque that is dependent on microwave detuning allowing either cooling of heating of this motional degree of freedom. The net result is a pronounced cooling (heating) of the diamond motion when the microwave field was red (blue) detuned from the spin resonance.  Application of an external magnetic field has been shown to lock the particle's orientation with an embedded NV center~\cite{perdriat2022Angle} while the torque from the magnetic field can be enhanced by tunable dipolar interactions~\cite{Pellet-Mary2021Magnetic}.
As discussed in this review, levitated particles can rotate at speeds faster than those achievable with larger mechanical rotators. This enables observing effects such as the Berry phase of embedded NV centers due to mechanical rotations~\cite{jin2024quantum}. Here nanodiamonds were levitated at high vacuum in a surface Paul trap with additional electrodes which when high AC voltage was applied, enabled rotation speeds up to 20 MHz. At these stable high speeds, the Berry phase generated by the embedded NV center electron spins was observed, generating a pseudomagnetic field of 0.71 mT for the electron spin in the rotating frame. 

Strong coupling between a single nitrogen-vacancy spin and the rotational mode of diamonds levitating within a uniform magnetic field has been studied theoretically, opening possibilities for the creation of torsional superpositions~\cite{ma2017proposal}. In the same year strong coupling was subsequently realized experimentally in a weak magnetic field while levitated in a Paul trap \cite{delord2017strong, Delord2018Ramsey}.
Rotation of symmetric nanorotors can be strongly affected by a small number of intrinsic spins. The resulting dynamics are observable with freely rotating nanodiamonds with embedded NV centers and persist for realistically shaped near-symmetric particles. This works opens the door towards torque free schemes to control rotation at the quantum level~\cite{Ma2021Torque}.

\subsection{Rotational-translational control in related systems}\label{othersystems}
Rotational control of molecular systems using strong optical fields is a well established field~\cite{koch2019quantum}. It is therefore important to put roto-translational control in levitated optomechanics in the context of these works and to outline the differences, similarities and opportunities that they offer to levitated optomechanics. 
Optical-induced adiabatic and nonadiabatic alignment of molecules, where angular momentum is transferred from an optical field to untrapped ensembles of molecules in the gas phase, has been extensively studied~\cite{friedrich1995alignment,larsen1999aligning,stapelfeld2003aligning,larsen2000three}. The resulting librational/pendular motion of these systems in both linearly and elliptically polarized light has been demonstrated for linear, symmetric and asymmetric molecular rotors~\cite{filsinger2009quantum, friedrich1991alignment,Rosca-Pruna2001experimental, peronne2003nonadiabatic}. As molecular systems are many orders of magnitude smaller and less massive than levitated micro and nanoparticles, their motion is most often always described quantum mechanically and therefore the theory developed for these systems can be directly applied to roto-translational motion and control in levitated quantum optomechanics~\cite{stickler2021quantum}. In particular, the aim to observe revivals of rotational wavepackets, following optically induced alignment has been explored theoretically, and while this is commonly performed experimentally in molecular systems, it is challenging to observe in levitated nanorotors as the revival time is typically longer than the motional decoherence. This leads to a contrast loss in the temporal interference of the evolving wavepackets making observation of these quantum features a challenge. At temperatures of only a few Kelvin, only a few rotational states are populated in simple molecular systems, while the much larger moments of inertia of even small nanorotors means that a large number of rotational states can be populated, again obscuring the quantum nature of this motional degree of freedom in nanorotors. In molecular systems, circularly polarized light has also been used to confine molecules within a plane, while elliptically polarized fields can provide full 3-D molecular alignment of even asymmetric top rotors~\cite{purcell2009tailoring,sun2015rotational}. These features are now routinely observed in many roto-translational trapping experiments. By utilizing alignment within a linearly polarized field and then rapidly accelerating the polarization as a function of time, extreme centrifugal forces have been demonstrated on molecules. This has allowed the study of highly rotational states and even molecular dissociation~\cite{karczmarek1999optical,korobenko2018control,koch2019quantum}. Large rotation rates in the GHz range have been achieved by transfer of the spin angular momentum to nanorotors trapped in circularly polarized fields. While much of this work has been performed with optical fields, orientational control (as opposed to alignment) has also been accomplished for molecules which possess electric and magnetic moments~\cite{koch2019quantum,Hogan2011deceleration,vandeMeerakker2008}. These state dependent forces can not only be used for orientation, but they can also be used for separation of molecular ensembles based on the state dependent Zeeman and Stark shifts they experience in inhomogeneous fields and represent a possible future source of control in levitated optomechanics. 

Rotational control has also been well developed in the overdamped regime within the optical trapping community, where nanoparticles and microparticles are usually trapped in liquids~\cite{Ashkin1987Optical,Friese1998Optical,Bruce2021Initiating,Jia2023Optical,Bustamante2021Optical,Roy2014Simultaneous,Maragò2013Optical}. Here, transfer of optical angular moment has also been extensively studied. Unlike molecular systems where only gradient forces need to be accounted for, both scattering and gradient forces must be considered. Beth's experiment~\cite{Beth1936Mechanical} which demonstrated the transfer of spin angular momentum to a waveplate was replicated on the microscale using a birefringent optically trapped particle. The transfer of orbital angular momentum has also been demonstrated in more exotic fields~\cite{Almeida2023Trapping,Rubinsztein-Dunlop2017Roadmap} thus opening up a large tool box of optical interactions for roto-translational control of trapped nano and microparticles which have been explored in overdamped environments~\cite{Jon2008Light,Donato2018Optical}. The full toolbox is yet to be fully explored within the undamped motion of levitated optomechanical systems.

\section{Challenges and future directions} \label{sec:discussion}

The development and application of roto-translational control and cooling in levitated optomechanics marks a significant evolution in the field of levitated optomechanics. It offers not just a richer set of dynamics to explore, but also an expanded platform for applications in sensing, quantum control, and fundamental physics as outlined above. In this section we describe promising new area of research within this field. 
    
The field still faces several persistent theoretical and experimental challenges. A significant barrier to the realization of long lived optomechanical quantum states is environmental decoherence. Even in high vacuum, residual gas collisions and photon scattering remain prominent sources of decoherence for optical levitation. Gas collisions can be reduced by performing experiments under ultrahigh vacuum~\cite{Dania2024Ultrahigh} or in a cryogenic environment~\cite{tebbenjohanns2021quantum}, but the dominant decoherence is photon scattering~\cite{jain2016direct}. This is a compound effect as the system suffers from scattering from the trap beam but also from the emission of thermal radiation~\cite{Schafer2024Decoherence,hebestreit2018measuring}. Proposals to mitigate recoil and thermal radiation effects include the use of squeezed light to reduce back-action~\cite{gonzalezballestero2023surpressing}, back-action suppression utilizing reflective boundaries~\cite{Gajewski2025Backaction}, hybrid systems that reduce reliance on continuous-wave laser trapping~\cite{Bonvin2023Hybrid}, and laser-free traps such as magnetic or Paul traps~\cite{latorre2023superconducting}. Ultimately, decoherence from internal phonons may define the limit of coherence time for many quantum experiments~\cite{Henkel2024Universal}.

One of the unique features of levitated systems, when compared to atomic and molecular quantum systems, is the macroscopic nature of the levitated object. This offers both advantages and challenges. For example, a major limitation is the inability to deterministically load particles at will but also to load particles with the same properties that are desirable for a particular experiment. This is important for all experiments but particularly so if we are to develop robust and reliable sensors. The stochastic nature of particle trapping and random particle size from one loading event to another has limited the reproducibility in many systems. However, recent advances in hollow-core delivery~\cite{Lindner2024Hollow}, direct loading~\cite{Nikkhou2021Direct} and particle characterization offer promising paths to deterministic loading of tailored particles. The macroscopic nature enables the fabrication of unique shapes and materials to explore a rich tapestry of dynamics~\cite{Seberson2020Simulation,rahman2019large,rahman2017laser,steiner2024pentacene,aggarwal2020searching,Laplane2024Inert,Brown2023Superfluid,Arita2023Cooling}. The use of engineered shapes and novel materials could enable new optomechanical coupling mechanisms and improved rotational sensing fidelity.

Levitated optomechanics allows us to work in a new macroscopic quantum regime. Because these levitated nanoparticles have masses in the $10^7$–$10^9$ atomic units range, preparing them in spatial superpositions or entangled states probes the quantum-classical transition. Proposals have emerged that use these systems to generate massive spatial superpositions as tests of collapse models or gravitationally induced decoherence~\cite{romero2011large,roda-llordes2024macroscopic}. For example, entanglement of a pair of nearby levitated nanospheres through their gravitational interaction, if both are prepared in a quantum superposition tests the quantumness of gravity~\cite{bose2017spin}. However, such experiments still require significant isolation from environmental noise sources as well as further development to control decoherence, characterization and full 6D control. Rudolph \textit{et al.} \cite{rudolph2020entangling} described using coherent scattering to entangle two nanoparticles, which could also be a stepping stone to observing nonclassical gravity-induced correlations at the macroscale. Other theoretical work has explored using nonlinear optics or active feedback to generate non-Gaussian quantum states (e.g. Fock states or Schrödinger-cat states of the particle’s motion). For instance, Weiss \textit{et al.} \cite{weiss2021large} proposed a periodic modulation scheme to repeatedly expand and contract a nanoparticle wave function, allowing it to briefly delocalize over much larger distances before decoherence. Such ideas aim at creating spatial superpositions large enough to test objective collapse theories or decoherence mechanisms. 

In some scenarios, rotational motion is a hindrance rather than a resource. For example, in dark matter searches using optically levitated particles~\cite{carney2021ultralight,moore2021searching,Kilian2024Dark} or in high-frequency gravitational wave detectors~\cite{arvanitaki2013detecting,winstone2022optical}, uncontrolled rotations introduce noise or complicate signal analysis. In these cases, suppressing or stabilizing unwanted rotational modes is essential. Rotations can also obscure sensing signals when trying to isolate pure translational responses. In such setups, anisotropic particle design and polarization control may be used to suppress parasitic torques. In magnetic levitation systems, rotational instabilities can emerge from trap asymmetries, requiring active feedback or trap redesign. For NV-center–based quantum control, uncontrolled rotation destroys spin readout fidelity, making torque damping and angular stabilization imperative~\cite{delord2020spin,neukirch2015multi-dimensional}.

Finally, realizing strong optomechanical coupling for all degrees of freedom remains a central goal~\cite{aspelmeyer2014cavity}. While rotational–translational coupling enriches the dynamics and allows for exotic interactions, it also complicates control. Cross-mode interactions and mode hybridization must be carefully managed to avoid dark modes and maintain effective cooling and readout. Advanced feedback schemes, cavity detuning strategies, and tailored trapping geometries will all be key to unlocking full 6D quantum control.

\section{Conclusions} \label{sec:conclusion}
%
The field of levitated optomechanics has undergone a remarkable transformation, evolving from early demonstrations of translational control to a broader understanding of the full six degrees of freedom accessible in vacuum. This review has outlined the foundations and current frontiers of roto-translational levitated optomechanics, a subfield that is still in its early stages but already showing considerable promise across theory, experiment, and application.

We have presented a unified picture of the classical and quantum dynamics of anisotropic particles in electromagnetic traps, with a particular focus on optical levitation, and discussed how the inclusion of rotational degrees of freedom enriches the optomechanical landscape. These additional degrees of freedom introduce new types of couplings, stability regimes, and dissipation channels, which must be accounted for in precise modeling and experimental control. The resulting dynamics, especially in asymmetric top geometries, are highly non-trivial and open up novel pathways for sensing, thermodynamics, and quantum control.

Experimentally, we are approaching the regime where the coherent manipulation of all six degrees of freedom is becoming viable. Recent results have demonstrated full 6D cooling, torque sensing at unprecedented levels, and first steps towards the preparation of non-classical states of librational motion. At the same time, efforts are underway to translate these capabilities to hybrid and light-free platforms, which could address some of the known limitations of optical levitation, such as recoil heating and photothermal instability.

However, many challenges remain. These include robust mitigation of decoherence in all degrees of freedom, the identification and suppression of dark modes in multimode systems, and the extension of quantum control techniques to rotational motion. Furthermore, the role of particle morphology, internal structure, and material-specific properties introduces a degree of complexity that is only beginning to be understood, especially when aiming for quantum-enhanced or quantum-limited operation.

Looking ahead, roto-translational optomechanics is well positioned to make key contributions to several open questions in physics: from precision measurements and inertial sensing to fundamental tests of quantum mechanics and gravity. By fully embracing the rich dynamical structure of levitated particles and continuing to develop both the theoretical tools and experimental platforms needed to explore it, we expect this field to serve as a fertile ground for both foundational and technological advances in the coming years.

\section*{Acknowledgments}
We would like to thank Julian H. Iacoponi for comments and discussions.
MT acknowledges funding from the Slovenian Research and Innovation
Agency (ARIS) under contracts N1-0392, P1-0416, SN-ZRD/22-27/0510
(RSUL Toro\v{s}). AP, MR, JMHG and PFB acknowledge funding from the United Kingdom's EPSRC and STFC via Grant Nos. EP/N031105/1, EP/S000267/1, EP/W029626/1, EP/S021582/1 and ST/W006170/1.

\newpage
 \appendix

\section{Recovering the classical results\label{app:Derivation}}

In this section we  briefly discuss how to recover the key classical expressions derived in Sec.~\ref{sec:model}, starting from a quantum model of an anisotropic polarized particle in an elliptically polarized Gaussian beam discussed in Sec.~\ref{sec:quantum_model}. The tools that we will provide in this section can be applied in principle to any dynamics with Hamiltonian, Lindblad, and stochastic terms.

The classical equations of motion can be obtained by consider the semi-classical limit. A shorthand prescription is to formally replace the operators by complex-values and commutators with Poisson brackets~\cite{Stickler2016Rotranslational}, i.e. 
\begin{equation}
     \hat{O}\rightarrow O,\quad\quad-\frac{i}{\hbar}\left[\,\cdot\,,\,\cdot\,\right]\rightarrow\left\{ \,\cdot\,,\,\cdot\,\right\}, \label{eq:prescription}
\end{equation}
which can be seen heuristically as reversing the canonical quantization procedure. In addition, we will assume the the expectation value of the quantum operator, when appearing in the dynamics, can be replaced by the corresponding classical value, i.e.,
\begin{equation}
    \langle\hat{O}\rangle\rightarrow O. \label{eq:prescription2}
\end{equation}

We now discuss separately the Hamiltonian, the Lindblad, and the stochastic terms.  We start by considering the Hamiltonian evolution encoded in the von-Neumann equation:
\begin{equation}
    d \hat{\rho}=  -\frac{i}{\hbar}[\hat{H},\hat{\rho}]dt, \label{eq:vN} 
\end{equation}
where $\hat{H}$ is the Hamiltonian operator, and $\hat{\rho}$ is the statistical operator. We obtain the dynamics for the observable $\hat{O}$ by inserting Eq.~\eqref{eq:vN} in $d\langle \hat{O} \rangle=\text{tr}[\hat{O}d\hat{\rho}]$, and using the cyclic property of the trace. In particular, we find the following equation of motion:
\begin{equation}
    d \hat{O}=  \frac{i}{\hbar}[\hat{H},\hat{O}]\,dt. \label{eq:Hamiltonian}
\end{equation}
Applying the prescription in Eq.~\eqref{eq:prescription} to \eqref{eq:Hamiltonian} we then finally find:
\begin{equation}
    d O=  \{O,H\}\,dt, \label{eq:HamiltonianF}
\end{equation}
which is the well-known classical Hamiltonian equation of motion.

We next consider the Lindblad dynamics given by:
\begin{equation}
    d \hat{\rho}=  \gamma \left(\hat{K} \rho \hat{K}^\dagger -\frac{1}{2}(\hat{K}^\dagger\hat{K} \hat{\rho}
    +\hat{\rho} \hat{K}^\dagger\hat{K})\right)dt, \label{eq:Lindblad} 
\end{equation}
where $\gamma$ is a characteristic rate, and $\hat{K}$ is a generic Lindblad operator. Using the same procedure as in the Hamiltonian case (see text below Eq.~\eqref{eq:vN}) we can obtain the equation of motion for an observable $\hat{O}$:
\begin{equation}
    d \hat{O}=  \gamma \left(\hat{K}^\dagger \hat{O} \hat{K} -\frac{1}{2}(\hat{K}^\dagger \hat{K}\hat{O}
    +\hat{O}\hat{K}^\dagger\hat{K})\right)dt, \label{eq:LindbladQ} 
\end{equation}
which can be rewritten using commutators as
\begin{equation}
    d \hat{O}=  \frac{\gamma}{2} \left(\hat{K}^\dagger [\hat{O}, \hat{K}] - [\hat{O}, \hat{K}^\dagger] \hat{K}\right)\,dt. \label{eq:LindbladQ2} 
\end{equation}
Applying the prescription from Eq.~\eqref{eq:prescription} to Eq.~\eqref{eq:LindbladQ2} we then find the following classical equation of motion:
\begin{equation}
    d O=  \frac{i\hbar\gamma}{2} \left(K^* \{O, K\} - \{O, K^*\}K\right)\,dt, \label{eq:LindbladC} 
\end{equation}
which can be written compactly as
\begin{equation}
    d O= - \hbar\gamma \,\, \text{Im}\left(K^* \{O, K\}\right)\,dt. \label{eq:LindbladF} 
\end{equation}

We now discuss the stochastic terms of the Belavkin form~\cite{wiseman2009quantum}:
\begin{equation}
    d \hat{\rho}= \sqrt{\gamma} \left(\hat{K} \hat{\rho} dZ^* +\hat{\rho} \hat{K}^\dagger dZ
    -\text{tr}(\hat{K} dZ^*+ \hat{K}^\dagger dZ)\hat{\rho}
    \right) \label{eq:stochastic} 
\end{equation}
where 
\begin{equation}
    dZ=i\frac{dX+idP}{\sqrt{2}} \label{eq:dZ}
\end{equation} 
is a complex-valued stochastic process. The  part $dX$ and $dP$ will be referred to as the amplitude and phase noise, respectively, in accordance with the chosen conventions of the main text. In particular, the two noises satisfy the following properties:
\begin{equation}
    \mathbb{E}[dX]=0,\quad \mathbb{E}[dP]=0,\quad\mathbb{E}[dXdP]=0,
    \quad\mathbb{E}[dXdX]= dt,\quad\mathbb{E}[dPdP]= dt, \label{eq:masterConvention}
\end{equation}
where $\mathbb{E}[.\cdot.]$ denotes the expectation value over different noise realizations. We also note that the Lindblad operator can be decomposed as 
\begin{equation}
    \hat{K}=\hat{K}^\text{(+)}+ i \hat{K}^\text{(-)}, \label{eq:decom}
\end{equation}
where $\hat{K}^\text{(+)}$ and $\hat{K}^\text{(-)}$ denote the hermitian and antihermitian part.

We can then proceed to find the equation of motion for the observable $\hat{O}$ (see text below Eq.~\eqref{eq:vN}):
\begin{equation}
    d \hat{O}= \sqrt{\gamma} \left(\hat{O} \hat{K}  dZ^* + \hat{K}^\dagger \hat{O} dZ
    -\text{tr}(\hat{K} dZ^*+ \hat{K}^\dagger dZ)\hat{O}
    \right) . \label{eq:stochasticQ} 
\end{equation}
 In particular, using Eq.~\eqref{eq:dZ} and \eqref{eq:decom} we rewrite Eq.~\eqref{eq:stochasticQ} as 
\begin{alignat}{1}
    d \hat{O}=-& \sqrt{\frac{\gamma}{2}} \left(i[\hat{O},\hat{K}^\text{(+)}]dX
    +i[\hat{O},\hat{K}^\text{(-)}]dP\right.\nonumber\\ 
    &\left.+(\hat{O}\hat{K}^\text{(+)} +\hat{K}^\text{(+)}\hat{O})dP  
    -(\hat{O}\hat{K}^\text{(-)} +\hat{K}^\text{(-)}\hat{O})dX 
    -2 \text{tr}( \hat{K}^\text{(+)} dP - \hat{K}^\text{(-)} dX)\hat{O}
    \right). \label{eq:stochasticQ2} 
\end{alignat}
We now apply the prescriptions in Eqs.~\eqref{eq:prescription} and \eqref{eq:prescription2} to Eq.~\eqref{eq:stochasticQ2} to find the classical equation of motion:
\begin{equation}
    d O= \hbar\sqrt{\frac{\gamma}{2}} \left(\, \{O,\text{Re} ( K)\}dX +\{O,\text{Im} ( K)\}dP \, \right).\label{eq:stochasticF} 
\end{equation}

Using suitable adaptations of Eqs.~\eqref{eq:HamiltonianF}, \eqref{eq:LindbladF} and \eqref{eq:stochasticF} we can recover the conservative, non-conservative deterministic, and non-conservative stochastic terms of the classical model, respectively (see Sec.~\ref{sec:model}). While the derivation in this appendix should only be understood as a shorthand prescription, i.e., without any pretense of a rigorous classical limit, it nonetheless illustrates that the classical results can emerge from the quantum modeling. We sketch below how one could apply the obtained formulae to the quantum model outlined in Sec.~\ref{sec:quantum_model}.

To obtain the classical optomechanical Hamiltonian in Eq.~\eqref{eq:Hgradient} we apply the procedure in Eq.~\eqref{eq:prescription} to the corresponding quantum Hamiltonian operator defined below Eq.~\eqref{eq:quantization}. To recover the non-conservative optical terms we introduce the family of operators $\hat{A}_{\bm{n},\nu}^\dagger$,  describing the scattering of the optical field off the nanoparticle (see the corresponding classsical observable defined in Eq.~\eqref{eq:nano} and below Eq.~\eqref{eq:p3}), i.e., in place of the generic observable $\hat{K}$ appearing in \eqref{eq:Lindblad}. Furthermore, we  need to introduce integrations and summations over the directions, $\bf{n}$, and polarizations,  $\nu$, respectively, in Eqs.~\eqref{eq:LindbladF} and \eqref{eq:stochasticF}. Finally, to connect the noises defined by the expectation values in Eq.~\eqref{eq:masterConvention},  with the classical input noises characterized in Eq.~\eqref{eq:noises2}, we need to set $dX^\text{(in)}= dX/\sqrt{2}$ and $dP^\text{(in)}= dP/\sqrt{2}$ (see discussion in Eqs.~\eqref{eq:noises2A1}-\eqref{eq:noises2}). Following these prescriptions, and using Eq.~\eqref{eq:stochasticF}, we can obtain the stochastic optical forces and torques given in Eq.~\eqref{eq:p15}. 

As an example of the correspondence between the quantum and classical modeling of gas collisions, let us consider the dissipative Caldeira-Leggett model~\cite{caldeira1983path}. In place of the generic operator $\hat{K}$ introduced in Eq.~\eqref{eq:Lindblad}, we need to include the operators describing the effect of gas collisions appearing in Eq.~\eqref{eq:moperators}.
Using ~\eqref{eq:LindbladF} we can readily recover the deterministic damping terms in Eqs.~\eqref{eq:Pd2} and  \eqref{eq:Pid2}. Furthermore, from Eq.~\eqref{eq:stochasticF} we find the following classical stochastic terms:
\begin{alignat}{1}
d\boldsymbol{P}_{k}^{\text{(sc)}} & =\sqrt{2k_{B}T\,m\,\gamma_{c}}\,dV_{k},\label{eq:Pnc2}\\
d\boldsymbol{{\pi}}_{k}^{\text{(sc)}} & =\sum_{\zeta,j=1}^{3}\sqrt{2k_{B}T\tilde{D}_{\zeta}\gamma_{c}}\,(\partial_{\mathbf{\mathbf{\phi}}_{k}}R)_{j,\zeta}\,dZ_{\zeta,j},\label{eq:Pinc2}
\end{alignat}
where $k_{B}$ is Boltzmann's constant, $T$ is the temperature of the gas, and $\tilde{D}_{\zeta}=\frac{1}{2}\text{tr}(I)-I_{\zeta}$ ($I$ is the moment of inertia tensor). $V_{k}$ and $Z_{\zeta,j}$ are zero mean independent Wiener processes, which are defined to model translational and rotational noises, respectively (in place of $dX$ and $dP$ introduced in Eq.~\eqref{eq:dZ}). One can show that Eqs.~\eqref{eq:Pnc2} and \eqref{eq:Pinc2}  give rise to the same noise correlation functions that also emerge from the classical modeling (see Eqs.~\eqref{eq:stoc1} and \eqref{eq:stoc2}, respectively). This shows that the classical limit of the dissipative Caldeira-Leggett model yields a reasonable classical model. 


\section{Correlated noises\label{app:Correlated-noises}}

In this appendix we briefly illustrate how to transform the equations of motions of $M$ degrees of freedom (vector $\bm{S}$), but containing $N$ independent noise terms (vector $d\bm{W}$), to a description with $M$ independent noises (vector $d\bm{V}$). Note that the calculation in this section is generic and the letters used are not associated with the notation used in other parts of the text.

Consider the $M$-dimensional
vector of zero-mean $\mathbb{R}$-valued independent Wiener processes
$\boldsymbol{W}$ and the $M$-dimensional vector process $\boldsymbol{S}$
defined as
\begin{equation}
d\boldsymbol{S}=Gd\boldsymbol{W},
\end{equation}
where $G$ is a $M\times N$ matrix with $\mathbb{R}$-valued elements.
We then define the $M$ dimensional covariance matrix $\sigma$ associated
to the process $\boldsymbol{S}$:

\begin{equation}
\sigma dt=\mathbb{E}[dS\otimes dS]=GG^{\top}dt,\label{eq:cholesky}
\end{equation}
where $\mathbb{E}$ denotes the average over different realization
of the noises. We now consider the Cholesky decomposition of the correlation
matrix $\sigma$, i.e. $\sigma=CC^{\top}$, where $C$ is a $M\times M$
matrix. Using this decomposition we have

\begin{equation}
d\boldsymbol{S}=Cd\boldsymbol{V},
\end{equation}
where $\boldsymbol{V}$ is a $M$-dimensional vector of zero-mean
$\mathbb{R}$-valued independent Wiener processes. The Cholesky decomposition can be implemented numerically or with a symbolic manipulation software once we provide the correlation matrix $\sigma$~\cite{trefethen2022numerical}.

For example, such a situation occurs for the photon recoil terms given by Eqs.~(\ref{eq:Pns}) and (\ref{eq:Pins}), where we have an infinite number of independent noises. With the procedure outlined above one can reduce the description to six independent noises (i.e., one noise per degree of freedom of the nanoparticle).

\section{Higher order approximation for the tweezer field}\label{app:gauss}

At the focus of a diffraction-limited beam, the asymmetry for the waists means that the typical $1/e$ circumference of a Gaussian beam is transformed into an ellipse. Assuming linear polarization, for example, the ellipse is always aligned such that its long axis is along the polarization vector regardless of the vector orientation in the polarization plane. When considering elliptical polarization, the orthogonal components of the electric field will have the waist ellipse orthogonally oriented as well. Now, the definition of the tweezer field in Eq.~\ref{eq:Ed} assumes that the vector and scalar part of the electric field can be factored. This greatly simplifies the notation but fail to represent a basic behavior, i.e., that for circular polarization, the trap frequencies become degenerate. A better approximation of the tweezer field requires defining two mode functions as
\begin{equation}
\begin{split}
u_x(\boldsymbol{r})&=\frac{w_{0}}{w(z)}\text{exp}\left(-\frac{x^{2}/ a_{1}+y^{2}a_{1}}{w(z)^{2}}\right)e^{-i(k_L z +\Phi)}\\
u_y(\boldsymbol{r})&=\frac{w_{0}}{w(z)}\text{exp}\left(-\frac{x^{2} a_{1}+y^{2}/a_{1}}{w(z)^{2}}\right)e^{-i(k_L z+\Phi)},\label{eq:umodexy}
\end{split}
\end{equation}
\noindent where we have also added the Gouy phase $\Phi$ which is given by
\begin{equation}
    \Phi\simeq\arctan(z/z_{R})-\frac{k_L z}{2}\frac{x^2+y^2}{z^2+z_R^2}.
\end{equation}

The tweezer field now reads
\begin{equation*}
    \bm{E}_d= i\sqrt{\frac{P}{2\epsilon_0c\sigma_L}}\left[\begin{pmatrix}b_x u_x\\i b_y u_y\\0\end{pmatrix}e^{i\omega_L t}-\begin{pmatrix}b_x u_x^*\\-i b_y u_x^*\\0\end{pmatrix} e^{-i\omega_L t}  \right].
\end{equation*}

To obtain the trap frequencies we expand the Hamiltonian of Eq.~\ref{eq:gradient} to second order in $q$ around the steady state value $q_s=(x_s,y_s,z_s,\alpha_s,\beta_s,\gamma_s)$, so that we have $H_{\text{gradient}}=\sum_i g_i\, q_i+\sum_{ij} g_{ij} \,q_i q_j$. The deterministic scattering force must also be expanded to an equivalent order again around $q_s$, so that we can write its contribution to the equation of motion as $\dot{\mathsf{p}}_i=\kappa_i+\sum_j \kappa_{ij} \, q_j$. Naturally, for both expansions not all coefficients will be non-vanishing and all are dependent on $q_s$. The trap frequencies are then given by
\begin{equation}
    \omega_q=\sqrt{\frac{2 g_{xx}-\kappa_{xx}}{m_q}}.\label{eq:zs}
\end{equation}
The steady state value is obtained by solving $g_q=\kappa_q$ for $q_s$.

We will now write explicitly the steady state solution and the trap frequencies for an asymmetric top particle under the assumption that the tweezer field is not sufficiently elliptic to spin the particle. For the parameters explored in Sec.~\ref{sec:Numerical-analysis} the only relevant non-vanishing steady state values are a displacement along $z$ due to the deterministic scattering force and a rotation of $\beta$ to $\pi/2$ to bring $\chi_3$ to the polarization plane. Then, the steady state vector simplifies to $q_s=(0,0,z_s,0,\pi/2,0)$ with $z_s$ given by
\begin{equation}
    z_s=\frac{6 \pi \left(\chi_3 b_x^2+\chi_2 b_y^2\right)-\frac{\sqrt{\Delta }}{2}}{k^4 V \left(\chi_2^2 \left(1-b_x^2+b_y^2\right)+\chi_3^2 \left(1+b_x^2-b_y^2\right)\right)}
\end{equation}
\noindent where
\begin{equation}
\Delta=144 \pi \left(\chi_3 b_x^2+\chi_2 b_y^2\right){}^2-4 k^7 V^2 z_R \left(k z_R-1\right) \left(\chi_2^2 \left(1-b_x^2+b_y^2\right)+\chi_3^2 \left(1+b_x^2-b_y^2\right)\right){}^2.
\end{equation}
The trap frequencies are then
\begin{equation}
    \begin{split}
     \omega_x^2 &=\frac{I_{0}}{c\,\rho}\left( \frac{2 z_R^4 \left(a_1^2 \chi _2 b_y^2+\chi _3 b_x^2\right)}{a_1 \omega _0^2 \left(z_R^2+z_s^2\right){}^2}-\frac{k^4 V \chi _2^2 \chi _3^2 z_R^2 z_s \left(b_x^2-b_y^2+1\right) \left(-b_x^2+b_y^2+1\right)}{12
   \pi  \left(z_R^2+z_s^2\right){}^2}  \right)\\
   \omega_y^2 &=\frac{I_{0}}{c\,\rho}\left(\frac{2 z_R^4 \left(a_1^2 \chi _3 b_x^2+\chi _2 b_y^2\right)}{a_1 \omega _0^2 \left(z_R^2+z_s^2\right){}^2}-\frac{k^4 V \chi _2^2 \chi _3^2 z_R^2 z_s \left(b_x^2-b_y^2+1\right) \left(-b_x^2+b_y^2+1\right)}{12
   \pi  \left(z_R^2+z_s^2\right){}^2}   \right)\\
   \omega_z^2 &=\frac{I_{0}}{c\,\rho} \frac{z_R^2}{6 \pi  \left(z_R^2+z_s^2\right){}^3}\bigg(  k^3 V \left(\chi _2^2+\chi _3^2\right) z_s \left(k \left(z_R^2+z_s^2\right)-2 z_R\right)-3 \pi  \left(\chi _2+\chi _3\right) \left(3 z_s^2-z_R^2\right) \\
   &\,\,\,\,\,\,+\left(\chi _2-\chi _3\right) \left(b_y^2-b_x^2\right) \left(k^3 V z_s \left(\chi _2+\chi _3\right)  \left(k \left(z_R^2+z_s^2\right)-2 z_R\right)+3 \pi  \left(z_R^2-3 z_s^2\right)\right)\bigg)\\
   \omega_\alpha^2 &=\frac{I_{0}}{c}\left(\frac{V \left(\chi _3-\chi _2\right) z_R^2 \left(b_x^2-b_y^2\right)}{\text{J1}
   \left(z_R^2+z_s^2\right)}   \right)\\
    \omega_\beta^2 &=\frac{I_{0}}{c}\left(\frac{V \left(\chi _3-\chi _1\right) b_x^2 z_R^2}{\text{J2} \left(z_R^2+z_s^2\right)}   \right)\\
    \omega_\beta^2 &=\frac{I_{0}}{c}\left( \frac{V \left(\chi _2-\chi _1\right) b_y^2 z_R^2}{\text{J3} \left(z_R^2+z_s^2\right)}  \right).\\
    \end{split}\label{eq:trapappendix}
\end{equation}

Expressions for the other geometries considered can be obtained from Eq.~\ref{eq:zs}-\ref{eq:trapappendix} by taking the relevant limit for the susceptibilities, i.e., $\chi_2\rightarrow\chi_1$ for a prolate ellipsoid (rod-like) and $\chi_2\rightarrow\chi_3$ for an oblate ellipsoid (disk-like).

\bibliography{ref}

\end{document}